\documentclass[12pt]{article}

\usepackage{amscd}
\usepackage{amsmath}
\usepackage[cp866]{inputenc}
\usepackage[russian]{babel}

\begin{document}

\thispagestyle{empty}

\section*{V.M. RED'KOV \\
On solutions of  Schr\"{o}dinger and Dirac equations in spaces of
constant curvature, spherical  and elliptical   models}

\begin{quotation}

Exact solutions of the Schr\"{o}dinger and Dirac equations in generalized
cylindrical coordinates of the 3-dimensional space of positive constant curvature,
spherical model, have  been obtained. It is shown that all basis Schr\"{o}dinger's  and Dirac's
wave functions are finite, single-valued, and continuous everywhere in spherical space model
$S_{3}$. The used coordinates $(\rho, \phi,z)$ are simply referred to Eiler's
angle variables $(\alpha, \beta, \gamma)$, parameters on the unitary group $SU(2)$,
 which permits to express the constructed wave solutions $\Psi (\rho, \phi,z)$
   in terms of Wigner's
 functions $D_{mm'}^{j}(\alpha, \beta, \gamma)$.
Specification of the analysis to the case of
elliptic, $SO(3.R)$ group space, model  has been done.
In so doing, the results substantially depend upon the spin of the particle.
In scalar case, the part of the Schr\"{o}dinger wave solutions must be excluded
by continuity considerations, remaining functions are continuous everywhere
in the elliptical 3-space. The latter is in agrement  with the known
statement: the Wigner functions $D_{mm'}^{j}(\alpha, \beta, \gamma)$ at $j =0,1,2,...$
make up a correct  basis in $SO(3.R)$ group space.
For the fermion case, it is shown that no Dirac solutions, continuous everywhere in elliptical
space, do exist. Description of the Dirac particle in elliptical space of positive
constant curvature cannot be correctly in the sense of continuity adjusted with its
topological structure.

\end{quotation}

\underline{Key words:}
 spherical, elliptical, geometry,   Shcr\"{o}dinger and
Dirac equation,

Wigner functions, Eiler angles, continuity, curvature, spin, topology.

\subsection*{1. Introduction  }

\hspace{5mm}
The main quantum mechanical equations
(say, for definiteness, scalar Schr\"{o}dinger and fermion
Dirac ones)  are assumed to have the corresponding wave functions
be given in the whole space. So, the global structure of the space model, its topological
properties, may be of substantial  significance in the quantum mechanical context.

In the present paper, two canonical examples, Schr\"{o}dinger's and Dicac's equations on the
manifolds of the unitary $SU(2)$ and orthogonal $SO(3.R)$ groups have been examined in detail.
These manifolds provide us with topologically different variants of the most simple
Riemannian space,  the 3-space of  constant positive  curvature, spherical
 $S_{3}$  and elliptical  $\tilde{S}_{3}$ models respectively.
These two groups,  $SU(2)$ and  $SO(3.R)$, are widely used  in many applications,
in the same time the groups give simple and familiar realizations
of these two geometrical models, that can be followed in detail by quite
elementary mathematical  tools.  So, in these simple systems, the influence of the topology
in quantum mechanics can be traced in  terms of exact solutions.

In the paper, solutions of the main equations  in the space $S_{3}$ and $\tilde{S}_{3}$
will be constructed in the coordinates  $(\rho,\phi,z)$ which can be considered as
extension of the familiar cylindrical coordinates in flat space.
These curved coordinates $(\rho,\phi,z)$ can be referred to the Eiler angle variables:
$
\alpha  = \phi  + z\; , \;\; \beta  = 2 \rho \; , \;\; \gamma  =
\phi  - z \; $.
The latter gives possibility to express any  given wave solution
(Schr\"{o}dinger's or Dirac's) in terms of Wigner's functions
 $D_{mm'}^{j}(\alpha, \beta, \gamma)$.

There exists one other point making  the quantum mechanical  problem under consideration
of interest in connection with the old and eminent theorem in the  theory of compact groups.
The  theorem by Peter and Weyl ([1], see its
formulation in [2] states:

\begin{quotation}

{\it Теорема I:
Let $\hat{G} = \{ T^{s} \}$  be the set of all irreducible unitary
representations
$
\sqrt{d_{s}}\; D^{s}_{jk}(g) \; ,\;
s \in  \hat{G} \; , \;  1 \le  j,k \le  d_{s} \; ,
$
$d_{s}$  stands for the dimension of the representation.
Matrix elements    $D^{s}_{jk}$  of the representation  $T^{s}$
make up the full orthogonal basis in $L^{2}(G)$.  For any function
$U(g) \in  L^{2}(G)$  there exists decomposi\-tion
$$
U(g) = \sum_{s \in  \hat{G}}  \sum_{j,k=1}^{s}\;
 C^{s}_{jk} \; D^{s}_{jk}(g)  \; , \;\;
\int_{G} \mid U(g) \mid ^{2 } dg  = \sum _{s \in  \hat{G}} d_{s}
 \sum_{j,k=1}^{d_{s}} \mid C^{s}_{jk} \mid ^{2}\; .
$$

}

\end{quotation}

\noindent
The Theorem I can be significantly  strengthened.

\begin{quotation}

{\it Теорема II :
Let  $f(g)$ be a function continuous in $G$. For any $\epsilon  > 0$ there exists
a linear combination of matrix elements such that
$$
\mid f(g) - \sum_{ s = 1}^{N_{\epsilon }}  [ \sum_{j,k = 1}^{d_{s} }
C^{s}_{jk} \; D^{s}_{jk}(g)  ] \mid  \;\;
\le  \epsilon \;\;  для \;\; всех \;\; g \in  G \;.
$$
}
\end{quotation}

This Theorem   II  states that any  function $f(g)$  continuous  in  $G$
can be represented by the series in terms of  $D^{s}_{jk}(g)$, and this series
coincides with  the initial function $g(g)$ in all the points of $G$.
By the Theorem I, the  function $f(g)$, discontinuous somewhere in $G$, can be
exactly approximated by the series only at the points of continuity.

It should be noted that in spite of the wide use the Eiler's parameters in the theory of
groups $SU(2)$ and $SO(3.R)$ in the  literature there exist  full confusion in pointing out
a domain in which the  variables $(\alpha, \beta, \gamma)$  should change
to be correct in context of the
Peter-Weyl theorem II -- see [3-20].
In that sense, the most interesting analysis of the problem  had been given in [21].

As a rule, they operate only  with the explicit form of matrix elements
$D^{s}_{jk}(\alpha, \beta, \gamma)$ and with the orthogonality conditions for them.
At this any clear distinction between different  variants $G(\alpha, \beta, \gamma)$
has not bee achieved. In the literature we can see each of the following domains:
$$
 G = 2 \pi \otimes \pi \otimes \pi \; , \qquad
 G =  \pi \otimes \pi \otimes 2\pi \; , \qquad
 G = 2 \pi \otimes \pi \otimes 2\pi\; ,
$$
$$
 G = 4 \pi \otimes \pi \otimes 4\pi \; , \qquad
 G = 2 \pi \otimes \pi \otimes 4\pi \; , \qquad
 G = 4 \pi \otimes \pi \otimes 2\pi\; , \qquad
 G = 2 \pi \otimes 2\pi \otimes 2\pi \; .
$$

 As a rule,  without any special examining, it is tacitely assumed that
 all functions involved  are continuous.
 In the  essence, two different concepts, a function continuous in
 the manifold $SU(2)$ or $SO(3.R)$,  and a function
 continuous with respect to  three variables $(\alpha, \beta, \gamma)$, are not
distinguished. Till now, no special treatment joining different points of view
does not exist.  So  other additional aim of the present work is to give such an
analysis that  might give  possibility to use the Peter-Weyl theorem II correctly.
In the same time, the question of parametrization of unitary and orthogonal groups by Eiler variables
must be fully  clarified in the framework of the  main goal of the present work --
to find exact solutions of Schr\"{o}dinger and Dirac equations in the  spaces of
constant positive curvature. Because it is evident in advance that the
main requirements for these are their finiteness and continuity in all the points of
spherical   $S_{3}$ and elliptic  $\tilde{S}_{3}$ spaces.

In the beginning (\S 1 - \S 2) we are consider solutions of the scalar Schr\"{o}dinger
equation in the spherical space
$$
n_{0}^{2} + n_{1}^{2} +  n_{1}^{2} + n_{1}^{2} = 1 \; , \qquad n_{a}= n_{a}(\rho,\phi,z)\; ,
$$

\noindent
in the the generalized cylindrical coordinates
$$
n_{0} = \cos  \rho  \cos  z \; , \;\;  n_{3} = \cos  \rho  \sin  z
\; , \;\; n_{1} = \sin  \rho  \cos  \phi  \; , \;\; n_{2} = \sin  \rho
\sin  \phi \; ,
$$
$$
G(\rho,\phi,z)\;: \qquad \rho \in [0, \pi/2 ],
\qquad \phi, z \in [-\pi , + \pi ]
$$

\noindent
That cylindrical parametrization has some peculiarities.
The domain $G$ may be  divided into three parts:
$G_{1}, G_{2},G_{3}$ dependently on  the  values of $\rho  \; : \; \rho \in (0, \pi/2),
\rho =0 , \rho =\pi/2$. At this the domains  $G_{2}(0,\phi,z)$ and
$G_{3}(\pi/2,\phi,z)$ correspond to 1-dimensional closed curves:
$$
G_{2}: \qquad n_{0} = \cos z ,  \qquad n_{3} = \sin z , \qquad n_{0}^{2} +n_{3}^{2} = 1 \; ,
$$
$$
G_{3}: \qquad n_{1} = \cos \phi ,  \qquad n_{2} = \sin \phi , \qquad n_{1}^{2} +n_{2}^{2} = 1 \; .
$$

\noindent
Therefore, any  wave function  $\Psi (\rho, \phi,z)$  continuous in the space  $S_{3}$
 must obey a number of definite restrictions:
$$
\left. \begin{array}{ll}
(G_{1}): \qquad &
\Psi ( 0 ,\rho < \pi/2 , \phi, z)=  \Psi (0 ,\rho < \pi/2, \phi \pm 2\pi, z \pm 2\pi)\; ,\\[3mm]
(G_{2}): \qquad & \Psi (0, \phi, z)=  \{ \Psi (z)=\Psi(z \pm 2\pi), \;\; \mbox{or} \;\; const\; \} ,\\[3mm]
(G_{3}): \qquad  & \Psi (0, \phi, z)=  \{ \; \Psi (\phi)=\Psi(\phi \pm 2\pi),
 \;\; \mbox{or} \;\; const\; \} .
\end{array} \right.
$$

Taking the elliptical model  (\S 3) will much change  the all treatment.
As "Cartesian" \hspace{2mm} may be used the known Gibbs [22] coordinates
 $\vec{c}= (c_{1},c_{2},c_{3})$ \footnote{For detailed
and systematic exposition of the theory of the rotation  group in terms of $\vec{c}$-parameters
and much  further extensions see the book [10]. }
$$
O(\vec{c}) =
 I + 2 {\vec{c}^{\;\times}  + (\vec{c}^{\;\times})^{2} \over 1 +
 \vec{c}^{\;2}}   \; ;
$$

\noindent
 as seen to any pair of inversely directed vectors of infinite length corresponds
 the same matrix,
 point in elliptic space:
$$
\vec{c}^{\;+} = + \infty \; \vec{c}_{0} \; , \qquad
\vec{c}^{\; -} = - \infty \; \vec{c}_{0} \; , \qquad \vec{c}^{\;2}_{0} = 1\; , \;
O(\vec{c}^{\;+}) = O(\vec{c}^{\; -}) =  I + 2 (\vec{c}^{\;\times}_{0})^{2} \; .
$$

Cylindrical parametrization of the space $\tilde{S}_{3}$
has its own peculiarities. The domain $\tilde{G}(\rho, \phi,z)$ should be
decrease twice  as much  and new identification rule for the boundary points  should be
accepted (here only one way to chose the domain $\tilde{G}$ is detailed):
$$
 \mbox{Identification in }  \;\;
\tilde{G}^{(+1)}_{1}(\phi,z) : \;  \rho \neq 0, \pi/2
$$


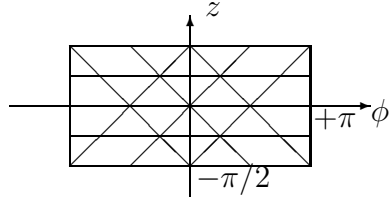
\unitlength=0.4 mm
\begin{picture}(160,60)(-140,-30)
\special{em:linewidth 0.4pt} \linethickness{0.4pt}

\put(-60,0){\vector(+1,0){120}}
\put(+60,-5){$\phi $} \put(0,-30){\vector(0,+1){60}}
\put(+5,+30){$z$}

\put(+41,-7){$+\pi$} \put(-40,-20){\line(+1,0){80}}
\put(-40,-20){\line(0,+1){40}} \put(+40,+20){\line(-1,0){80}}
\put(+40,+20){\line(0,-1){40}} \put(+40,+20){\line(-1,-1){40}}
\put(-40,-20){\line(+1,+1){40}} \put(-20,-20){\line(+1,+1){40}}
\put(-40,+20){\line(+1,-1){40}}
\put(-20,+20){\line(+1,-1){40}} \put(0,+20){\line(+1,-1){40}}
\put(-40,-10){\line(+1,0){80}}
\put(-40,+10){\line(+1,0){80}}

\put(+2,-27){$-\pi /2$}

\end{picture}
$$
\left. \begin{array}{llll}
\tilde{G}_{2}(0,\phi,z): \qquad  & \rho =0 \; , \qquad  & \vec{c} = (0,0, \tan z ),
\qquad  & \phi -\mbox{немая} \; ;\\[2mm]
\tilde{G}_{3}(0,\phi,z): \qquad  & \rho =\pi/2\;  , \qquad  &\vec{c} =
\infty \; (\cos \phi, \sin \phi, 0 ),
\qquad &
 z -\mbox{mute } \; .
 \end{array} \right.
$$

\noindent The requirement of continuity in  elliptic
space $\tilde{S}_{3}$  must be agreed   with these identification rules.
In that way it is shown that the obtained  solutions of Schr\"{o}dinger equation in
$S_{3}$  can be divided into two classes as follows:
the first type functions turn out to be discontinuous in space $\tilde{S}_{3}$ and therefore
they should be excluded from the list of "good"\hspace{2mm} basis functions.
The second type solutions are continuous both in spherical and elliptic spaces
and they make up the full basis system in  $\tilde{S}_{3}$.
In view of referring the coordinates  $(\rho, \phi,z)$
to Eiler variables  $SU(2)$ и $S0(3.R)$, the above can be formulated (\S 4 -\S 5) as follows:
basis for continuous functions on the group  $SU(2)$ consists of
$D^{j}_{m\sigma}(\alpha, \beta, \gamma)$ with  $j = 0,1/2,2,3/2,...$;
basis of  continuous functions on the group  $S0(3)$ consists of
 $D^{j}_{m\sigma}(\alpha, \beta, \gamma)$
with  $j=0,1,2,...$.

It is natural to examine in analogous way the case of Dirac particle
in spherical and elliptical spaces. In  \S 6
the Dirac equation is specified in spherical space; cylindrical coordinates and tetrad are used.
The wave solutions  $\Psi_{\epsilon,\lambda,m,k}(t,\rho, \phi,z)$, eigenfunction of four operators
have  been looked for.
 The  problem is reduced to a linear system of radial equations for
two functions  $f_{1}(\rho),\; f_{2}(\rho)$.
In    \S 7 a special transformation is found
$
f_{1}(\rho),\; f_{2}(\rho)   \; \Longrightarrow  \; G_{1}(\rho),\; G_{2}(\rho)\; ,
$
which permits to solve the problem  for   $G_{1}(\rho),\; G_{2}(\rho)$
in terms of hypergeometric  functions.
Further (\S 8- \S 11 )  the quantization rules for $\epsilon, \lambda,m,k$
are to be derived and corresponding wave functions are to be constructed.

The main requirement is: these wave functions must be finite and continuous
in the space $S_{3}$.
The formulation of the latter condition in the curved space
is much more complex in case of fermion. The  matter is that besides
the simple technical complication --  the number of components of a bispinor is four --
the mathematical formalism itself,
by Tetrode-Weil-Fock-Ivanenko [23-25],
involves an additional local gauge transformations -- for the case under consideration
those are local tetrad $SU(2)$ rotations. These  gauge  transforms connected with the freedom in
specifying tetrads  are singular so that the explicit form of continuity requirement depends
upon the occasional choice of the tetrad.

Strictly speaking, the same problem is in flat space as well. Indeed, the wave functions
being single-valued in Cartesian tetrad  do not look single-valued in the cylindrical
tetrad. Therefore, if one starts with that cylindrical tetrad one will face the problem --
which solutions are continuous and which ones are  not.

In the case of spherical space $S_{3}$ the all situation is more complex
 though principally the same.
The role of a Cartesian tetrad basis in $S_{3}$  belongs to the
conformally flat tetrad. In Supplement A, an explicit form of the spinor  gauge matrix
$B(\rho,\phi,z)$ referring cylindrical and conformally-flat tetrads is calculated.
The main point is that the fermion wave function may be discontinuous  in the cylindrical basis
but it must become continuous in conformally-flat basis.

In particular, because the gauge matrix $B(\rho,\phi,z)$ contains the trigonometrical functions
of the half angles $\phi/2$ and$z/2$ the quantum  numbers
  $m$ and  $k$  must take the  the half-integer values.
 In quantum number $\lambda$, the proper value of the curved helicity operator, the energy
 levels are twice degenerated: $
\epsilon = \sqrt{M^{2} + \lambda^{2}} \; .
$
Finally,  all hypergeometric function involved become polynomials when
the $\lambda$ is quantized as follows:
$
\lambda = \pm 3/2, 5/2, 7/2, ...
$
In \S 11 the special analysis, showing that all the constructed fermion wave functions
are continuous in all the space $S_{3}$, is given.
In addition (\S 9), all the solutions with
 $\lambda = \pm 1/2$   must be excluded by continuity reason.

In  \S 12 the Dirac equation in elliptic space $\tilde{S}_{3}$ has been considered.
Application  of the tetrad formalism in this model has its own peculiarities.
In particular, because of ($2 \rightarrow 1 $)-correspondence between points of
spherical and elliptic manifolds any tetrad given in the space $S_{3}$ provides us with two
different tetrads in elliptical space $\tilde{S}_{3}$. For instance,
the conformally flat retrad after transforming to $c^{i}$-coordinates in $\tilde{S}_{3}$
lead us to a pair of those

$$
e^{(\delta)}_{(i)j}(c)   ={ \delta_{ij} \over  \delta \sqrt{1+ c^{2} }} -
{  c^{i} c^{j} \over (1+c^{2}) \;  (1 + \delta \sqrt{1+c^{2} } ) } ,
 \qquad \delta = \pm 1 \; \;,
$$

\noindent Such a doubling is a consequence of the following:
elliptic manifold  can be constructed  from spherical one by two ways:
on the  base of the  half-space $S_{3}$ containing the point $n_{a}=(+1,0,0,0)$ or
on the base of the half-space $s_{3}$ containing another  point  $n_{a}=(-1,0,0,0)$.

In the same manner,  cylindrical coordinates and tetrads of spherical space produce
a pair of coordinate systems and tetrad in elliptic space which manifests  itself as
freedom to use any of two different domains:
$
(\phi,z) \in \tilde{G}^{(+1)}(\phi,z)$ and $(\phi,z) \in  \tilde{G}^{(-1)}(\phi,z) $.
We have been calculated the tetrad gauge matrix  $L_{(\delta)}$ referring
cartesian and cylindrical tetrads, then calculate the corresponding
spinor gauge matrices (for both cases  $\delta = +1,-1$):
$$
L_{(\delta)}  (\rho, \phi, z) \;e^{(\delta)}_{cart} (c) =  e^{(\delta)}_{cyl} (\rho, \phi, z)
\;,
\qquad
\psi^{(\delta)}_{cart} = \pm B_{(\delta)} ^{-1} (\rho, \phi, z) \; \psi^{(\delta)}_{cyl} \;.
$$

\noindent
So there exist  two cartesian representations for fermion wave  functions
and any of these  can be used in considering the  requirement of continuity
for  fermion wave  functions\footnote{Two other possibilities resulting from combination
of the above two have been considered as well.}.

It is shown that all Dirac wave solutions  constructed continuous in  spherical space
model become discontinuous when specifying them to elliptic  model.
Therefore, description of the Dirac particle in elliptical space of positive
constant curvature cannot be correctly in the sense of continuity adjusted with its
topological structure.

The paper is divided into three parts. In the first part (\S 2 -\S 5), we consider Schr\"{o}dinger
equation, in spherical and then in elliptical space models. In the second part (\S 6 - \S 11)
construct  exact solutions of the Dirac equation on the sphere $S_{3}$, all the functions
founded are continuous in the space.  Third part  (\S 12 -\S 13)
treats  the case of Dirac equation in elliptical space. It is shown that
no Dirac solutions, continuous everywhere in elliptical space, do exist.

\begin{center}
{\bf PART I. SHCR\"{O}DINGER EQUATION IN SPHERICAL AND ELLIPTICAL SPACES,
EILER PARAMETRIZATION AND WIGNER $D$-FUNCTIONS}

\end{center}

\subsection*{2.  Schr\"{o}dinger equation in spherical space  $S_{3}$.}

\hspace{5mm}
The cylindrical coordinates  on the sphere  $S_{3}$ are defined by
$$
n_{0} = \cos  \rho  \cos  z \; , \;\;  n_{3} = \cos  \rho  \sin  z
\; , \;\; n_{1} = \sin  \rho  \cos  \phi  \; , \;\; n_{2} = \sin  \rho
\sin  \phi \; ,
$$
$$
dl^{2} = [\; d \rho ^{2}  + \sin ^{2} \rho  d \phi ^{2} + \cos
^{2} \rho  dz^{2} \; ] \;  , \qquad  \rho  \in  [ 0, \pi /2 ] \; , \;
\phi , z \in  [ - \pi  ,   \pi  ] \;  .
\eqno(2.1)
$$

\noindent For a small part of the space  $S_{3}$
in vicinity of  $n_{a} = (1, 0, 0, 0 ) $  the quantities  $(\rho , \phi , z)$
become ordinary (flat) cylindrical coordinates.
The   Schr\"{o}dinger  generally covariant  equation
$$
H\; \Phi  = \epsilon \; \Phi  \; , \;\; H = - { 1 \over 2 } \; {1
\over \sqrt{g}}\; (\partial /\partial x^{i}\; \sqrt{g}\; g^{ij}
\partial / \partial x^{j} ) \; ,
$$

\noindent where $\epsilon  = E/(\hbar ^{2} / MR^{2})$ stands for the particle   energy,
being specified in the system (2.1) will take the  form
$$
[\; \sin ^{-1} \rho \; \cos ^{-1} \rho \; (\partial _{\rho} \sin
\rho \; \cos \rho \; \partial_{ \rho} )  + \sin ^{-2} \rho\;
\partial ^{2}_{\phi} +
 \cos ^{-2} \rho \;  \partial ^{2}_{z} + 2 \epsilon  \; ]\;
 \Phi  = 0     \; .
\eqno(2.2a)
$$

\noindent Solutions can be searched as cylindrical waves
$
\Phi (\rho ,\phi ,z) = e^{+i M \Phi } \; e^{+iKz} \; R(\rho ) \; ;
$
for  $R(\rho )$  one has the equation
$$
[ \; d^{2} /d\rho ^{2} + (\cos \rho / \sin  \rho  - \sin  \rho / \cos
\rho ) d/d\rho  +
$$
$$
+ \;  2 \epsilon  - M^{2} / \sin ^{2} \rho - K^{2}/\cos ^{2} \rho \;
] \; R(\rho ) = 0 \; . \eqno(2.2b)
$$

\noindent The function  $R(\rho )$  reduces to hypergeometric one
$$
R(\rho ) = \sin ^{a} \rho  \; \cos ^{b} \rho  \;  F ( A , B , C ;
\cos ^{2}\rho  ) \; , \;\; \eqno(2.2в)
$$
$$
A = ( a + b + 1 - \sqrt{2 \epsilon  + 1 } ) / 2  \; , \; B = ( a +
b + 1 + \sqrt{2 \epsilon  + 1} ) / 2 \; , \; \; C = ( b + 1 ) \; .
$$

\noindent The wave function  $\Phi (\rho ,\phi ,z)$  will be  single-valued in
$S_{3}$ (in the part $G_{1}$)  when
$$
M  =  0, \pm 1, \pm 2, \ldots  \; ,  \;\; N  =  0, \pm 1,
\pm 2, \ldots     \; , \;\; a  =  + \mid M \mid \; , \;\; b  =
\mid K \mid \;  . \eqno(2.3a)
$$

\noindent The wave function will be finite if the hypergeometric function becomes  a
polynomial:
$$
A  =  - n  , \;\;  n = 1, 2, 3,\ldots  \; ,
$$

\noindent which  provides us with the spectrum for energy:
$$
\epsilon _{N} = {1 \over 2} \; (N^{2}- 1)\; , \;\; N = a + b + 1
+ 2 n \; ; \eqno(2.3b)
$$

\noindent where the number  $N$  takes on the values  $+  1,  +  2,  + 3,
\ldots $. The energy level are degenerated
as much as $g_{N} =
N^{2}$. For instance, let  $N = 3$  then  $ g_{3} =9 $:
$$
\left. \begin{array}{lllll}
a = 0 &   b = 2   &  M =  0    & K = \pm 2  &  \\
a = 1 &   b = 1   &  M = \pm 1 & K = \pm 1  &  n = 0 \\
a = 2 &   b = 0   &  M = \pm 2 & K =  0   &    \\ [3mm] a = 0 &
b = 0   &  M =  0    & K =  0   & n = 1 \; \; .
\end{array} \right.  \;
$$

\noindent Now, let us show that the constructed wave  functions
$$
\Phi _{\epsilon MK}\;  = \; C^{\epsilon }_{MK} \; e^{+iM\phi } \;
e^{+iKz}\;  \sin ^{+\mid M \mid} \rho  \; \cos ^{+\mid K \mid}
\rho  \;\;  F (A, B, C; \cos  ^{2} \rho  ),
\eqno(2.4)
$$

\noindent are continuous in space $S_{3}$.
The whole space $S_{3} -G(\rho ,\phi ,z )$   let be divided into three parts:
$ G = (G_{1} \cup G_{2} \cup  G_{3})$:
$$
G_{1} = G( 0< \rho  < \pi /2) \; , \;\; G_{2} = G ( \rho  = 0
)\; \; , \;\; G_{3} = G ( \rho  = \pi /2 )\;  . \eqno(2.5)
$$

\noindent For the part $G_{1}$ which can be sketched graphically as
\begin{center}
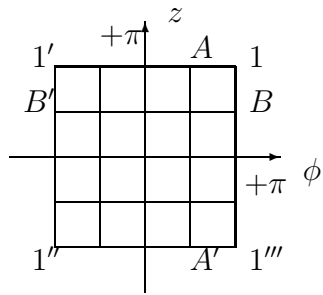
\mbox{Fig} \;  1  \;\mbox{The domain}  $\;\; G_{1}$
\end{center}

\unitlength=0.6 mm
\begin{picture}(160,60)(-115,-25)
\special{em:linewidth 0.4pt} \linethickness{0.4pt}

\put(-30,0){\vector(+1,0){60}} \put(+35,-5){$\phi $}
\put(0,-30){\vector(0,+1){60}} \put(+5,+30){$z $}
\put(-20,+10){\line(+1,0){40}} \put(-20,+20){\line(+1,0){40}}
\put(-20,-10){\line(+1,0){40}} \put(-20,-20){\line(+1,0){40}}
\put(-20,+20){\line(0,-1){40}} \put(-10,+20){\line(0,-1){40}}
\put(+10,+20){\line(0,-1){40}} \put(+20,+20){\line(0,-1){40}}

\put(-27,+10){$B' $}       \put(+23,+10){$B$} \put(+10,-25){$A' $}
\put(+10,+22){$A$} \put(+22,-8){$+\pi $}      \put(-10,+25){$+\pi
$} \put(-25,+20){$1'$}        \put(+23,+20){$1$}
\put(-25,-25){$1''$}       \put(+23,-25){$1'''$}

\end{picture}
\vspace{5mm}

\noindent here  pairs  $(A,A'), (B,B')$  and so on, and also $(1, 1', 1'', 1''')$
represent  respectively only one point in $S_{3}$. It is evident that integer values for
$M$  and  $K$ will lead us to have the same values of $\Phi _{\epsilon MK}$
 in all identified points on the  plane $(\phi ,z)$. So the  wave functions are continuous on
 the  $G_{1}$-part   $n_{a} = (n_{0}, n_{i})$ of $S_{3}$.
The domain $G_{2}$ representing the closed curve
$$
n_{0} = \cos  z \; , \;\; n_{1} = 0 \; , \;\; n_{2} = 0 \; , \;\;
n_{3} = \sin  z\; , \eqno(2.6a)
$$

\noindent
can be  described  by  the scheme
$$
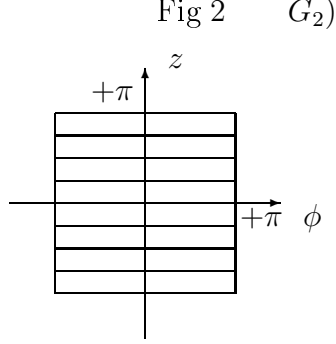
\mbox{Fig} \; 2  \qquad  G_{2}
координата)
$$
\unitlength=0.6mm
\begin{picture}(160,60)(-115,-25)
\special{em:linewidth 0.4pt} \linethickness{0.4pt}

\put(-30,0){\vector(+1,0){60}}    \put(+35,-5){$\phi $}
\put(0,-30){\vector(0,+1){60}}    \put(+5,+30){$z$}
\put(-20,+10){\line(+1,0){40}}

\put(-20,+20){\line(+1,0){40}}    \put(-20,-20){\line(0,+1){40}}
\put(-20,-10){\line(+1,0){40}}    \put(+20,-20){\line(0,+1){40}}
\put(-20,-20){\line(+1,0){40}} \put(-20,+5){\line(+1,0){40}}
\put(-20,+15){\line(+1,0){40}} \put(-20,-5){\line(+1,0){40}}
\put(-20,-15){\line(+1,0){40}}

\put(+21,-5){$+\pi $}             \put(-12,+23){$+\pi $}
\end{picture}

\vspace{5mm}

\noindent So that any function $f(\rho =0,\phi ,z)$ depending on  the mute variable
$\phi $ will be discontinuous in $S_{3}$. Evidently,
the  wave function being specified on $G_{2}$
$$
M \neq  0 \; \Longrightarrow  \; \Phi _{\epsilon MK} = 0 \;\;  ;
$$
$$
M = 0 \; \Longrightarrow  \; \Phi _{\epsilon 0K} = e^{+iKz} \;
F(A, B, C; 1)  = \Phi (n) \; . \eqno(2.6b)
$$

\noindent is continuous  about the curve $G_{2}$. Analogously, for the domain $G_{3}$
$$
n_{0}  = 0 \; , \; n_{1}  = \cos  \phi \; , \; n_{2} = \sin  \phi
\; , \; n_{3}  = 0 \; ; \eqno(2.7a)
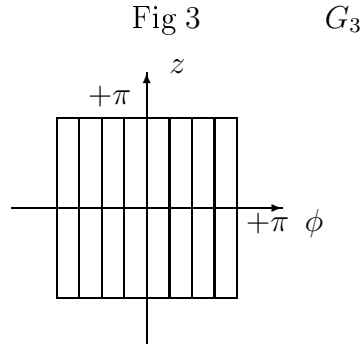
$$
$$
\mbox{Fig}\;  3 \qquad \qquad G_{3}
$$

\unitlength=0.6mm
\begin{picture}(160,60)(-105,-25)
\special{em:linewidth 0.4pt} \linethickness{0.4pt}

\put(-30,0){\vector(+1,0){60}}    \put(+35,-5){$\phi $}
\put(0,-30){\vector(0,+1){60}}    \put(+5,+30){$z$}

\put(-20,-20){\line(0,+1){40}}     \put(-20,-20){\line(+1,0){40}}
\put(-15,-20){\line(0,+1){40}}     \put(-20,+20){\line(+1,0){40}}
\put(-10,-20){\line(0,+1){40}} \put(-5,-20){\line(0,+1){40}}
\put(+5,-20){\line(0,+1){40}} \put(+10,-20){\line(0,+1){40}}
\put(+15,-20){\line(0,+1){40}} \put(+20,-20){\line(0,+1){40}}

\put(+22,-5){$+\pi $}              \put(-13,+23){$+\pi $}

\end{picture}

\vspace{5mm}

\noindent  and
$$
K \neq  0 \; \Longrightarrow  \;  \Phi _{\epsilon MK}  = 0 \;\;
$$
$$
\;\; K = 0 \; \Longrightarrow  \; \Phi _{\epsilon M0} = e^{+iM\phi
} \; F(A, B, C; 0) = \Phi (n) \; . \eqno(2.7b)
$$

Now let us consider the orthogonality condition for
the wave function   $\Phi _{\epsilon MK}(\rho ,\phi ,z)$.
To this end, one should instead of $\rho$ coordinate  introduce
another variable $x$ = $\cos  2\rho $ and take into accont
defining relations for Yacobi  polynomials  [17]
$$
P^{(a,b)}_{n}(x) = N^{(a,b)}_{n} \; F (- n , a + b + 1 + n, b + 1
; {1 + x \over 2})\;  , \; \; N^{(a,b)}_{n} = (-1) {n (n + b)!
\over b!} \; ,
$$

\noindent then the functions  $\Phi _{\epsilon MK}$  read as
$$
\Phi _{\epsilon MK}(\rho ,\phi ,z)  = {1 \over 2\pi  }\;
e^{+iM\phi }\; e^{+iKz} \;{ C^{\epsilon }_{MK} \over N^{(a,b)}_{n}
} \; {1 \over 2^{a/2}\; 2^{b/2} }\;  (1 - x)^{a/2}\; (1 + x)^{b/2}
\; P_{n} ^{(a,b)}\; .
\eqno(2.8a)
$$

\noindent So we have relationship
$$
\int_{S_{3} } \Phi ^{*}_{\epsilon MK} \; \Phi  _{\epsilon ' M'
K'}\; dV = \delta _{MM'}\; \delta _{KK'} \;  {1 \over 2 ^{a+b} } \; \times
\; ,
$$
$$
\times \; ( C^{\epsilon *} _{MK} C^{\epsilon ' }_{M' K} ) \; { N_{n}
^{(a,b)}   \over N ^{(a',b')}_{n'} } {1 \over 4} \; \int_{-1}^{+1}
(1-x)^{a} (1+x)^{b} \; P^{(a,b)}_{n} \; P^{(a,b)}_{n'} \;dx \; .
\eqno(2.8b)
$$

\noindent Taking in mind the orthogonality
conditions for Yacobi polynomials [17]
$$
\int_{-1}^{+1} (1-x)^{a}\; (1+x)^{b}\; P_{n} ^{(a,b)}\;
P^{(a,b)}_{n'} \;dx  = \delta _{nn'} \; { 2^{a+b+1} (n + a)! (n +
b)!  \over n!(a + b + n)! (a +b + 1 + 2n)} \; , \eqno(2.8c)
$$

\noindent from (2.8b) it follows
$$
\int_{S_{3}} \Phi ^{*}_{\epsilon MK} \; \Phi _{\epsilon ' M' K'}
\;dV  = C\;  \delta _{\epsilon \epsilon' }\; \delta _{MM'}\;
\delta _{KK' }\;   \eqno(2.8d)
$$

\noindent where  $C$ is defined by
$$
C = { \mid \; C^{\epsilon }_{MK} \; \mid^{2} \over \mid \;
N^{(a,b)}_{n} \;\mid ^{2}   } \; {(n + a)! ( n + b)!  \over 2(a +
b + 1 + 2n) n! (a + b + n) }\; .
$$

\noindent Demanding  $C  =  1$  we arrive at the normalized  wave solutions -- see (2.8d):

$$
\Phi _{\epsilon MK} = {e^{iM\phi } \over \sqrt{2\pi }} \; {e^{iKz}
\over \sqrt{2\pi }} \; A^{(a,b)}_{n}\; (1 - x)^{a/2} \; (1 +
x)^{b/2} \; P^{(a,b)}_{n}(x)\; ,
 \eqno(2.9)
$$
$$
A_{n} ^{(a,b)} = 2 \; \sqrt{{ n!(a + b + n)! (a + b + 1 + 2n)
\over (n + a)! (n + b)! 2^{a+b+1} } }   \; .
$$

\noindent It is convenient to have the following table
(restricting ourselves to $N  = 1, 2, 3, 4$):

$$
\left. \begin{array}{lllll}
\underline{N = 1}\;\; & b = 0 & & & \\ [5mm]
a=0 \;\; & \sqrt{1 \; 1}\times & & & \\
& P^{(0.0)}_{0}(x) & & & \\  [8mm]
\underline{N = 2 }  & b = 0   &  b=1 &    &  \\[5mm]
a = 0  &  & \sqrt{ 1\; (1+x)^{1}} \times & & \\
&  &  P^{(0.1)}_{0}(x) & & \\ [5mm]
a = 1  &  \sqrt{(1-x)^{1}\; 1 }\times &  & &  \\
& P^{(1.0)}_{0}(x)  &  &  &
\end{array}\right.
$$
$$
\left. \begin{array}{lllll}
\underline{N = 3 } & b = 0  &  b=1 & b=2 &  \\[5mm]
a = 0 & \sqrt{1 \; 1} \times  &  & \sqrt{1 (1+x)^{2}} \times & \\
& P^{(0.0)}_{1}(x)  &  & P^{(0.2)}_{0}(x)  & \\   [5mm]
a=1 &  & \sqrt{(1-x)(1+x)} \times &  & \\
&  & P^{(1.1)}_{1}(x) & & \\[5mm]
a=2  & \sqrt{(1-x)^{2} \; 1}\times &  &  & \\
& P^{(2.0)}_{0}(x)  &  &  & \\       [8mm]
\underline{N = 4} &  b = 0  &  b=1  & b=2 & b=3 \\[5mm]
a=0  &  & \sqrt{1 (1+x)}\times &  & \sqrt{1 (1+x)^{3}}\times \\
&  & P^{(0.1)}_{1}(x)  &  & P^{(0.3)}_{0}(x) \\   [5mm]
a=1  & \sqrt{(1-x)^{1}\; 1}\times & & \sqrt{(1-x)(1+x)^{2}}\times & \\
& P^{(1.0)}_{1}(x)  & & P^{(1.2)}_{0}(x)  & \\ [5mm]
a=2  &  & \sqrt{(1-x)^{2}(1+x)}\times &   &  \\
&  & P^{(2.1)}_{0}(x)  &   &  \\           [5mm]
a=3 & \sqrt{ (1-x)^{3}\; 1}\times  & & & \\
& P^{(3.0)}_{0}(x) & & &
\end{array}\right.
$$

\begin{quotation}

{\it It should be given special attention  to the  following:
the system of all $\{ \Psi _{\epsilon MK} \}$ can be divided into two parts depending
of  $N$,  even  or odd.  Orthogonality of the functions of opposite classes
is achieved by integration over in $(\phi,z)$-variables.

}

\end{quotation}

\subsection*{3. The case of elliptical space }

\hspace{5mm}
Now let us turn to the elliptical case.
It is a Riemmanian space of constant positive curvature as well, though different in its
topological structure   $S_{3}$  is 1-connected whereas $\tilde{S}_{3}$ is 2-connected.
These space models have simple realizations as parametric manifolds of the unitary and orthogonal
groups,  $SU(2)$ and $S0(3)$:
$$
B =
 \sigma ^{0}\; n_{0}\; - \;i \;\sigma ^{k} \; n_{k} \; ,\qquad
 \mbox{det}\; B= + 1  \;,
\eqno(3.1a)
$$
$$
0(\vec{c}) =  \;I \;+\; 2 \; {\vec{c}^{\;\times } +
(\vec{c}^{\;\times })^{2}
 \over 1 + \vec{c}^{\;2}}\;  , \;\;
(\vec{c}^{\; \times })_{kl}  = - \; \epsilon _{klj} \; c_{j}   \;.
 \eqno(3.1b)
$$

\noindent The matrix  $O(\vec{c}\;)$ represents a point in elliptic space.
To every pair of different vector of infinite length
$
\vec{c}^{\; \pm}_{\infty } = \pm\;  \infty \; \vec{c}_{0}\;  ,  \;\;
\vec{c}^{\; 2}_{0} = 1 \;
$ corresponds one the same matrix, point in  elliptical space
$$
0(\vec{c}^{\; \pm \; \infty  }) = \; I \; + \; 2\; (\vec{c}^{\;
\times }_{0})^{2}  \; . \eqno(3.1c)
$$

\noindent
The mapping  $2\; \Longrightarrow \; 1$  from  $SU(2)$  to
$S0(3)$
$
\{ + n_{a} , \; - n_{a} \} \;    \Longrightarrow \;  c^{i} =
  n_{i} / n_{0}    \;
$ will be used to establish a domain $\tilde{G}$ of cylindrical coordinates
of elliptical model on the  base of the  known that for spherical one.

At searching the wave Schr\"{o}dinger solutions in elliptical space
one does not need to repeat all the calculation, instead it is sufficient
to separate continuous solutions in space $\tilde{S}_{3}$ among these continuous in $S_{3}$.
So, we start with the relationships
$$
c_{1}  = { \tan  \rho  \over \cos  z } \; \cos  \phi  \; , \; c_{2}
= { \tan  \rho \over  \cos  z }  \;  \sin  \phi  \; , \; c_{3} = \tan
z\;  \eqno(3.2a)
$$

 \noindent where the coordinates  $(\rho ,\phi ,z)$ change in the  limits:
$$
 \tilde{G }^{(+1)} = \{ \; \rho  \in  [ 0 , \pi /2 ] \; , \; \phi  \in  [-
\pi  ,\;  + \pi  ]\;  ,  \; z \in  [-\pi /2, +\pi /2 ] \; \}\; .
\eqno(3.2b)
$$

From the very beginning, one point should be emphasized:
the domain  $\tilde{G}$ according to  (3.2b) is the most simple and
evident variant, though not unique several other might be used as well.
For instance, the following  $\tilde{G}'$ is appropriate:
$$
\tilde{G }^{+1)} = \{ \; \rho  \in  [ 0 , \pi /2 ] \; , \; \phi  \in  [-
\pi  ,\;  + \pi  ]\;  ,  \; z \in  [-\pi , -\pi/2 ] \oplus [+\pi /2, +\pi  ] \; \}\; .
\eqno(3.2c)
$$

Transition from elliptical curved model to a flat space model
looks differently for cases $\tilde{G}^{(\delta = \pm 1)}$:

$
\underline{(\delta= +1 )}
$

$$
\rho \rightarrow  0 , z \rightarrow  \pm 0 :
\qquad
 c_{1} = \rho \cos \phi \;, \;\; c_{2} = \rho \sin \phi \; , \;\;  c_{3} = z \;;
$$

$
\underline{( \delta = -1)}
$
$$
 \rho \rightarrow  0 \; , \;\;
z>0\; ,\;  z =\pi - Z \rightarrow \pi  \;  : \qquad
- c_{1} = \rho \cos \phi \; ,  \;\; -c_{2} = \rho \sin \phi \;, \; \;  -c_{3} = Z  > 0  \; ;
$$
$$
 \rho \rightarrow  0\;  , \;\;
z<0 \; ,  \; z = -\pi - Z \rightarrow \pi \; : \qquad
 -c_{1} = \rho \cos \phi \;,\;\;   -c_{2} = \rho \sin \phi \;,\;\;   -c_{3} = Z \; < 0 \;  .
$$

The variant with $\delta = +1$ is simpler and it will be used below\footnote{
However, these alternative variants should be taken in mind; in the  following
they  will help to understand some peculiarity of tetrad formalism
in elliptical space.}.
One should find identification rules adjusted with topological structure of the group
$SO(3.R)$.
Again, the domain   $\tilde{G}$  is to be divided into three parts:
$$
\tilde{G}_{1} = \tilde{G} (\rho  \neq  0 , \pi /2) \;,\;\;
\tilde{G}_{2} = \tilde{G} (\rho = 0) \; , \;\; \tilde{G}_{3} =
\tilde{G} (\rho = \pi /2) \; .
$$

First, let us consider the domain  $\tilde{G}_{1}$.
The topological structure of the group
$SO(3.R)$ leads us to the following identification scheme:
$$
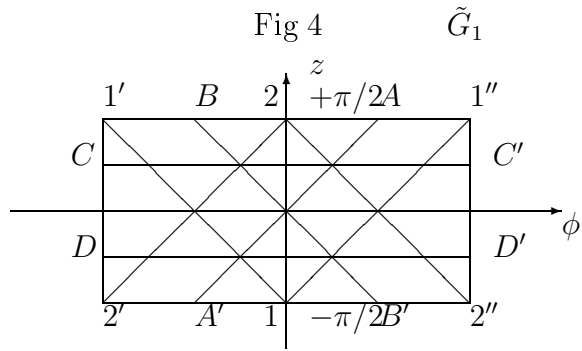
\mbox{Fig}\; 4 \qquad \qquad \tilde{G}_{1}
$$
\vspace{5mm} \unitlength=0.61mm
\begin{picture}(160,60)(-115,-30)
\special{em:linewidth 0.4pt} \linethickness{0.4pt}

\put(-60,0){\vector(+1,0){120}} \put(+60,-5){$\phi $}
\put(0,-30){\vector(0,+1){60}} \put(+5,+30){$z$}

\put(-40,-20){\line(+1,0){80}}
\put(-40,-20){\line(0,+1){40}} \put(+40,+20){\line(-1,0){80}}
\put(+40,+20){\line(0,-1){40}} \put(+40,+20){\line(-1,-1){40}}
\put(-40,-20){\line(+1,+1){40}} \put(-20,-20){\line(+1,+1){40}}
\put(-40,+20){\line(+1,-1){40}}
\put(-20,+20){\line(+1,-1){40}} \put(0,+20){\line(+1,-1){40}}
\put(-40,-10){\line(+1,0){80}}
\put(-40,+10){\line(+1,0){80}} \put(-40,-25){$2'$}
\put(-20,-25){$A'$}   \put(-5,-25){$1$}

\put(+5,-25){$-\pi /2$} \put(+20,-25){$B'$} \put(+40,-25){$2''$}

\put(-40,+23){$1'$}   \put(-20,+23){$B$}   \put(-5,+23){$2$}

\put(+5,+23){$+\pi /2$} \put(+20,+23){$A$} \put(+40,+23){$1''$}

\put(-47,+10){$C$}     \put(+45,+10){$C'$} \put(-47,-10){$D$}
\put(+45,-10){$D'$}

\end{picture}


\noindent here $(A, A')\; , \; (B, B')$  and so on, also
 $(1, 1', 1'')$  and $(2, 2 ', 2'')$ represent one the same respective points in $\tilde{S}_{3}$.
Let us give some details.

Consider the  vicinity of the point  $1'$ ; one can approach it by different directions
$$
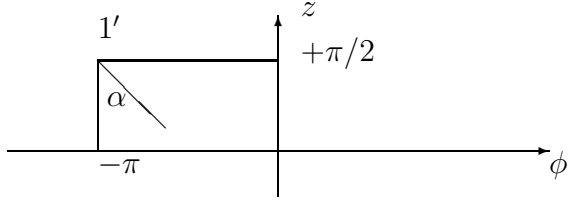
{\bf \mbox{Fig}\; 5 \;\; \mbox{Vicinity of }\;  1' }
$$
\unitlength=0.6mm
\begin{picture}(160,60)(-115,-30)
\special{em:linewidth 0.4pt} \linethickness{0.4pt}

\put(-60,0){\vector(+1,0){120}} \put(+60,-5){$\phi $}
\put(0,-10){\vector(0,+1){40}} \put(+5,+30){$z$}

\put(-40,-5){$-\pi $} \put(-40,0){\line(0,+1){20}}
\put(-40,+20){\line(+1,0){40}}

\put(-40,+25){$1'$}  \put(+5,+20){$+\pi /2$}
\put(-40,+20){\line(+1,-1){15}}  \put(-38,+10){$\alpha$}

\end{picture}

\vspace{-10mm}

\noindent The $\alpha$-direction is described by the equation
$$
\phi   = \; ( - \pi  + {\pi \over 2}\; \tan \alpha ) \;+\; z\;
\tan (\pi  - \alpha ) \;  \;  . \eqno(3.3a)
$$

\noindent Near to  $1'$  one has  $z = (\pi /2 - \delta )\;$,
$\delta $ is a very small positive  number, then eq. (3.3a) reads
 $$
\phi  = \; -\pi  + \delta \; \tan \alpha \; \; , \; \cos \phi  = -
\cos (\delta \; \tan \alpha  )\; , \; \sin \phi  = - \sin (\delta
\; \tan \alpha ) \; ;
$$

\noindent and therefore  $c^{j}$  in vicinity of $1'$ looks
$$
c^{1}  = \tan \rho \; {- \cos (\delta \tan  \alpha ) \over \sin
\delta  } \; ,\; c^{2}  = \tan  \rho\; {- \sin  (\delta  \tan
\alpha ) \over \sin  \delta }\; ,\; c^{3}  = {\cos  \delta  \over
\sin \delta  }\;  . \eqno(3.3b)
$$

\noindent The limit of the above when  $\delta\; \rightarrow \; 0$ (  $0 \le  \alpha  < \pi /2 )$
is one the same without any dependence of $\alpha$:
$$
\underline{1'\;: } \qquad  \vec{c} = \infty \; (- \tan \rho \;,\;
0\; ,\; 1 )\;  . \eqno(3.3c)
$$

\noindent Take notice that $\alpha  \neq  \pi /2;$.
Analogously< for the vicinity of the point $1''$ one has
$$
\underline{1''\; :} \qquad  \vec{c} = \infty \;  ( - \tan  \rho \;
, \; 0\; ,\; 1 )\; .
\eqno(3.4)
$$

\noindent And finally, for the  point $1$:
$$
\phi  = \; - { \pi  \over 2} \; \tan  g \;+\; z \; \tan (\pi
-\alpha )\; ,
$$

\vspace{-5mm}

$$
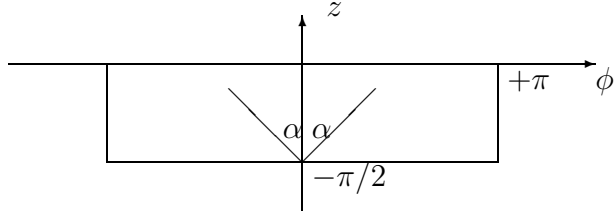
 \mbox{Fig } \;6\;\; \mbox{The vicinity of }  \; \; 1
$$

\unitlength=0.65 mm
\begin{picture}(160,40)(-115,-30)
\special{em:linewidth 0.4pt} \linethickness{0.4pt}

\put(-60,0){\vector(+1,0){120}} \put(+60,-5){$\phi $}

\put(0,-30){\vector(0,+1){40}} \put(+5,+10){$z$}

\put(-40,-20){\line(+1,0){80}}   \put(-40,-20){\line(0,+1){20}}
\put(+40,-20){\line(0,+1){20}} \put(0,-20){\line(+1,+1){15}}
\put(0,-20){\line(-1,+1){15}} \put(+2,-15){$\alpha$}
\put(-4,-15){$\alpha$}

\put(+2,-25){$-\pi /2$}  \put(+42,-5){$+\pi$}
\end{picture}


\noindent here   $z = ( -\pi /2 + \delta )$  и $\phi = - \delta \;
\tan \alpha $, and
$$
c_{1} = \tan  \rho \; {+ \cos (\delta \tan \alpha ) \over \sin
\delta  }\; , \; c_{2} = \tan  \rho  {- \sin  (\delta \tan  \alpha
) \over \sin  \delta }\; ,\; c_{3} = {- \cos  \delta \over \sin
\delta  } \; .
$$

\noindent When  $\delta $ changes to zero, one has
$$
\underline{1 \; : } \qquad \vec{c} = - \; \infty\;  (- \tan \rho
\;,\; 0\; ,\; 1 ) \; . \eqno(3.5)
$$

\noindent Therefore,  the  $(1, 1', 1'')$  indeed represent
one the same point in elliptical space. In the same  manner, three points $(2
, 2', 2'')$  correspond to
$$
\underline{2 \; :} \qquad \vec{c} = + \; \infty \;  (\tan \rho\;
,\; 0 \; ,\; 1 ) \; ,
$$
$$
\underline{2'\; :} \qquad \vec{c} = + \; \infty  \; ( \tan  \rho
\; ,\; 0\; ,\; 1 )\; ,
$$
$$
\underline{2'' \; :} \qquad \vec{c} = - \; \infty \; (\tan  \rho
\; ,\; 0\; ,\; 1 )\; . \eqno(3.6)
$$

\noindent For instance. let us detail four other points:
$$
\underline{A \; : } \qquad \vec{c} = +\; \infty \;  ( 0\; ,\; \tan
\rho \; ,\; 1 )\; ,  \;\; \underline{A' :} \qquad  \vec{c}=  ( 0
\; , \; \tan  \rho \; ,\; 1 )\;  ,
$$
$$
\underline{B\; : } \qquad \vec{c} = + \; \infty \; ( 0\; ,\;-\tan
\rho \; ,\; 1 )\; , \;\; \underline{B' \; :} \qquad  \vec{c} = -
\; \infty \;   ( 0\; , \; -\tan  \rho \; ,\; 1 )\; . \eqno(3.7)
$$

Identification on the boundaries  $( 1' C D 2' )$  and $( 1'' C' D'
2'')$ is shown in the Fig 4;  here only finite vector ${\vec c}$ are presented.
Consideration of the boundary of   $\tilde{G}_{1}$ is  finished.
Transition from $SU(2)$ to $SO(3.R)$ group spaces can be characterized by the scheme

$$
 \mbox{Fig} \; 7 \qquad \qquad  G_{1} \Longrightarrow \tilde{G}_{1}
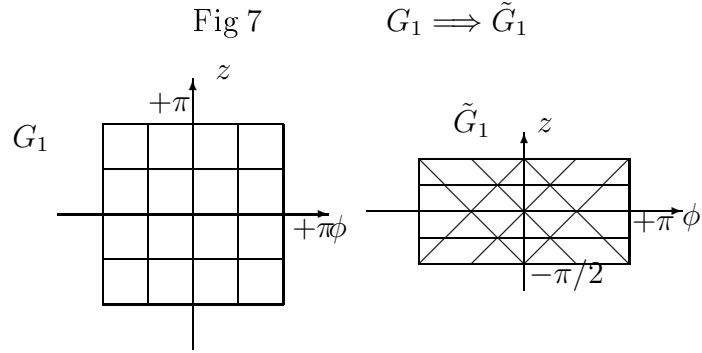
$$

\unitlength=0.6mm
\begin{picture}(160,60)(-90,-25)
\special{em:linewidth 0.4pt} \linethickness{0.4pt}

\put(-40,+15){$G_{1}$} \put(-30,0){\vector(+1,0){60}}
\put(+30,-5){$\phi $} \put(0,-30){\vector(0,+1){60}}
\put(+5,+30){$z$} \put(-20,+10){\line(+1,0){40}}
\put(-20,+20){\line(+1,0){40}} \put(-20,-10){\line(+1,0){40}}
\put(-20,-20){\line(+1,0){40}} \put(-20,+20){\line(0,-1){40}}
\put(-10,+20){\line(0,-1){40}} \put(+10,+20){\line(0,-1){40}}
\put(+20,+20){\line(0,-1){40}}

\put(+22,-5){$+\pi $}      \put(-10,+23){$+\pi $}

\end{picture}

\vspace{-28mm}

\unitlength=0.35mm
\begin{picture}(160,60)(-280,-25)
\special{em:linewidth 0.4pt} \linethickness{0.4pt}

\put(-27,+30){$\tilde{G}_{1}$} \put(-60,0){\vector(+1,0){120}}
\put(+60,-5){$\phi $} \put(0,-30){\vector(0,+1){60}}
\put(+5,+30){$z$}

\put(+41,-7){$+\pi$} \put(-40,-20){\line(+1,0){80}}
\put(-40,-20){\line(0,+1){40}} \put(+40,+20){\line(-1,0){80}}
\put(+40,+20){\line(0,-1){40}} \put(+40,+20){\line(-1,-1){40}}
\put(-40,-20){\line(+1,+1){40}} \put(-20,-20){\line(+1,+1){40}}
\put(-40,+20){\line(+1,-1){40}}
\put(-20,+20){\line(+1,-1){40}} \put(0,+20){\line(+1,-1){40}}
\put(-40,-10){\line(+1,0){80}}
\put(-40,+10){\line(+1,0){80}}

\put(+2,-27){$-\pi /2$}

\end{picture}

\vspace{+10mm}

Now we can examine two remaining domains.
In the  $\tilde{G}_{2}$ for right  point  $P$ we have
$$
\tilde{G}_{2} \;\;(\phi \;\; - \mbox{любое}) \; ,  \;\;
$$
$$
\rho  =  \pi /2\; \tan \alpha  + z\; \tan \alpha  \;
\Longrightarrow
 z = -\pi /2 + \delta \;  , \rho  = \delta \; \tan \alpha   ,
$$
$$
{\bf  \mbox{Fig} \;8   \;\; \mbox{граница} \;\; \tilde{G}_{2}}
$$
\vspace{7mm} \unitlength=0.65 mm
\begin{picture}(160,60)(-115,-30)
\special{em:linewidth 0.4pt} \linethickness{0.4pt}


\put(-40,0){\vector(+1,0){80}} \put(+40,-5){$z$}

\put(0,-10){\vector(0,+1){40}} \put(+5,+30){$\rho $}

\put(-20,0){\line(0,+1){20}}    \put(+20,0){\line(0,+1){20}}

\put(-20,+20){\line(+1,0){40}}  \put(-20,+1){\line(+1,0){40}}

\put(-20,0){\line(+1,+1){10}} \put(-13,+3){$\alpha $} \put(-28,+2){$P'$}
\put(+20,0){\line(-1,+1){10}} \put(+12,+3){$\alpha $} \put(22,+2){$P$}

\put(-20,-5){$-\pi /2$} \put(+20,-5){$+\pi /2$} \put(+5,+25){$+\pi
/2$}

\end{picture}
\vspace{-20mm}

\noindent Therefore,
$$
c_{1} = { \tan  (\delta  \tan  \alpha ) \over \sin ( \delta  \tan
\alpha )} \;{ \cos  \phi \over \sin \delta }\;  , \;\; c_{2} = {
\tan  (\delta  \tan  \alpha ) \over \sin ( \delta \tan \alpha )}
\; {\sin  \phi  \over \sin \delta }\; ,\;\; c_{3} = - {\cos
\delta  \over \sin  \delta } \; ,
$$

\noindent in the  limit  $\delta \rightarrow 0\; ( \alpha  \neq
0)$ it follows
$$
\underline{\tilde{G}_{2} :} \qquad P \; , \qquad \vec{c} = + \;
\infty \;  ( 0 \; ,\;  0\;  , \; - 1 ) \; ; \eqno(3.8a)
$$

\noindent In the same manner, for the left point  $P'$ аwe have
$$
\underline{G_{2}\; :} \qquad  P' \; ,  \qquad \vec{c} = - \;
\infty \;  ( 0 \;, \;0 \;,\;- 1 ) \; . \eqno(3.8b)
$$

\noindent So, all the domain $\tilde{G}_{2}$ determines
the  point of elliptical space in accordance with the formula
$$
\underline{G_{2}\; :} \qquad    \qquad \vec{c} = - \;
\infty \;  ( 0 \;, \;0 \;,\; \tan z ) \;
\eqno(3.8c)
$$

\noindent and $\phi$ is a mute variable.
Mow consider third domain  $\tilde{G}_{3}$:

$$
 \mbox{Fig} \; 9 \qquad \qquad \tilde{G}_{3}  $$

\vspace{2mm} \unitlength=0.65mm
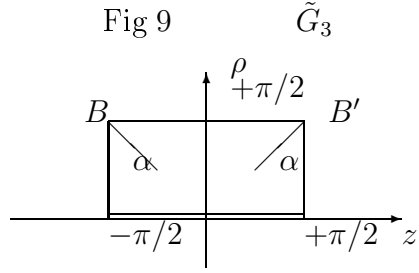
\begin{picture}(160,60)(-115,-30)
\special{em:linewidth 0.4pt} \linethickness{0.4pt}

\put(-25,+20){$B$}       \put(+25,+20){$B'$}
\put(-40,0){\vector(+1,0){80}} \put(+40,-5){$z$}
\put(0,-10){\vector(0,+1){40}} \put(+5,+30){$\rho $}

\put(-20,0){\line(0,+1){20}}    \put(+20,0){\line(0,+1){20}}
\put(-20,+20){\line(+1,0){40}}  \put(-20,+1){\line(+1,0){40}}

\put(-20,+20){\line(+1,-1){10}} \put(-15,+10){$\alpha $}

\put(+20,+20){\line(-1,-1){10}} \put(+15,+10){$\alpha $}

\put(-20,-5){$-\pi /2$} \put(+20,-5){$+\pi /2$} \put(+5,+25){$+\pi
/2$}

\end{picture}
\vspace{-15mm}

\noindent Near the point  $B$ we have
$$
\rho = \; - z\; \tan \alpha + ({\pi \over 2} - {\pi \over 2}\;\tan
\alpha )\; \;\Longrightarrow \;
 \; z = (- {\pi \over 2} + \delta  )\; ,\; \rho  = ({\pi \over 2}
 - \delta \; \tan \alpha  )\;   \; ;
$$

\noindent and
$$
c_{1} = { \cos  (\delta \; \tan  \alpha  ) \over \sin (\delta \tan
\alpha ) } \;{ \cos \phi \over \sin \delta }\; , \;\; c_{2} = {
\cos (\delta \tan \alpha ) \over \sin (\delta \tan \alpha )} \; {
\sin  \phi  \over \sin \delta }\; ,\;\; c_{3} = - { \cos  \delta
\over \sin  \delta } \;.
$$

\noindent From where, in the limit $ \delta \; \rightarrow \; 0 ,
\alpha  \neq  \pi /2$ , it follows
$$
\underline{B\; :} \qquad \vec{c} = \infty  \; (\cos \phi ,
\; \sin \phi , \; 0 ) \; . \eqno(3.9a)
$$

\noindent In the same  manner, for the point $B'$  we have
$$
\underline{B' \;  :}
 \qquad \vec{c} = \infty  \; (\cos \phi , \; \sin \phi , \;+0  ) \; .
\eqno(3.9b)
$$

All the domain $\tilde{G}_{3}$ parameterizes the point in $\tilde{S}_{3}$
in accordance  with the relation

$$
\underline{B' \;  :}
 \qquad \vec{c} = {\infty \over cos z}
  \; (\cos \phi , \; \sin \phi , \;+ {\sin z \over \infty}  )  \sim
 \infty    \; (\cos \phi , \; \sin \phi , \;0   ) \; ,
\eqno(3.9c)
$$

\noindent and $z$ is a mute variable.

Now we are ready to establish which of the wave functions $\Phi _{\epsilon MK}$  founded in
 \S 2 are continuous in elliptical space and which are not.
From the equation (see Fig 4)
$$
f(1)  = f(1' )  = f(1'' )
$$

\noindent  one can reduce
$$
e^{-iK (\pi /2)}   = e^{-iM\pi } \; e^{+iK (\pi /2)}   = e^{+iM\pi
}\; e^{+iK(\pi /2)} \; ,
$$

\noindent from where it follows
$$
e^{i2M\pi } = 1\; ,\;\;  e^{i(K-M) \pi } = 1\; ,\;\; e^{i(K+M) \pi
} = 1 \; ;
$$

\noindent that is  $( K - M )$ and  $( K + M )$  must be integer and even.
Analogous relations for the points  $(2 ,2', 2'')$ results in the same.

\begin{quotation}

{\it Therefore,
the functions $\phi _{\epsilon MK}(\rho ,\phi ,z)$
founded in \S 2 will be  single-valued  and continuous in the  $\tilde{G}_{1}$ if
$M$ and $K$ are  both even or both odd. Correspondingly, the main quantum number
 $N$ takes on the values  $ N = 1,\; 3,\; 5,\; \ldots $

}

\end{quotation}

In the domain $\tilde{G}_{2}$
the function  $\Phi _{\epsilon KM}(\rho,\phi,z)$   given by
$$
G_{2}\; : \qquad \phi _{\epsilon MK} = \left \{ \begin{array}{l}
0 , \;\; если \;\; M \neq  0 ; \\
e^{iKz}\; F (A, B, C; 1) \; , \;\; если \;\; M = 0\;
\end{array} \right.  \; ,
\eqno(3.10)
$$

\noindent is a continuous function on $\tilde{S}_{3}$.
In the   $\tilde{G}_{3}$ we have  behavior correct in the sense of continuity as well
$$
\tilde{G_{3}} = \left \{ \begin{array}{l}
 0 \;   ,\;\;  если \;\; K \neq  0\;  , \\
e^{iMz}\; F (A, B, C; 0) \;  , \;\; если K = 0\;\;  .
\end{array} \right.
\eqno(3.11)
$$

Thus, the functions  $\Phi _{\epsilon MK}(\rho
,\phi ,z)$ ,  $N = 1,\;3,\; 5,\; \ldots $  provide us with the system of single-valued
and continuous in the elliptical space $S0(3.R)$ -- one formal change must be done:
$$
{1 \over \sqrt{2\pi } }\;  e^{iKz} \; \Longrightarrow \; {1 \over
\sqrt{\pi }}\; e^{iKz} \;  ,
$$
$$
\int_{S0(3)} \Phi ^{*}_{\epsilon MK} \; \Phi _{\epsilon ' M' K' }
\; dV = \delta _{MM'}\;  \delta _{KK'}\; \delta _{NN'}\; , \;\; N = 1,3,5,...
\eqno(3.12)
$$

\noindent
At the same time, the remaining functions
 $$
 \Phi _{\epsilon MK}(\rho ,\phi ,z) \;\; , \;\;
N = 2, \;4, \; \ldots \;
 \eqno(3.13)
$$

\noindent
are discontinuous in elliptical  space
 $S0(3.R)$ and they  do not consist of a basis in the space of
 continuous function on the $SO(3)$ though obey the same orthogonality conditions  as (3.12).

\subsection*{4. Eiler variables as coordinates on the group  $SU(2)$.}

\hspace{5mm}
Now we are to refer the above  Schr\"{o}dinger's wave solutions
 $\Phi _{\epsilon MK}(\rho ,\phi ,z)$ to the known Wigner's $D$-functions [17].
$$
\Phi _{\epsilon MK} = { e^{iM\phi } \over \sqrt{2\pi }}\;
{ e^{iK\phi } \over \sqrt{2\pi }}  \;   \left [\;
A^{(a,b)}_{n} \; (1 -x)^{a/2}\; (1+x)^{b/2}\; P^{(a,b)}_{n}(x) \;\right ]  =
$$
$$
= { e^{iM\phi } \over \sqrt{2\pi }}\; { e^{iK\phi } \over
\sqrt{2\pi }} \; d ^{\;j}_{m\sigma }(\beta ) \;=\;
D^{\;j}_{m\sigma }(\alpha ,\beta ,\gamma )\; ; \eqno(4.1a)
$$

\noindent where
$$
 N = (2j + 1), \qquad
 \epsilon _{N} = {1 \over 2} (N^{2} - 1)  = 2j(j + 1) \; ,
$$
$$
 m =
-{1 \over 2} (M + K ) \;  , \;\; \sigma  = -{1\over 2} (M - K)\; ,
\eqno(4.1b)
$$

\noindent and  coordinates $(\rho ,\phi ,z)$  are connected with Eiler variables
$(\alpha ,\beta ,\gamma )$ as follows:
$$
\alpha  = \phi  + z\; , \;\; \beta  = 2 \rho \; , \;\; \gamma  =
\phi  - z \;. \eqno(4.1c)
$$

\noindent
This transition can be traced also in differential equations
$$
 [\; \sin ^{-1} \rho \; \cos ^{-1} \rho \;
 \partial _{\rho } \; \sin  \rho  \; \cos  \rho  \;
\partial _{\rho }   \; +
\sin  ^{-2} \rho \;  \partial ^{2} _{\phi}  +
\cos  ^{-2} \rho  \; \partial ^{2}_{z} + 2 \epsilon \;  ] \;
\Phi  = 0 \; ,
$$
$$
-i \partial_{ \phi } \; \Phi  = M\; \Phi \; , \qquad  -i \partial
_{z} \; \Phi = K \; \Phi \; \eqno(4.2a)
$$

\noindent which in  $(\alpha ,\beta ,\gamma )$ variables become
$$
[\; - \sin  ^{-1} \beta  \; \partial_{ \beta } \sin  \beta \;
\partial _{\beta }  +
 \; \sin  ^{-2} \beta \;
(- \partial ^{2}_{\alpha}  +
2 \cos  \beta \; \partial_{ \alpha } \; \partial_{\gamma }
 - \partial ^{2}_{\gamma } ) \; ] \; \Phi  = (\epsilon /2) \; \Phi \; ,
$$
$$
i \partial_{\alpha }\; \Phi  = -{1\over 2} (M + K) \; \Phi \;
,\;\; i \partial_{\gamma }\; \Phi  = - {1\over 2} (M - K) \; \Phi
\; . \eqno(4.2b)
$$

\noindent and  coincide  with determining relations for $D$-functions:
$
\Phi _{\epsilon MK} (\rho ,\phi ,z)  = D^{j}_{m\sigma } (\alpha ,\beta ,\gamma ) \;.
$

All specifics of the Eiler angles as parameters on the unitary group  can be
found with absolute fullness from the previous analysis.
First, these variables are non-orthogonal coordinates in  $S_{3}$:
$$
dl^{2}  = {1 \over 4 } \; [\; d  \beta ^{2} \; +  \; d  \alpha
^{2}\;  - \; 2\; \cos  \beta \; d \alpha \; d \gamma\;  + \;d
\gamma ^{2} \; ] \; . \eqno(4.2c)
$$

\noindent
Again, the whole domain $G(\alpha, \beta,\gamma)$ should be divided into
three parts  $G  = (G_{1} \cup  G_{2} \cup  G_{3}) \;$:
$$
G_{1} = G(\beta  \neq  0, \pi  ) \; ,  \;\; G_{2} = G(\beta  = 0 )
\; ,  \;\; G_{3} = G ( \beta  = \pi  )\;  .
\eqno(4.3)
$$

\noindent The domain  $G_{1}(\alpha ,\beta )$ is characterized by
the scheme
$$
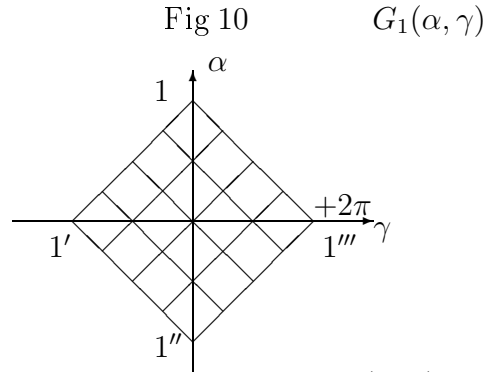
\mbox{Fig} \; 10 \qquad \qquad G_{1}(\alpha ,\gamma )
$$

\vspace{+8mm} \unitlength=0.4 mm
\begin{picture}(160,100)(-140,-70)
\special{em:linewidth 0.4pt}
\linethickness{0.4pt}

\vspace{-10mm}
\put(-60,0){\vector(+1,0){120}}    \put(+60,-5){$\gamma$}
\put(0,-50,0){\vector(0,+1){100}}  \put(+5,+50){$\alpha$}

\put(-40,0){\line(+1,+1){40}}  \put(+40,0){\line(-1,+1){40}}
\put(0,-40){\line(-1,+1){40}}  \put(0,-40){\line(+1,+1){40}}

\put(-30,-10){\line(+1,+1){40}} \put(-20,-20){\line(+1,+1){40}}
\put(-10,-30){\line(+1,+1){40}}
\put(+30,-10){\line(-1,+1){40}}   \put(+20,-20){\line(-1,+1){40}}
\put(+10,-30){\line(-1,+1){40}}

\noindent \put(-48,-12){$1'$}         \put(+43,-12){$1'''$}
\put(-13,+40){$1$}  \put(-13,-45){$1''$} \put(+40,+2){$+2\pi $}
\end{picture}
\vspace{-10mm}

\noindent The closed curve associated with the domain
$G_{2}(\alpha ,\gamma )$
$$ n_{0} = \cos
{\alpha - \gamma  \over 2}\; ,\;\; n_{1} = 0 \; , \;\; n_{2} = 0
\; , \;\; n_{3} = \sin {\alpha -\gamma \over 2} \; ; \eqno(4.4a)
$$

\noindent from where it follows that $(\alpha , \beta )$ and $(\alpha ', \beta ')$
represent the same point in the curve if
$(\alpha - \gamma ) = (\alpha ' - \beta ')$. In accordance with this
the  domain  $G_{2}(\alpha ,\gamma )$ is characterized by the scheme
$$
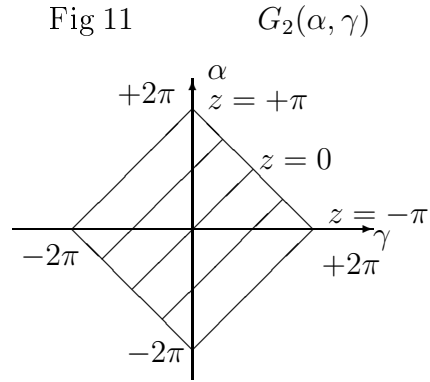
\mbox{Fig} \;11 \qquad \qquad G_{2}(\alpha ,\gamma )
$$

\vspace{+25mm}
\unitlength=0.4 mm
\begin{picture}(160,50)(-185,-60)
\special{em:linewidth 0.4pt}
\linethickness{0.4pt}

\put(-60,0){\vector(+1,0){120}}  \put(+60,-5){$\gamma$}
\put(0,-50,0){\vector(0,+1){100}}  \put(+5,+50){$\alpha$}

\put(-40,0){\line(+1,+1){40}}  \put(+40,0){\line(-1,+1){40}}
\put(0,-40){\line(-1,+1){40} } \put(0,-40){\line(+1,+1){40}}

\put(-30,-10){\line(+1,+1){40}} \put(-20,-20){\line(+1,+1){40}}
\put(-10,-30){\line(+1,+1){40}}

\put(+5,+40){$z =+\pi $}      \put(+45,+2){$z =- \pi $}
\put(+22,+20){$z = 0$}

\put(-25,+42){$+2\pi $}   \put(-57,-12){$-2\pi $}
\put(+43,-15){$+2\pi $}   \put(-22,-44){$-2\pi$}
\end{picture}
\vspace{-3mm}

\noindent here the full segment  $(\alpha ,\beta )_{z}$ given by equation
 $\alpha =  \gamma  + 2z $ corresponds to the single  point $n =
( \cos z,  0,  0,  \sin z )$. Wigner functions in the
$G_{2}(\alpha ,\gamma )$
$$
D^{j}_{m\sigma } (\alpha , 0, \gamma ) = \left  \{  \begin{array}{l}
0  , \;\; если\;\;  (m + \sigma ) \neq  0 \; , \\[2mm]
F (A, B, C; 1) \;, \;\; если \;\; m = 0 \; , \; \sigma  = 0 \; , \\[2mm]
e^{i m (\alpha -\gamma )} \; F (A, B, C; 1)\; ,\;\; если \;\;
 (m + \sigma ) = 0 \;
\end{array}  \right.
\eqno(4.4b)
$$

\noindent are continuous in the curve. Consideration of the domain
$G_{3}$  will is the same:
$$
n_{0}  = 0\; , \;\; n_{1} = \cos {\alpha + \gamma  \over 2} \; ,
\;\; n_{2}  = \sin  {\alpha + \gamma  \over 2} \; , \;\; n_{3}  =
0\;  ; \eqno(4.5a)
$$
$$
 \mbox{Fig} \; 12 \qquad \qquad G_{3}(\alpha ,\gamma )
$$

\vspace{+7mm}
\unitlength=0.4 mm
\begin{picture}(160,100)(-160,-60)
\special{em:linewidth 0.4pt}
\linethickness{0.4pt}

\vspace{-3mm}
\put(-60,0){\vector(+1,0){120}}     \put(+60,-5){$\gamma$}
\put(0,-50,0){\vector(0,+1){100}}   \put(+5,+50){$\alpha$}

\put(-40,0){\line(+1,+1){40}}     \put(+40,0){\line(-1,+1){40} }
\put(0,-40){\line(-1,+1){40}}     \put(0,-40){\line(+1,+1){40}  }
\put(+30,-10){\line(-1,+1){40}}   \put(20,-20){\line(-1,+1){40}}
\put(+10,-30){\line(-1,+1){40}}
\put(-30,+10){\circle*{2}}        \put(-77,+3){$\phi = - \pi$}
\put(-20,+20){\circle*{2}}        \put(-48,+20){$\phi = 0 $}
\put(-10,+30){\circle*{2}}        \put(-45,+37){$\phi = + \pi$}

\put(-57,-12){$-2\pi$}             \put(+43,-10){$+2\pi$}
\put(+5,+41){$+2\pi$}             \put(-25,-45){$-2\pi$}

\end{picture}

\noindent and Wigner function
$$
D^{j}_{m\sigma } (\alpha ,\pi ,\gamma ) =  \left \{ \begin{array}{l}
0 \; ,\;\;  если\;\; (m - \sigma ) \neq  0 \; ,  \\[2mm]
F (A, B, C; 0) \;, \;\;  если \;\; m = 0 \; , \; \sigma = 0  \; ,  \\[2mm]
e^{- i m (\alpha + \gamma )}\; F (A, B, C; 0)\;,\;\; если\;\; (m -
\sigma ) = 0 \; ; \end{array} \right. \eqno(4.5b)
$$

\noindent  depending on the argument $( \alpha  + \beta  )$ are continuous in the curve.

Now  let us reformulate  the orthogonality conditions (2.8):
$$
\int_{S_{3} } \Phi ^{*}_{\epsilon MK} \; \Phi _{\epsilon 'M'K'} dV  =
\delta _{\epsilon \epsilon '} \; \delta _{MM'}\; \delta _{KK'}\; ,
\;\;\;dV  =  \sin \rho \; \cos \rho \; d \rho  d \phi  dz \;  ,
$$
$$
\mbox{Fig}  \; 13 \qquad \qquad G(\phi ,z)
$$

\vspace{5mm}
\unitlength=0.4 mm
\begin{picture}(160,80)(-160,-50)
\special{em:linewidth 0.4pt}
\linethickness{0.4pt}

\put(-40,0){\vector(+1,0){80}}    \put(+40,-7){$\phi$}
\put(0,-40){\vector(0,+1){80}}    \put(-15,+30){$z$}
\put(+22,-7){$+\pi$}  \put(+3,+22){$+\pi$}
\put(-20,-20){\line(+1,0){40}}     \put(-20,-20){\line(0,+1){40}}
\put(+20,+20){\line(-1,0){40}}     \put(+20,+20){\line(0,-1){40}}

\end{picture}


\noindent which after transition to Eiler variable swill looks as follows:
$$
\int_{SU(2)} D^{j*} _{m\sigma } (\alpha ,\beta ,\gamma )\;
D ^{j'}_{m' \sigma '} (\alpha ,\beta ,\gamma ) \;d \omega = \delta _{mm'}\;
\delta _{\sigma \sigma '}\; \delta _{jj '}\; ,
$$
$$
d \omega = {1 \over 8}\; \sin \beta d\alpha d\beta d \gamma \; ; \;\;
0 \le  \beta  \le  \pi  \; ,
$$
$$
\mbox{Fig}  \;\; 14  \qquad \qquad G(\alpha ,\gamma )
$$


\unitlength=0.4 mm
\begin{picture}(160,80)(-140,-30)
\special{em:linewidth 0.4pt}
\linethickness{0.4pt}

\vspace{-5mm}
\put(-60,0){\vector(+1,0){120}}    \put(+63,-5){$\gamma$}
\put(0,-50,0){\vector(0,+1){100}}  \put(-10,+50){$\alpha$}

\put(-40,0){\line(+1,+1){40}}  \put(+40,0){\line(-1,+1){40}}
\put(0,-40){\line(-1,+1){40}}  \put(0,-40){\line(+1,+1){40}}

\put(-67,-13){$-2\pi '$}
\put(+5,+42){$+2\pi $}  \put(-25,-42){$-2\pi $}  \put(+42,-10){$+2\pi $}

\end{picture}
\vspace{+13mm}

\noindent
Indeed, taking the following  order in integrating
$$
\int_{G(\alpha ,\gamma )} \; d \alpha \; d \gamma \;
F(\alpha ,\gamma )\;   =
$$
$$
 = \int_{-2\pi }^{0} d \gamma \;
\int_{\alpha  = - (\gamma + 2 \pi )}^{ \alpha  = + (\gamma + 2 \pi
)} \; d \alpha  \; F(\alpha ,\gamma )\;  + \int_{0}^{+2\pi } \;
d\gamma \int_{\alpha  = - (- \gamma  + 2 \pi ) }^{\alpha = + (-
\gamma  + 2 \pi )}\; d \alpha \; F(\alpha ,\gamma )  \; .
\eqno(4.6)
$$

\noindent it is the matter of simple calculation to obtain the identity
$$
{1 \over 8 \pi ^{2}} \int _{G(\alpha ,\gamma )} d \alpha \; d
\gamma\;  [\; e^{ i(m-m')\alpha  } \; e^{i(\sigma - \sigma ')
\gamma  } \; ]  = \delta _{mm'}\; \delta _{ \sigma \sigma '} \; ,
\eqno(4.7)
$$

One can easily see that the same symbol $( \delta _{mm'}\; \delta _{\sigma \sigma '}) $
might be  achieved in integrating over two other domains,
$\Delta (\alpha ,\gamma )$   or  $\Delta ' (\alpha ,\gamma )$ determined in Fig 15 and Fig 16 below.
$$
 \mbox{Fig}  \; 15 \qquad \qquad \Delta (\alpha ,\gamma )
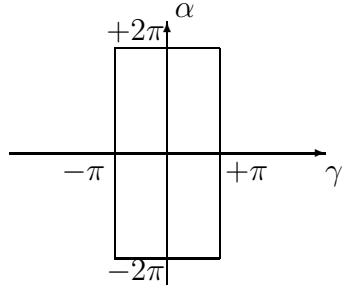
$$

\vspace{5mm}

\unitlength=0.35mm
\begin{picture}(160,80)(-160,-40)
\special{em:linewidth 0.4pt}
\linethickness{0.4pt}
\vspace{-10mm}

\put(-60,0){\vector(+1,0){120}}   \put(+62,+5){$\phi$}
\put(0,-40){\vector(0,+1){80}}    \put(+5,+42){$z$}

\put(-40,-20){\line(0,+1){40}}   \put(-40,-20){\line(+1,0){80}}
\put(+40,+20){\line(0,-1){40}}    \put(+40,+20){\line(-1,0){80} }
\put(-63,-10){$-2\pi$}             \put(+41,-10){$+2\pi$}
\put(-18,-27){$-\pi$}              \put(-18,+23){$+\pi$}

\end{picture}

$$
\mbox{Fig}  \; 16 \qquad \qquad \Delta ' (\alpha ,\gamma )
$$

\vspace{5mm}

\unitlength=0.35mm
\begin{picture}(160,80)(-160,-40)
\special{em:linewidth 0.4pt}
\linethickness{0.4pt}

\put(-60,0){\vector(+1,0){120}}   \put(+60,-10){$\gamma$}
\put(0,-50){\vector(0,+1){100}}    \put(+3,+52){$\alpha$}

\put(-20,-40){\line(0,+1){80}}    \put(-20,-40){\line(+1,0){40}}
\put(+20,+40){\line(0,-1){80}}    \put(+20,+40){\line(-1,0){40}}

\put(-40,-10){$-\pi$}              \put(+22,-10){$+\pi$}
\put(-23,-48){$-2\pi$}             \put(-23,+43){$+2\pi$}

\end{picture}
\vspace{5mm}

This fact is not occasional. Indeed, now we will show that
a more general statement is true: {\em if
 $\Phi (\alpha ,\beta ,\gamma ) = \Phi (n)$ is any single-valued and
 continuous  function on the sphere $S_{3}$, then identity holds
 (or $\Delta \; \rightarrow \; \Delta '$)}

$$
\int_{R(\beta ) \otimes G(\alpha ,\gamma )} d \omega \; \Phi (n)
= \int_{R(\beta ) \otimes \Delta (\alpha ,\gamma )}   d \omega  \;
\Phi (n)\; . \eqno(4.8)
$$

\noindent Indeed, any that  function
$\Phi (n)$,  being expressed in coordinates  $(\rho ,\phi ,z)$,
gives a function  $2\pi $-periodical in  $\phi $  и $z$:
$$
\Phi (\rho ,\phi ',z')  =  \phi (\rho ,\phi '' ,z'' ) \;
\Longleftrightarrow   \; (\phi '' - \phi ' )  =  2 \pi \nu \;
,\;\; (z' - z'' )  = 2 \pi \mu\; . \eqno(4.9a)
$$

\noindent After translating to variables $(\alpha ,\gamma )$ one has
$$
{1 \over 2} \; (\alpha  - \gamma ) = \phi \; , \;\;
{1 \over 2} \; (\alpha  + \gamma ) = z \;  ,
$$
$$
F(\rho ,\alpha ,\gamma ) = \Phi ( \rho , { \alpha  + \gamma  \over 2},
{ \alpha  - \gamma  \over 2}) \; .
$$

\noindent Therefore, any function on $S_{3}$,
 single-valued and continuous, must obey the following periodicity-condition:
 $$
F(\rho ,\alpha ', \gamma ' )  = F (\rho ,\alpha '' , \gamma '' ) \; ,
$$
$$
(\alpha '' - \alpha ') = 2\pi (\nu + \mu )\; ,\;\; (\gamma '' -
\gamma ' )  = 2 \pi (\nu - \mu )\; . \eqno(4.9b)
$$

\vspace{5mm}

\noindent Now let us compare two domains, $G(\alpha ,\gamma )$ and
$\Delta (\alpha ,\gamma )$:
$$
\mbox{Fig}  \; 17
$$

\unitlength=0.35 mm
\begin{picture}(160,120)(-170,-50)
\special{em:linewidth 0.4pt}
\linethickness{0.4pt}

\vspace{-30mm}
\put(-70,0){\vector(+1,0){140}} \put(+72,+5){$\phi$}
\put(0,-60){\vector(0,+1){120}}  \put(+5,+60){$z$}

\put(-40,-20){\line(0,+1){40}}    \put(-40,-20){\line(+1,0){80}}
\put(+40,+20){\line(0,-1){40}}    \put(+40,+20){\line(-1,0){80}}

\put(-40,0){\line(+1,+1){40}}     \put(-40,0){\line(+1,-1){40}}
\put(+40,0){\line(-1,-1){40}}     \put(+40,0){\line(-1,+1){40}}

\put(-53,+5){$D'$}                \put(+45,+5){$C'$}
\put(-53,-15){$B'$}                \put(+45,-15){$A'$}
\put(-25,+28){$A$}                \put(+20,+28){$B$}
\put(-30,-30){$C$}                \put(+20,-30){$D$}

\end{picture}
\vspace{5mm}

\noindent It is easily seen that the parts   $A, B, C, D$  and  $A', B', C' , D'$
can be moved  to each other by respective   shifts:$$
A\; \Longrightarrow   \;
A' (\alpha ' = \alpha - 2 \pi  , \gamma ' = \gamma + 2\pi  )\; ,
$$
$$
B\; \Longrightarrow   \;
B'\; (\alpha '= \alpha - 2\pi  , \gamma '= \gamma - 2\pi  )\; ,
$$
$$
C \; \Longrightarrow   \;
C'\; (\alpha ' = \alpha  + 2 \pi  , \gamma ' = \gamma  + 2\pi  )\; ,
$$
$$
D\;  \Longrightarrow   \;
D'\;   (\alpha ' = \alpha  + 2 \pi  , \gamma ' = \gamma - 2\pi  ) \; .
$$

\noindent From where, taking in  mind
$$
G  = [ G_{0}  +  ( A  + B  + C  + D ) ] \; , \;\;
\Delta   = [ G   + ( A' + B ' + C'  +  D'  ) ] \;
$$

\noindent and  the above  $2\pi $-periodicity of  $F(\rho ,\alpha ,\gamma )$ in variables
$\alpha $   and $\gamma $,  one  arrives at (4.8).  Consideration of the  domain
$\Delta '$   is the same.

Thus, to cover the group  $SU(2)$, instead of $ [ R (\rho ) \otimes G(\alpha ,\gamma )]$, one  may use
 other variants,
$[R (\rho )  \otimes \Delta (\alpha ,\gamma ) ] $
or $[R(\rho ) \otimes  \Delta ' (\alpha ,\gamma ) ] $. However, one should remember on changing
the  identification rule for the boundary points.
For instance, tracing the all identical points in the  replacement
 $G \; \rightarrow \; \Delta $  we  get a new scheme:

$$
 \mbox{Fig} \;\; 18 \qquad \qquad \Delta _{1}(\alpha ,\gamma )
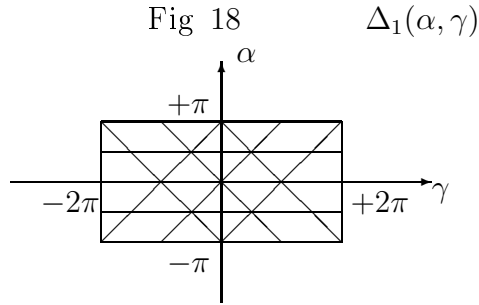
$$

\vspace{1mm}

\unitlength=0.4 mm
\begin{picture}(160,80)(-160,-40)
\special{em:linewidth 0.4pt}
\linethickness{0.4pt}

\put(-70,0){\vector(+1,0){140}}   \put(+70,-5){$\gamma$}
\put(0,-40){\vector(0,+1){80}}    \put(+5,+40){$\alpha$}

\put(-40,-20){\line(+1,0){80}}    \put(-40,-20){\line(0,+1){40}}
\put(+40,+20){\line(-1,0){80}}    \put(+40,+20){\line(0,-1){40}}
\put(-40,-10){\line(+1,0){80}}    \put(-40,+10){\line(+1,0){80}}

\put(-40,-20){\line(+1,+1){40}}   \put(-20,-20){\line(+1,+1){40}}
\put(0,-20){\line(+1,+1){40}}
\put(-40,+20){\line(+1,-1){40}}   \put(-20,+20){\line(+1,-1){40}}
\put(0,+20){\line(+1,-1){40}}

\put(-60,-10){$-2\pi$}   \put(+43,-10){$+2\pi$}
\put(-18,-30){$-\pi$}    \put(-18,+23){$+\pi$}
\end{picture}

\noindent It formally coincides with that for the cylindrical variables
$(\phi,z)$ in the case of orthogonal group $SO(3.R)$ group.

\subsection*{5. Eiler angles, coordinates on the group $S0(3.R)$.}

\hspace{5mm}
Transition to the Eiler variables in the  theory of
orthogonal group can be done as  before.
Below let us written down only the main graphical schemes.

Instead of Fig. 9  there is
$$
\mbox{Fig}  \;\; 19 \qquad \qquad G_{1}(\alpha ,\gamma )
$$

\vspace{15mm}
\unitlength=0.5 mm
\begin{picture}(160,60)(-120,-40)
\special{em:linewidth 0.4pt}
\linethickness{0.4pt}

\put(-60,0){\vector(+1,0){120}}     \put(+60,-5){$\gamma$}
\put(0,-40){\vector(0,+1){80}}      \put(+5,+40){$\alpha$}

\put(-30,-10){\line(+1,+1){40}}     \put(-10,-30){\line(+1,+1){40}}
\put(-10,-30){\line(-1,+1){20}}     \put(+10,+30){\line(+1,-1){20}}
\put(-30,-10){\line(+1,0){40}}      \put(-10,+10){\line(+1,0){40}}
\put(-10,-30){\line(0,+1){40}}      \put(+10,-10){\line(0,+1){40}}

\put(-28,+2){$-\pi $} \put(-15,-20){$-\pi $}
\put(-17,+20){$+\pi $}   \put(+20,-7){$+\pi $}

\end{picture}

\noindent Instead of Fig 11 there is

$$
 \mbox{Fig}  \;\; 20 \qquad    G_{2}(\alpha ,\gamma )  \qquad
    \vec{c} = (0, \;0,\; \tan  z )
$$

\vspace{5mm}
\unitlength=0.5 mm
\begin{picture}(160,60)(-120,-20)
\special{em:linewidth 0.4pt}
\linethickness{0.4pt}

\put(-60,0){\vector(+1,0){120}}   \put(+60,-5){$\gamma$}
\put(0,-40){\vector(0,+1){80}}
\put(+5,+42){$\alpha$}

\put(-30,-10){\line(+1,+1){40}}     \put(-10,-30){\line(+1,+1){40}}
\put(-10,-30){\line(-1,+1){20}}     \put(+10,+30){\line(+1,-1){20}}

\put(-20,-20){\line(+1,+1){40}}     \put(-25,-15){\line(+1,+1){40} }
\put(-15,-25){\line(+1,+1){40} }    \put(-10,-30){\line(+1,+1){40}}

\put(-35,+2){$-\pi $}              \put(+2,-25){$-\pi $}

\put(+10,+30){$z=\pi /2$}           \put(+30,+10){$z=-\pi /2$}
\put(+20,+20){$z= 0$}
\end{picture}
\vspace{+10mm}

\noindent Instead of Fig 12 there is

$$
\mbox{Fig} \;\; 21 \qquad \qquad  \vec{c} = \infty \; (\cos \phi ,\; \sin , \; 0)
$$

\vspace{10mm}
\unitlength=0.5 mm
\begin{picture}(160,60)(-120,-30)
\special{em:linewidth 0.4pt}
\linethickness{0.4pt}

\put(-60,0){\vector(+1,0){120}}   \put(+60,-5){$\gamma$}
\put(0,-40){\vector(0,+1){80}}
\put(+5,+40){$\alpha$}

\put(-30,-10){\line(+1,+1){40}}     \put(-10,-30){\line(+1,+1){40}}
\put(-10,-30){\line(-1,+1){20}}     \put(+10,+30){\line(+1,-1){20}}
\put(-20,0){\line(+1,-1){20}}       \put(-10,+10){\line(+1,-1){20} }
\put(0,+20){\line(+1,-1){20}}

\put(-75,-10){$\phi = -\pi$}
\put(-44,+10){$\phi = 0 $}
\put(-40,+25){$\phi = +\pi$}

\put(+25,-7){$+\pi $}   \put(+2,-25){$-\pi $}

\end{picture}

\vspace{+10mm}

\noindent Instead of Fig  17

$$
\mbox{Fig}  \;\; 22 \qquad \qquad
G(\alpha ,\gamma ) \; \rightarrow \;  \Delta (\alpha ,\gamma )
$$

\vspace{5mm}
\unitlength=0.5 mm
\begin{picture}(160,60)(-120,-20)
\special{em:linewidth 0.4pt}
\linethickness{0.4pt}

\put(-60,0){\vector(+1,0){120}}   \put(+60,-5){$\gamma$}
\put(0,-40,){\vector(0,+1){90}}   \put(+5,+50){$\alpha$}

\put(-30,-10){\line(+1,+1){40}}     \put(-10,-30){\line(+1,+1){40}}
\put(-10,-30){\line(-1,+1){20}}     \put(+10,+30){\line(+1,-1){20}}


\put(-20,-20){\line(+1,0){40}}       \put(-20,-20){\line(0,+1){40}}
\put(+20,+20){\line(-1,0){40}}       \put(+20,+20){\line(0,-1){40}}
\put(-20,+20){\line(+1,-1){10}}      \put(+20,-20){\line(-1,+1){10}}
\put(-15,+32){$A'$}                  \put(-10,-40){$A$}
\put(+10,+32){$C$}                   \put(+10,-40){$C'$}
\put(-43,-12){$B$}    \put(+32,-12){$B'$}
\put(-43,+10){$D'$}    \put(+32,+10){$D$}
\end{picture}
\vspace{+10mm}

\noindent Instead of Fig 18 there is

$$
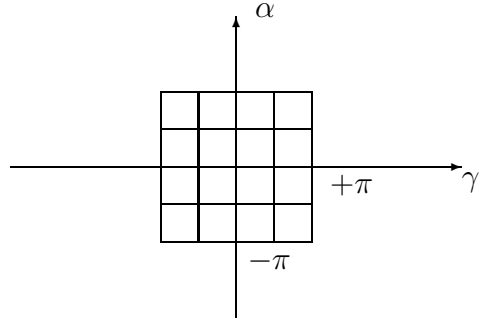
\mbox{Fig}  \;\; 23 \qquad \qquad
\Delta _{1}(\alpha ,\gamma )
$$

\vspace{5mm}
\unitlength=0.5 mm
\begin{picture}(160,60)(-120,-20)
\special{em:linewidth 0.4pt}
\linethickness{0.4pt}

\put(-60,0){\vector(+1,0){120}}   \put(+60,-5){$\gamma$}
\put(0,-40){\vector(0,+1){80}}    \put(+5,+40){$\alpha$}

\put(-20,-20){\line(+1,0){40}}  \put(-20,-20){\line(0,+1){40}}
\put(+20,+20){\line(-1,0){40}}  \put(+20,+20){\line(0,-1){40}}

\put(-20,-10){\line(+1,0){40}}     \put(-20,+10){\line(+1,0){40}}
\put(-10,-20){\line(0,+1){40}}     \put(+10,-20){\line(0,+1){40}}

 \put(+25,-7){$+\pi $}  \put(+3,-27){$-\pi $}
\end{picture}
\vspace{15mm}

{\bf PART II. DIRAC EQUATION ON THE SPHERE $S_{3}$}

\subsection*{6. Dirac equation in cylindrical coordinates, separating the variables}

\hspace{5mm}
Let us turn to the Dirac equation in cylindrical coordinates
and tetrad on the sphere $S_{3}$:
$$
n_{1} = \sin \rho \; \cos \phi \; , \;\; n_{2} = \sin \rho \; \sin
\phi \; , \;\;
$$
$$
 n_{3} = \cos \rho \; \sin z  \;   , \;\; n_{4} =
\cos \rho \; \cos z  \; ;
$$
$$
dl^{2} (y) =  d\rho^{2} + \sin^{2} \rho  \; d\phi^{2} + \cos^{2}\rho
\; dz^{2}  \; ,
$$
$$
 e_{(a)}^{\alpha}(y) = \left |
\begin{array}{llll}
1  &  0  &  0  &  0  \\
0  &  1  &  0  &  0  \\
0  &  0  & \sin^{-1} \rho  &   0  \\
0  &  0  &  0  & \cos^{-1}\rho
\end{array}  \right |  \; ;
\eqno(6.1)
$$

\noindent where  $(\rho,\; \phi,\; z )$ change in the  domain
$$
G =  \{ \rho \in [ 0, \; +\pi /2 ] \; , \;\;
\phi \in [-\pi,\; +\pi ] \; , \;\; z    \in [-\pi,\;+\pi ] \; \}
\; .
$$

\noindent As in scalar case, here it is convenient to divide
 $G(\rho,\; \phi,\; z )$ into three parts:
  $
G_{1}(\rho \neq 0,\pi/2) \; , \;\; G_{2}(\rho =  0 ) \; , \;\;
G_{3}(\rho = \pi/2) \; . $
The Dirac equation in orthogonal coordinates and tetrad can be written in the form
$$
 [ \;
i\; \gamma^{a} \;  ( e_{(a)}^{\alpha}\;{\partial \over \partial x^{\alpha}} +
{1\over 2}\; ({1 \over \sqrt{-g}}{\partial \over \partial x^{\alpha}}
\sqrt{-g} \;  e_{(a)}^{\alpha}    ) )    \; - \; M \;   ] \;\Psi (x)  = 0 \; .
\eqno(6.2a)
$$

\noindent
At the given tetrad  (6.1) equation  (6.2a) will read
$$
\left [ \; \gamma^{0} {\partial \over \partial t} + \gamma^{1} (
 {\partial \over \partial \rho } + {\cos 2\rho \over \sin 2\rho }) \;+\;
\gamma^{2}\; {1 \over \sin \rho}\; {\partial \over \partial \phi } \; + \;
\gamma^{3} {1 \over \cos \rho} \; {\partial \over \partial z } \; + iM \; \right ]\;
\Psi  =0 \; .
\eqno(6.2b)
$$

\noindent Let us construct solutions, the eigenfunctions of
 $i\partial_{t},\; i\partial_{\phi},\;i\partial_{z}$:
 $$
\Psi (t; \rho, \phi,z) = e^{-i\epsilon t} \; e^{im\phi} \; e^{ikz} \;
\left | \begin{array}{r}
f_{1}(\rho) \\ f_{2}(\rho) \\   f_{3}(\rho) \\ f_{4}(\rho)
\end{array} \right | \; .
\eqno(6.3)
$$

\noindent With the  use of explicit form of the Dirac matrices in the spinor basis
from  (6.2b) it  follows equations for  $f_{i}$:
$$
-i\epsilon \; f_{3} \; -\; ({d \over d \rho} +
{\cos 2\rho \over \sin 2 \rho} ) \; f_{4} \; - \;
{m \over \sin \rho } \; f_{4} \; - \;
{ik \over \cos \rho } \; f_{3} + iM \; f_{1} = 0 \; ,
$$
$$
-i\epsilon \; f_{4} \; -\;  ({d \over d \rho} +
{\cos 2\rho \over \sin 2 \rho}  ) \; f_{3} \; + \;
{m \over \sin \rho } \; f_{3} \; + \;
{ik \over \cos \rho } \; f_{4} + iM \; f_{2} = 0 \; ,
$$
$$
-i\epsilon \; f_{1} \; +\;  ({d \over d \rho} +
{\cos 2\rho \over \sin 2 \rho}  ) \; f_{2} \; + \;
{m \over \sin \rho } \; f_{2} \; + \;
{ik \over \cos \rho } \; f_{1} + iM \; f_{3} = 0 \; ,
$$
$$
-i\epsilon \; f_{2} \; +\;  ({d \over d \rho} +
{\cos 2\rho \over \sin 2 \rho} ) \; f_{1} \; - \;
{m \over \sin \rho } \; f_{1} \; - \;
{ik \over \cos \rho } \; f_{2} + iM \; f_{4} = 0 \; .
$$
$$
\vspace{-2mm}
\eqno(6.4)
$$

To simplify the system (6.4) one other operator should be diagonalized.
In flat space, as  that can be  taken the helicity operator
$$
(\vec{\Sigma} \;\vec{P}) \; \Psi_{cart}  = \lambda \; \Psi_{cart} \; ,
\;\; \vec{P} = - i \vec{\nabla} \;  .
$$

\noindent Because the cartesian and cylindrical bases are related by
spinor gauge transformation over fermion wave functions:
$$
\Psi^{cyl} = S\; \Psi^{cart} \; , \;\; S = \left | \begin{array}{cc}
B  &  0  \\  0  &  B  \end{array} \right | \; , \;\;
B =  \left | \begin{array}{cc}
e^{+i\phi/2} & 0 \\ 0  & e^{-i\phi/2}
 \end{array} \right | \; ,
\eqno(6.5a)
$$

\noindent
for the helicity operator in cylindrical representation
$$
B \;\vec{\Sigma} \;\vec{P} \; B^{-1}   =
\left | \begin{array}{cc}
P_{3}  &   e^{+i\phi/2} \; ( P_{1} - iP_{2} ) \; e^{-i\phi/2}  \\
e^{+i\phi/2} \; ( P_{1} + iP_{2} ) \; e^{-i\phi/2}     & -P_{3}
\end{array} \right |   \; ,
$$
$$
( P_{1} \pm i P_{2} ) = -i e^{\pm i\phi/2}
 ( {\partial \over \partial \rho }  \pm  {i \over \rho}
{\partial \over  \partial \phi}  ) \; ,
$$
$$
e^{\mp i\phi/2} ( P_{1} \pm i P_{2} ) e^{\pm i\phi/2}  =
[ -i  ( {\partial \over \partial \rho} + {1 \over 2\rho} )
\pm {1 \over \rho } {\partial \over \partial \phi}  ] \; ,
\eqno(6.5b)
$$

\noindent one  produces the expression
$$
\Lambda^{0} =    \gamma^{2}\gamma^{3}\;
({\partial \over \partial \rho }  +  {1 \over 2\rho} ) \;   + \;
\gamma^{3}\gamma^{1} \;{1 \over \rho} \;{\partial \over \partial \phi } \; + \;
\gamma^{1} \gamma^{2} \;  {\partial \over \partial z} \;  \; .
\eqno(6.6)
$$

\noindent
Now, taking in mind eq. (6.6) and  the Dirac equation  (6.2b)  in space $S_{3}$,
one  may guess the form of a generalized helicity operator $\Lambda$ in the curved space:
$$
\Lambda =    \gamma^{2}\gamma^{3}\;
({\partial \over \partial \rho }  +  {\cos 2\rho  \over \sin 2\rho} ) \;  + \;
\gamma^{3}\gamma^{1} \;{1 \over \sin \rho} \;{\partial \over \partial \phi } \; + \;
\gamma^{1} \gamma^{2} \; {1 \over \cos \rho}\;  {\partial \over \partial z}  \;  \; .
\eqno(6.7)
$$

\noindent Equation on the proper values $ \Lambda \Psi = \lambda \; \Psi $ leads us to
$$
-i ( {d \over d \rho} + {\cos 2\rho \over \sin 2\rho}  )\;
f_{2} \; - \; {im \over \sin \rho }\;f_{2} \; + \;
{k \over \cos \rho} \; f_{1} = \lambda \; f_{1} \; ,
$$
$$
-i ({d \over d \rho} + {\cos 2\rho \over \sin 2\rho}  )\;
f_{1} \; + \; {im \over \sin \rho }\;f_{1} \; - \;
{k \over \cos \rho} \; f_{2} = \lambda \; f_{2} \; ,
$$
$$
-i ( {d \over d \rho} + { \cos 2\rho \over \sin 2\rho }  )\;
f_{4} \; - \; {im \over \sin \rho }\;f_{4} \; + \;
{k \over \cos \rho} \; f_{3} = \lambda \; f_{3} \; ,
$$
$$
-i ({d \over d \rho} + {\cos 2\rho \over \sin 2\rho}  )\;
f_{3} \; + \; {im \over \sin \rho }\;f_{3} \; - \;
{k \over \cos \rho} \; f_{4} = \lambda \; f_{4} \;      .
\eqno(6.8)
$$

\noindent
Considering eqs.   (6.8)  and     (6.4) together, one  produces a linear algebraic system
$$
-i\epsilon\; f_{3} \; - \; i\lambda \; f_{3} \; + \; iM \; f_{1} = 0 \; ,
$$
$$
-i\epsilon\; f_{4} \; - \; i\lambda \; f_{4} \; + \; iM \; f_{2} = 0 \; ,
$$
$$
-i\epsilon\; f_{1} \; + \; i\lambda \; f_{1} \; + \; iM \; f_{3} = 0 \; ,
$$
$$
-i\epsilon\; f_{2} \; + \; i\lambda \; f_{2} \; + \; iM \; f_{4} = 0 \; .
\eqno(6.9a)
$$

\noindent From  (6.9a) it follows
$$
\lambda = \pm \; \sqrt{\epsilon ^{2}  - M^{2}} \; ,
\qquad f_{3} = {\epsilon - \lambda \over M } \; f_{1} \; , \;
f_{4} = {\epsilon - \lambda \over M } \; f_{2} \; ,
$$
$$
\Psi (t; \rho, \phi,z) = e^{-i\epsilon t} \; e^{im\phi} \; e^{ikz} \;
\left | \begin{array}{r}
f_{1}(\rho) \\ f_{2}(\rho) \\   f_{3}(\rho) \\ f_{4}(\rho)
\end{array} \right | \; .
\eqno(6.9b)
$$

\noindent Taking into account (6.9b), four equations in  (6.4) give a simpler system for
 $f_{1},\; f_{2}$:
$$
( {d \over d \rho} \; + \; {\cos 2\rho \over
\sin 2\rho} \; + \;  {m \over \sin \rho }  ) \; f_{2} \; - \;
i\; (\lambda - {k \over \cos \rho }) \; f_{1} = 0 \; ,
$$
$$
( {d \over d \rho} \; + \; {\cos 2\rho \over
\sin 2\rho} \; - \;  {m \over \sin \rho } ) \; f_{1} \; - \;
i\; (\lambda + {k \over \cos \rho }) \; f_{2} = 0 \; .
\eqno(6.10)
$$

\noindent
By means of separating a special factor
$$
f_{1}(\rho) = { 1 \over \sqrt{\cos \rho \; \sin \rho }} \; \varphi_{1}(\rho) \; ,
\qquad
f_{2}(\rho) = { 1 \over \sqrt{\cos \rho \; \sin \rho }} \; \varphi_{2}(\rho) \; ,
$$

\noindent  one can exclude  the term $(\cos 2\rho / \sin 2\rho)$ from  eqs. (6.10):
$$
( {d \over d \rho}  + \;  {m \over \sin \rho }  )
\; \varphi_{2} \; - \; i\; (\lambda - {k \over \cos \rho }) \; \varphi_{1} = 0 \; ,\;\;
$$
$$
( {d \over d \rho} \; - \;  {m \over \sin \rho }  ) \;
\varphi_{1} \; - \;i\; (\lambda + {k \over \cos \rho }) \; \varphi_{2} = 0 \; .
\eqno(6.11)
$$

\subsection*{7. Some additional  transforming}

\hspace{5mm}
 Summing and subtracting  equations in  (6.11), one gets
$$
{d \over d \rho} (\varphi_{1} + \varphi_{2}) - \;
{m \over \sin \rho }   (\varphi_{1} - \varphi_{2})
\; - \; i\; \lambda\; (\varphi_{1} + \varphi_{2})    \; + \;
{k \over \cos \rho }  (\varphi_{1} - \varphi_{2}) = 0 \; ,
$$
$$
{d \over d \rho} (\varphi_{1} - \varphi_{2}) - \;
{m \over \sin \rho }   (\varphi_{1} + \varphi_{2})
\; + \; i\; \lambda\; (\varphi_{1} - \varphi_{2})    \; - \;
{k \over \cos \rho }  (\varphi_{1} + \varphi_{2}) = 0 \; .
\eqno(7.1)
$$

\vspace{5mm}

\noindent
Now, instead of  functions $(\varphi_{1},\;\varphi_{2})$ it will be helpful to define new
ones ($F_{1},\;F_{2}$)\footnote{Author is grateful to V.S. Otchik for pointing out this
additional transformation over functions  $\varphi_{1}, \varphi_{2}$.}:

$$
\varphi_{1} = \cos {\rho \over 2} \; F_{1} \; -  \; i \;
\sin {\rho \over 2} \; F_{2}  \; ,\qquad
\varphi_{2} = -i\; \sin  {\rho \over 2} \; F_{1} \; +  \;
\cos {\rho \over 2} \; F_{2}  \; ,
\eqno(7.2a)
$$
$$
(\varphi_{1} + \varphi_{2}) = e^{-i\rho /2} (F_{1} + F_{2}) \; ,
\qquad (\varphi_{1} - \varphi_{2}) = e^{+i\rho /2} (F_{1} - F_{2}) \;  .
\eqno(7.2b)
$$

\vspace{5mm}

\noindent With  (7.2b), eqs.  (7.1)  give
$$
{d \over d \rho} (F_{1} + F_{2}) \; - \;
{i \over 2}\; (F_{1} + F_{2}) \;- \;
{m \over \sin \rho } e^{+i\rho}  (F_{1} - F_{2})
\; - \; i\; \lambda\; (F_{1} + F_{2})    \;  +  \;
{ik \over \cos \rho } e^{+i\rho}\; (F_{1} - F_{2}) = 0 \; ,
$$
$$
{d \over d \rho} (F_{1} - F_{2}) \;+ \;
{i \over 2}\; (F_{1} - F_{2}) \;- \;
{m \over \sin \rho } e^{-i\rho}  (F_{1} + F_{2})
\; + \; i\; \lambda\; (F_{1} - F_{2})    \; - \;
{ik \over \cos \rho } e^{-i\rho}\; (F_{1} + F_{2}) = 0 \; .
$$

\vspace{5mm}

\noindent Summing and subtracting them we  will arrive  at
$$
({d \over d \rho} \; -\; m\;{\cos \rho  \over \sin \rho} \;-\; k\;
{\sin \rho \over \cos \rho } )\; F_{1} = i\; (\lambda + {1\over 2} - m + k )\; F_{2} \; ,
$$

$$
({d \over d \rho} \; + m\;{\cos \rho  \over \sin \rho} \;+\; k\;
{\sin \rho \over \cos \rho } )\; F_{2} = i \; (\lambda + {1\over 2} + m - k )\; F_{1} \; .
\eqno(7.3)
$$

\noindent
From  (7.3), for  $F_{1}(\rho)$ can be  produced  the second order equations:
(take notice of the symmetry $m \Longrightarrow -m, \; k  \Longrightarrow -k$):

$$
[ {d^{2} \over d \rho^{2}} \; - \;
{m(m-1) \over \sin^{2} \rho } -{k(k+1) \over \cos^{2} \rho } \; + \;
(\lambda + {1\over 2} )^{2} ] \; F_{1} (\rho) = 0 \; ,
\eqno(7.4a)
$$
$$
[ {d^{2} \over d \rho^{2}} \; - \;
{m(m+1) \over \sin^{2} \rho } - {k(k-1) \over \cos^{2} \rho } \; + \;
(\lambda + {1\over 2} )^{2} ] \; F_{2} (\rho) = 0 \; .
\eqno(7.4b)
$$

\noindent
One  can again turn to the old functions
$$
F_{1} (\rho) = \sqrt{\sin \rho \cos \rho } \;\; G_{1} (\rho) \; ,
$$
$$
F_{2} (\rho) = \sqrt{\sin \rho \cos \rho } \; \;G_{2} (\rho) \; ,
\eqno(7.5)
$$

\noindent then, taking in mind identity
$$
 {d^{2} \over d\rho^{2}} \;
 \sqrt{\sin 2\rho } \;\; G
= \sqrt{\sin 2\rho } \; \left [ \; {d ^{2} \over d \rho^{2}}
+ 2 \; {\cos 2\rho \over \sin2 \rho  } \; {d \over d \rho } \;
- 1   -  { 1  \over  \sin^{2} 2\rho } \;
 \; \right ] \;  G \; .
$$

\vspace{5mm}
\noindent
one  produces
$$
\left [\;  {d^{2} \over d \rho^{2}} \;
+\;
2 \;{\cos 2\rho \over \sin2 \rho  } \;  {d \over d \rho } \;
- 1   -  { 1  \over  \sin^{2} 2\rho } -  \;
{m(m-1) \over \sin^{2} \rho }  -{k(k+1) \over \cos^{2} \rho } \; + \;
(\lambda + {1\over 2} )^{2}
 \; \right ] \; G_{1} = 0 \; ,
\eqno(7.6a)
$$
$$
\left [\;  {d^{2} \over d \rho^{2}} \;\;
+\;
2\;  {\cos 2\rho \over \sin2 \rho  } \;  {d \over d \rho } \;
- 1   -  { 1 \over  \sin^{2} 2\rho } \;
- \;
{m(m+1) \over \sin^{2} \rho }  -{k(k-1) \over \cos^{2} \rho } \; + \;
(\lambda + {1\over 2} )^{2}
 \;  \right ] \; G_{2}  = 0 \; .
\eqno(7.6b)
$$

\vspace{5mm}
Again one can note the symmetry $ m \Longrightarrow -m, \; k \Longrightarrow -k\;$.
These equation can be rewritten as follows:
$$
\left [\;  {d^{2} \over d \rho^{2}} \;
+\;
2 \;{\cos 2\rho \over \sin2 \rho  } \;  {d \over d \rho } \; -\;
{(m -1/2)^{2}  \over \sin^{2} \rho }  -{(k +1/2)^{2}  \over \cos^{2} \rho } \; + \;
(\lambda + {1\over 2} )^{2} -1
 \; \right ] \; G_{1} = 0 \; ,
\eqno(7.7a)
$$

$$
\left [\;  {d^{2} \over d \rho^{2}} \;
+\;
2 \;{\cos 2\rho \over \sin2 \rho  } \;  {d \over d \rho } \; - \;
{(m +1/2)^{2}  \over \sin^{2} \rho }  -{(k -1/2)^{2}  \over \cos^{2} \rho } \; + \;
(\lambda + {1\over 2} )^{2} -1
 \; \right ] \; G_{2} = 0 \; .
\eqno(7.7b)
$$

\noindent
The corresponding first order equations look
as
$$
({d \over d \rho} + {\cos 2\rho \over \sin 2\rho }  - m \;{\cos \rho  \over \sin \rho} \;-\; k\;
{\sin \rho \over \cos \rho } )\; G_{1} = i\; (\lambda + {1\over 2} - m + k )\; G_{2} \; ,
\eqno(7.8a)
$$

$$
({d \over d \rho} + {\cos 2\rho \over \sin 2\rho }  + m\;{\cos \rho  \over \sin \rho} \;+\; k\;
{\sin \rho \over \cos \rho } )\; G_{2} = i \; (\lambda + {1\over 2} + m - k )\; G_{1} \; .
\eqno(7.8b)
$$

\vspace{5mm}

The initial radial functions  $f_{1}(\rho),\;f_{2}(\rho) $  are related with
these  $G_{1},G_{2}$ by the formulas:
$$
 f_{1}(\rho)  = \cos {\rho \over 2}  \; G_{1} - i  \; \sin{\rho \over 2}  \; G_{2} \; , \qquad
f_{2} (\rho)  = -i  \; \sin {\rho \over 2} \;  G_{1} +  \cos{\rho \over 2} \; G_{2} \; .
 \eqno(7.9)
 $$
$$
\lambda = \pm \sqrt{\epsilon^{2} - M^{2}} \;,  \qquad
f_{3} = M^{-1}(\epsilon - \lambda)  \; f_{1} \; , \;
f_{4} =M^{-1}( \epsilon - \lambda )  \; f_{2} \; ,
$$
$$
\Psi_{\epsilon,m,k,\lambda} = e^{-i\epsilon t} \; e^{im\phi} \; e^{ikz} \;
\left | \begin{array}{r}
f_{1}(\rho) \\ f_{2}(\rho) \\   f_{3}(\rho) \\ f_{4}(\rho)
\end{array} \right | \; .
\eqno(7.10)
$$

\noindent
In the following it suffices to trace only by two first components of the fermion  wave function:
$$
\Psi_{\epsilon,m,k,\lambda}   =
 {e^{im\phi} \; e^{ikz} \over  \sqrt{2 }}
\left | \begin{array}{r}
\sqrt{1 + \cos \rho}   \; G_{1} - i  \; \sqrt{1 - \cos \rho}   \; G_{2}  \\[2mm]
-i  \; \sqrt{1 - \cos \rho}   \;  G_{1} +  \sqrt{1 + \cos \rho}   \; G_{2}
\end{array} \right | \; .
\eqno(7.11)
$$

\noindent
When  $\rho$ approaches the  values $0$ and $\pi/2$  the wave function $\psi$ reads respectively
$$
\underline{\rho \rightarrow 0 } \qquad \qquad
\Psi_{\epsilon,m,k,\lambda}   \;\; \rightarrow \;\;
 e^{im\phi} \; e^{ikz}
\left | \begin{array}{r}
 G_{1}(0) -    {i\rho \over 2}    \; G_{2} (0) \\[2mm]
- {i\rho \over 2} \;  G_{1}(0)  +      G_{2}(0)
\end{array} \right | \; ,
\eqno(7.12a)
$$

$$
\underline{\rho \rightarrow \pi/2 } \qquad
\Psi_{\epsilon,m,k,\lambda}  \;\;  \rightarrow\;\;
 e^{im\phi} \; e^{ikz}
\left | \begin{array}{r}
 G_{1}(\pi/2)  - i  \; G_{2}(\pi/2)   \\[2mm]
-i  \;  G_{1}(\pi/2)  +     \; G_{2}(\pi/2)
\end{array} \right | \; .
\eqno(7.12b)
$$

\subsection*{8.  Fermion in a  curved space, tetrad gauge symmetry and formulation of
continuity criterion    for Dirac  wave functions in  spherical space $S_{3}$   }

\begin{quotation}

The requirement of continuity for a fermion wave  function
in  any Riemannian space-time is much more complicated than  in the
flat space-time by Minkowski.
The matter is that the form of generally covariant Dirac equation  [...] itself
involves  the tetrad formalism. In other  words,
one may use any form of the Dirac equation in space-time with a given
metrical tensor $g_{\alpha\beta}(x)$ from many different ones depending on a chosen
tetrad, all such tetrad-based  fermion wave functions $\Psi_{tetrad} (x)$
are related to each other by  local gauge transformations from the group $SL(2.C)$:
 $$
\Psi_{tetrad ' } (x) = S(x) \; \Psi_{tetrad} (x) ,\qquad S(x) \in SL(2.c) \; .
$$

\noindent
As a rule, these $(4\times 4)$ gauge matrices S(x) are themselves singular.
So  the explicit embodiment of the general continuity criterion for fermion
wave functions depend on the occasional choice of  the tetrad.

\end{quotation}

For more clarity, one may turn to the familiar  case of  flat space.
It is tacitely  accepted that in ordinary representation of the Dirac equation
all  fermion wave  function must be single-valued and continuous.
However, if one wishes to employ any  other tetrad
\footnote{A more detailed treatment see in  [22,23].}, for instance let it be the  cylindrical
tetrad  (see (6.5a))

\vspace{2mm}
\underline{In the space  $E_{3}$ }
\vspace{2mm}

$$
\Psi^{cyl} = S\; \Psi^{cart} \; , \;\; S = \left | \begin{array}{cc}
B  &  0  \\  0  &  B  \end{array} \right | \; , \;\;
B =  \left | \begin{array}{cc}
e^{+i\phi/2} & 0 \\ 0  & e^{-i\phi/2}
 \end{array} \right | \; ,
\eqno(8.1a)
$$

\noindent because of the singular behavior of the gauge  matrix  $B(\phi)$ at the axis $z$
the wave  function $\Psi^{cyl} $  will become discontinuous at the axis $z$:
$$
\Psi _{cyl} = e^{-i\epsilon t} \; e^{ikz} \;
\left | \begin{array}{r}
e^{im\phi} f_{1}(\rho) \\e^{im\phi} f_{2}(\rho) \\  e^{im\phi} f_{3}(\rho) \\e^{im\phi} f_{4}(\rho)
\end{array} \right | \;  , \qquad
\Psi _{cart} = e^{-i\epsilon t} \; e^{ikz} \;
\left | \begin{array}{r}
 e^{im\phi}  e^{-i\phi/2} f_{1}(\rho) \\  e^{im\phi} e^{+i\phi/2} f_{2}(\rho) \\
  e^{im\phi}  e^{-i\phi/2}  f_{3}(\rho) \\  e^{im\phi}
e^{+i\phi/2} f_{4}(\rho)
\end{array} \right | \; .
\eqno(8.1b)
$$

\noindent From (8.1b) one may see that to the single-valued functions in Cartesian
basin correspond the non-single-valued ones in  the cylindrical basis.
In eq. (8.1b) the quantum number $m$  must be half-integer,
one should note  the gauge invariant character of the proper value  problem:
$$
cart:  \qquad \hat{J}_{3} = \hat{l}_{3} + \hat{S}_{3}\; ; \qquad
\hat{J}_{3} \; \Psi_{Cart}  = m \; \Psi_{Сart} \; ,
$$
$$
cyl: \qquad  \hat{l}_{3} \; \Psi _{cyl} = m\; \Psi_{cyl} \; .
\eqno(8.2)
$$

In the curved space $S_{3}$ all the situation is principally the same though more
complicated. There exists an analog of the Cartesian basis, here its role plays
 a conformally-flat basis (coordinates  $(x^{1},x^{2},x^{3})$ and the tetrad
 $e_{(a)}^{\;\;\alpha}(x) $)
$$
 n_{0}^{2} + n_{1}^{2} + n_{2}^{2} +  n_{3}^{2}  = 1 \; ,
$$
$$
x^{1} = {n_{1} \over 1 + n_{4}}\; , \;\;
x^{2} = {n_{2} \over 1 + n_{4}}\; , \;\;
x^{3} = {n_{3} \over 1 + n_{4}}\; ,
$$
$$
dl^{2}(x) = {1 \over f^{2}} \; [\; (dx^{1})^{2} +
(dx^{2})^{2} +    (dx^{3})^{2} \; ] \; , \;\; f = { 1 + x^{2}\over 2} \; ,
$$
$$
e_{(a)}^{\;\;\alpha}(x) = \left | \begin{array}{cccc}
1 &  0    &     0  & 0 \\
0 &  1/f &  0  &  0   \\
0 &  0 &  1/f  &  0   \\
0 &  0 &  0   & 1/f   \\
\end{array}  \right | \; .
\eqno(8.3a)
$$

\noindent
Spinor gauge transformation relating conformally-flat and cylindrical tetrads
is given by (details see in Supplement A)
$$
\Psi_{cyl} = S\; \Psi_{cart} \; , \qquad  S = \left | \begin{array}{cc}
B  &  0  \\  0  &  B  \end{array} \right | \; , \;\;
\eqno(8.3b)
$$
$$
B  =  \sigma \; \left | \begin{array}{rr}
A \; e^{+i\phi/2}   & C\;e^{-i\phi/2}     \\
- C\;e^{+i\phi/2}   &  A \; e^{-i\phi/2}  \end{array} \right | \;  , \;\; \sigma \pm 1 \; ,
$$
$$
A =  \sqrt{{ (1 + \cos \rho ) \over
(1 + \cos \rho \cos z ) }} \; \cos {z\over 2} \;\;, \;\;\;
C =  \sqrt{{ (1 - \cos \rho ) \over
(1 + \cos \rho \cos z ) }} \; \sin  {z\over 2} \; .
$$

One can calculate an explicit form of  $\hat{l}_{3}$ in conformally-flat  basis
$$
\hat{l}_{3} = -i
 \partial_{\phi} \qquad \Longrightarrow \qquad  B^{-1} \hat{l}_{3} \;B  =
-i  B^{-1}  \partial_{\phi} B\; =
$$
$$
=-i\partial_{\phi} + {1 \over 2} \;
\left | \begin{array}{rr}
A \; e^{-i\phi/2}   &- C\;e^{-i\phi/2}     \\
 C\;e^{+i\phi/2}   &  A \; e^{+i\phi/2}  \end{array} \right |
 \left | \begin{array}{rr}
+A \; e^{+i\phi/2}   & -C\;e^{-i\phi/2}     \\
- C\;e^{+i\phi/2}   &  -A \; e^{-i\phi/2}  \end{array} \right |=
$$
$$
= -i\partial_{\phi} + {1 \over 2} \left | \begin{array}{cc}
A^{2} +C^{2} & 0 \\
0 & -A^{2} -C^{2}
 \end{array} \right | = -i\partial_{\phi} + {1 \over 2} \left | \begin{array}{cc}
 1 & 0 \\0 &-1    \end{array} \right | = -i\partial_{\phi} + {1\over 2} \;\sigma^{3}
  \; ,
 $$

\noindent
so
$$
cyl \;\; \Longrightarrow \;\;cart : \qquad
B^{-1} \hat{l}_{3} \; B  =
-i\partial_{\phi} + {1\over 2} \;\sigma^{3}=\hat{J}_{3}
  \; .
 \eqno(8.4)
 $$

Now let us calculate the form of the wave function in conformally-flat basis:
$$
\psi_{cart} = B^{-1} \psi_{cyl} =
  e^{im\phi} \; e^{ikz} \;
\left | \begin{array}{r}
A \; e^{-i\phi/2}\;  f_{1}(\rho)  \; - \; C\;e^{-i\phi/2} \;   f_{2}(\rho)    \\
 C\;e^{+i\phi/2}  \; f_{1}(\rho)  \; + \;  A \; e^{+i\phi/2} \; f_{2}(\rho)
  \end{array} \right |=
\left | \begin{array}{c}
\psi_{1} \\ \psi_{2}   \end{array} \right | \; ;
\eqno(8.5a)
$$

\noindent
for   $\psi_{1}$  and  $\psi_{2}$ we  obtain expressions
$$
\psi_{1} = e^{i(m-1/2)\phi}  \left [
\; f_{1} (\rho)  \sqrt{{1+ \cos \rho \over 1 + \cos \rho \cos z}} \; \cos{z\over 2} \; e^{ikz} -
f_{2} (\rho)  \sqrt{{1- \cos \rho \over 1 + \cos \rho \cos z}} \; \sin{z\over 2} \;  e^{ikz}\; \right ]\;,
$$
$$
\psi_{2} = e^{i(m+1/2)\phi}  \left [
\; f_{1} (\rho)  \sqrt{{1- \cos \rho \over 1 + \cos \rho \cos z}} \; \sin{z\over 2}  \;e^{ikz} +
f_{2} (\rho)  \sqrt{{1+ \cos \rho \over 1 + \cos \rho \cos z}} \; \cos{z\over 2} \; e^{ikz}\; \right ]\;.
$$
$$
\eqno(8.5b)
$$

\noindent
From where, taking into account
$$
\cos{z\over 2}  e^{ikz}  = {e^{i(k+1/2)z} +  e^{i(k-1/2)z} \over 2} \; , \qquad
\sin{z\over 2}  e^{ikz}  = {e^{i(k+1/2)z} -  e^{i(k-1/2)z} \over  2i} \; ,
$$

\noindent  and expression for  $f_{1},f_{2}$ in terms of  $G_{1},G_{2}$),
it follows
$$
\psi_{1} = { e^{i(m-1/2)\phi } \; e^{i(k+1/2)z}  \over \sqrt{1 + \cos \rho \cos z } }
  \; \left [   \; (   1  + \cos \rho \;  e^{-iz} )
  \;   G_{1} (\rho)    -i \sin \rho \;  e^{-iz} \;    G_{2} (\rho)  \; \right ] \; ,
$$
$$
\psi_{2} = { e^{i(m+1/2)\phi } \; e^{i(k-1/2)z}  \over \sqrt{1 + \cos \rho \cos z } }
  \; \left [ \;   (   1  + \cos \rho  \; e^{+iz} )
  \;   G_{2} (\rho)    -i \sin \rho \;  e^{+iz} \;    G_{1} (\rho)  \;     \right ] \; .
$$
$$
\eqno(8.5c)
$$

From eqs. (8.5с)  one can conclude:
\begin{quotation}
{\em
only when  $m$  and   $k$ are half-integer, the constituents
 $\psi_{1}$  and  $\psi_{2}$  of the fermion  wave function in conformally-flat basis
 may be single-valued and continuous  in the spherical space $S_{3}$.}

\end{quotation}

\vspace{10mm}
It may be of interest the  form of   $\hat{N}_{3}^{cyl} = -i\partial_{z}$ after
transforming to a conformally-flat basis, $\hat{N}_{3}^{cart} = -i  B^{-1}  \partial_{z} B$:
$$
\hat{N}_{3}^{cart}  =
-i\partial_{z} -i
 \left | \begin{array}{rr}
0 & +e^{-i\phi} (A \partial_{z} C -C \partial_{z} A) \\[2mm]
-e^{i\phi} (A \partial_{z} C -C \partial_{z} A) &  0
\end{array} \right | \; ;
\eqno(8.6a)
$$

\vspace{5mm}
\noindent
from where taking into account
$$
A \partial_{z} C -C \partial_{z} A  = { \sin \rho \over
2 (1+\cos \rho \cos z )^{2}  }  \; ,
$$

\noindent we  get
$$
\hat{N}_{3}^{cart}  = -i\partial_{z} + { i\sin \rho \over
2 (1+\cos \rho \cos z )^{2}  }
 \left | \begin{array}{rr}
0 & -e^{-i\phi}  \\[2mm]
+e^{i\phi} &  0
\end{array} \right | \; ;
\eqno(8.6b)
$$

\noindent With the use of relations
$$
x^{1} = {\sin \rho \; \cos \phi \over 1 + \cos \rho \cos z} \; , \;
x^{2} = {\sin \rho \; \sin \phi \over 1 + \cos \rho \cos z} \; ,
\;x^{3} = {\cos \rho \; \sin \phi \over 1 + \cos \rho \cos z} \; , \;
f = {1 \over 1 + \cos \rho \cos z}   \; ,
$$

\vspace{5mm}
\noindent
$\hat{N}_{3}^{cart}$  from (8.6b) will look as
$$
\hat{N}_{3}^{cart}  =
-i\partial_{z} +  {i\over 2} \; f(x) \; \left | \begin{array}{cc}
0 & - (x^{1} -ix^{2} )  \\
(x^{1} +ix^{2} )&  0
\end{array} \right | \;.
\eqno(8.6c)
$$

\subsection*{9. Analysis of  the special cases $\lambda = \pm 1/2$ and  $ \lambda = \pm 3/2$}

\hspace{5mm}
The above relation between  differential equations of first order
(7.8) and  equations of second order (7.7) has some peculiarities.
Indeed, from (7.8)  it follows (7.7) only if

$$
\lambda + {1\over 2} - m + k  \neq 0 , \qquad
\lambda + {1\over 2} + m - k  \neq 0 \; .
\eqno(9.1a)
$$

\noindent
In opposite case
$$
\lambda + {1\over 2} - m + k  = 0 , \qquad
\lambda + {1\over 2} + m - k  = 0 \; ,
$$

\noindent which is possible when
$$
\lambda = -1/2, \qquad m= k \; ,
\eqno(9.1b)
$$

\noindent the system (7.8) turns to be a pair of two
separated first order equations:
$$
({d \over d \rho} + {\cos 2\rho \over \sin 2\rho }  - m \;{\cos \rho  \over \sin \rho} \;-\; m\;
{\sin \rho \over \cos \rho } )\; G_{1} = 0\; ,
\eqno(9.2a)
$$
$$
({d \over d \rho} + {\cos 2\rho \over \sin 2\rho }  + m\;{\cos \rho  \over \sin \rho} \;+\; m\;
{\sin \rho \over \cos \rho } )\; G_{2} = 0\; .
\eqno(9.2b)
$$

\vspace{5mm}
\noindent
Eqs. (9.2) may be quite easily solved.
Indeed,  they can be rewritten as
$$
(\; {d \over d \rho} +    {\cos 2\rho  -2m\over \sin 2\rho }\; )\; G_{1} = 0\; ,
\qquad ( \; {d \over d \rho} +   {\cos 2\rho  +2m \over \sin 2\rho }\;  )\; G_{2} = 0\; ,
$$

\noindent and further
$$
{G'_{1} \over G_{1}} = +{2m \over \sin 2\rho } - {\cos 2\rho \over  \sin 2\rho} \; ,
\qquad
{G'_{2} \over G_{2}} = -{2m \over \sin 2\rho } - {\cos 2\rho \over  \sin 2\rho} \; ,
$$

\noindent these  can be readily integrated:
$$
\ln G_{1} = \ln C_{1} + m \; \ln  (\tan \rho ) -{1\over 2} \ln (\sin 2\rho) \; ,
$$
$$
\ln G_{2} = \ln C_{2} - m \; \ln  (\tan \rho ) -{1\over 2} \ln (\sin 2\rho) \; .
$$

\noindent
Thus, the following class of solutions is found:
$$
\underline{\lambda = -1/2, \;\; k = + m \;}  , \qquad
G_{1} (\rho) = C_{1} \; { \tan^{m}  \rho  \over \sqrt{\sin 2\rho} } \; , \qquad
G_{2} (\rho) = C_{2} \; { \tan^{-m} \rho \over \sqrt{\sin 2\rho} } \; .
\eqno(9.3)
$$

\begin{quotation}

Take notice that the constructed wave functions (9.3)
are singular in the points  $\rho =0, \pi/2$. This makes necessary  some additional
analysis because  one must exclude all solutions not-finite or
discontinuous in the  space $S_{3}$.

\end{quotation}
For instance, let us consider the solution
$$
\underline{\lambda = -1/2,} \qquad  k =+1/2,   m =+1/2 \;  ,
$$
$$
G_{1} (\rho) = { C_{1}\over \sqrt{2} \cos \rho }  \; , \qquad
G_{2} (\rho) = {C_{2} \over \sqrt{2} \sin \rho }  \; .
\eqno(9.4a)
$$

It may be viewed as physical only if its singularities will be eliminated by
means of the gauge transformation involved.
In conformally-flat tetrad this solution
will read as
$$
\psi_{1} = {  e^{iz}  \over  \sqrt{2} \sqrt{1 + \cos \rho \cos z } }
  \; \left [   \; (   1  + \cos \rho \;  e^{-iz} )
  \;  { C_{1}\over  \cos \rho }    -i \sin \rho \;  e^{-iz} \;
    {C_{2} \over  \sin \rho }  \; \right ] \; ,
$$
$$
\psi_{2} = { e^{i\phi }    \over \sqrt{2} \sqrt{1 + \cos \rho \cos z } }
  \; \left [ \;   (   1  + \cos \rho  \; e^{+iz} )
  \; {C_{2} \over  \sin \rho }
      -i \sin \rho \;  e^{+iz} \;  { C_{1}\over  \cos \rho }  \;     \right ] \; .
$$
$$
\eqno(9.4b)
$$

When  $\rho =0 $   we have
$$
\psi_{1}  \qquad \rightarrow  \qquad  {  e^{iz}  \over \sqrt{2} \sqrt{1 + \cos z } }
  \; [   \; (   1  +   e^{-iz} )
  \;   C_{1}    -i \;  e^{-iz} \;     C_{2}   \;  ]   \; ,
$$
$$
\psi_{2}  \qquad \rightarrow  \qquad  { e^{i\phi }    \over  \sqrt{2} \sqrt{1 +  \cos z } }
  \;  [ \;   (   1  +  e^{+iz} )
  \; {C_{2} \over   0  }
      -0 \; i\;    e^{+iz} \;   C_{1}  \;     ]  \; .
      \eqno(9.4c)
      $$

\noindent
The solution must be considered on the closed curve
$n_{3} = \sin z , \; n_{4} = \cos z$,  so to be continuous in  на $S_{3}$  it
may depend only on the $z$, or it must
be a constant. From this it follows  $C_{2}=0$.
When  $\rho = \pi /2 $  we have
$$
\Psi_{1} \qquad \rightarrow  \qquad      {e^{iz} \over \sqrt{2}}
  \; [    \;  { C_{1}\over  0 }    -i  \;  e^{-iz} \;
    C_{2}  \; ] \; , \qquad
\Psi_{2} \qquad \rightarrow  \qquad     {e^{i\phi}  \over  \sqrt{2}}
  \;  [ \; C_{2}       -i  \;   e^{+iz} \;  { C_{1}\over 0 }  \;     ] \; .
\eqno(9.4d)
$$

\noindent Here,
the solution must be considered on the closed curve
$n_{1} = \cos \phi  , \; n_{2} = \sin \phi $,
so to be continuous in  на $S_{3}$  it
may depend only on the $\phi$ or it must
be a constant. From this it follows  $C_{1}=0$. Thus, the solution
 (9.4a)-(9.4b)  is a good in the sense of continuity only in trivial case
  $C_{1}=0,C_{2}=0$. Therefore, this solution must be excluded.

In the same manner one can treat another (symmetrical) case:
$$
\underline{\lambda = -1/2, } \qquad  k = -1/2,  m =-1/2 \;  ,
$$
$$
G_{1} (\rho) = { C_{1}\over \sqrt{2} \sin \rho }  \; , \qquad
G_{2} (\rho) = {C_{2} \over \sqrt{2} \cos \rho } ; ,
\eqno(9.5a)
$$

\noindent  in conformally-flat basis it looks as
$$
\psi_{1} = { e^{-i\phi }    \over  \sqrt{2} \sqrt{1 + \cos \rho \cos z } }
  \; \left [   \; (   1  + \cos \rho \;  e^{-iz} )
  \;   { C_{1}\over  \sin \rho }     -i \sin \rho \;  e^{-iz} \;
  {C_{2} \over \cos \rho }   \; \right ] \; ,
$$
$$
\psi_{2} = {  e^{-iz}  \over \sqrt{2} \sqrt{1 + \cos \rho \cos z } }
  \; \left [ \;   (   1  + \cos \rho  \; e^{+iz} )
  \;  {C_{2} \over  \cos \rho }
      -i \sin \rho \;  e^{+iz} \;    { C_{1}\over  \sin \rho }   \;     \right ] \; .
$$
$$
\eqno(9.5b)
$$

\noindent \underline{When $\rho =0 $}
$$
\psi_{1}  \qquad \rightarrow  \qquad
{ e^{-i\phi }    \over  \sqrt{2} \sqrt{1 +  \cos z } }
  \; [   \; (   1  +   e^{-iz} )
  \;    {C_{1} \over 0}    -i\;  0 \;  e^{-iz} \;
  C_{2} \; ] \; ,
$$
$$
\psi_{2}  \qquad \rightarrow  \qquad
{  e^{-iz}  \over \sqrt{2} \sqrt{1 + \cos z } }
  \; [ \;   (   1  +  e^{+iz} )
  \;  C_{2}
      -i  e^{+iz} \;     C_{1}   \;  ] \; .
\eqno(9.5c)
$$

\underline{When $\rho = \pi/2$}
$$
\psi_{1}  \qquad \rightarrow  \qquad
{ e^{-i\phi }    \over  \sqrt{2}  }
  \;  [   \;
  \;    C_{1}     -i \;  e^{-iz} \;
  {C_{2} \over 0 }   \; ] \; ,
$$
$$
\psi_{2} \qquad \rightarrow  \qquad
{  e^{-iz}  \over \sqrt{2}  }
  \; [ \;        {C_{2} \over  0 }
      -i  \;  e^{+iz} \;     C_{1}   \;      ] \; .
\eqno(9.5d)
$$

\noindent
Evidently, this solution must be excluded by continuity reason.

All the remaining solutions of the type (9.3)
$$
\underline{\lambda = -1/2,} \qquad  k =N +1/2,   m =N +1/2 \;,   \;\;(N = 1,2,... )\;,
$$
$$
G_{1} (\rho) = {\sin ^{N} \rho \over \cos^{N} \rho} \; \; { C_{1}\over \sqrt{2} \cos \rho }  \; , \qquad
G_{2} (\rho) = {\cos ^{N} \rho \over \sin^{N} \rho} \;\;  {C_{2} \over \sqrt{2} \sin \rho }  \; ;
$$
$$
\underline{\lambda = -1/2,} \qquad  k =-N -1/2,  m = -N -1/2 \;,   \;\;N = 1,2,...  \;,
$$
$$
G_{1} (\rho) =   {\cos ^{N} \rho \over \sin^{N} \rho} \;\; { C_{1}\over \sqrt{2} \sin \rho }  \; , \qquad
G_{2} (\rho) =  {\sin ^{N} \rho \over \cos^{N} \rho} \; \;{C_{2} \over \sqrt{2} \cos \rho }  \; .
$$

\noindent are not finite in all tetrad bases  and  they must be excluded.

\vspace{5mm}

There are two other classes of simplest solutions which
should be considered here as well: they.
Indeed, let it be
$$
G_{1}(\rho)  = C_{1} = \; const , \qquad    m = +1/2 \; , \;\; k=-1/2 \; .
\eqno(9.6a)
$$

\noindent then eq. (7.8)  take the form
$$
( {\cos 2\rho \over \sin 2\rho }  -  \;{\cos \rho  \over 2\sin \rho} \;+
{\sin \rho \over 2 \cos \rho } )\; C_{1} = i\; (\lambda  -1/2 )\; G_{2} (\rho)\; ,
$$
$$
({d \over d \rho} + {\cos 2\rho \over \sin 2\rho }  + {\cos \rho  \over 2\sin \rho} -
{\sin \rho \over 2 \cos \rho } )\; G_{2}(\rho) = i \; (\lambda + 3/2  )\; C_{1} \; ,
$$

\noindent or
$$
 0  = (\lambda  -1/2 )\; G_{2} (\rho) \; ,
 $$
$$
( \; {d \over d \rho} + 2 \; {\cos 2\rho \over \sin 2\rho } \; )\; G_{2} (\rho) =
i \; (\lambda + 3/2  )\; C_{1} \; ,
\eqno(9.6b)
$$

\noindent
there can be separated three possibilities:
$$
(1) \qquad \lambda \neq1/2,-3/2 : \qquad  \qquad   C_{1}(\rho) =0 ,\qquad G_{2}(\rho) =0\; ,
$$
$$
(2) \qquad \lambda = -3/2 : \qquad \qquad  G_{1} (\rho) = C_{1} , \qquad G_{2} (\rho) = 0 \; ,
\eqno(9.6c)
$$
$$
(3) \qquad
\lambda = +1/2:  \qquad \qquad
( \; {d \over d \rho} + 2 \; {\cos 2\rho \over \sin 2\rho } \; )\; G_{2}(\rho) =
2i \;  C_{1} \; .
$$

\noindent In the  last case  $\lambda = +1/2$  equations are readily integrated:
$$
G_{2} = {H (\rho)\over \sin 2\rho }, \qquad
({d \over d \rho} + 2 \; {\cos 2\rho \over \sin 2\rho } \; )\; {H (\rho)\over \sin 2\rho } =
 {1 \over  \sin 2\rho } {d H\over d \rho}  \; , \qquad \Longrightarrow
$$
$$
{d H \over d\rho }  =  2iC_{1}\; \sin 2\rho , \qquad H = C _{2}- iC_{1} \cos 2\rho \; ,
$$

\noindent
that is
$$
G_{2}(\rho)  = { C _{2}- iC_{1} \cos 2\rho  \over  \sin 2\rho } \; .
$$

Thus, two solutiona constaructed:
$$
(2): \qquad  \underline{ \lambda = -3/2 }\;, \;  m = +1/2 \; , \; k=-1/2 \; ,
\qquad \qquad
$$
$$
 \qquad \qquad \qquad G_{1} (\rho) = C_{1} , \qquad G_{2} (\rho) = 0 \; ; \qquad \qquad
$$
$$
\Psi_{1} = { 1   \over \sqrt{1 + \cos \rho \cos z } }
  \; (   1  + \cos \rho \;  e^{-iz} )
  \;   C_{1}  \; ,
$$
$$
\Psi_{2} = { e^{+i\phi }  e^{-iz}  \over \sqrt{1 + \cos \rho \cos z } }
  \; (  \;      -i \sin \rho \;  e^{+iz} \;    C_{1}  \;   ) \; ;
\eqno(9.7)
$$
$$
\rho=0 : \qquad \Psi_{1} = { C_{1}  (   1  +   e^{-iz} )
   \over \sqrt{1 +\cos z } }   \;   , \qquad
  \Psi_{2} =  0
   \; .
$$
$$
\rho=\pi/2 : \qquad \qquad \Psi_{1} =   \;    C_{1}  \; , \qquad
  \Psi_{2} =  -i C_{1} e^{+i\phi }  \; .
$$

{\em it is finite and continuous in  $S_{3}$}

$$
(3): \qquad \qquad
\underline{ \lambda = +1/2  } \; , \;  m = +1/2 \; , \; k=-1/2 \; , \qquad
$$
$$
\qquad \qquad   G_{1}(\rho)  = C_{1} \; , \qquad
G_{2}(\rho)  = { C _{2}- iC_{1} \cos 2\rho  \over  2\sin \rho \cos \rho } \; .
\eqno(9.8a)
$$
 $$
\Psi_{1} = { 1  \over \sqrt{1 + \cos \rho \cos z } }
  \; \left [   \; (   1  + \cos \rho \;  e^{-iz} )
  \;   C_{1}     -i \sin \rho \;  e^{-iz} \;
  { C _{2}- iC_{1} \cos 2\rho  \over  2\sin \rho \cos \rho }   \; \right ] \; ,
$$
$$
\Psi_{2} = { e^{i\phi }  e^{-iz}  \over \sqrt{1 + \cos \rho \cos z } }
  \; \left [ \;   (   1  + \cos \rho  \; e^{+iz} )
  \;  { C _{2}- iC_{1} \cos 2\rho  \over  2\sin \rho \cos \rho }
      -i \sin \rho \;  e^{+iz} \;    C_{1}   \;     \right ] \; ;
$$
$$
\eqno(9.8b)
$$
$$
\rho=0:
\qquad
\Psi_{1} = { 1  \over \sqrt{1 +  \cos z } }
  \; [   \; (   1  +   e^{-iz} )
  \;   C_{1}     -i \;  e^{-iz} \;
  { C _{2} - iC_{1}  \over  2  }   \; ] \; ,
$$
$$
\rho=0:
\qquad
\Psi_{2} = { e^{i\phi }  e^{-iz}  \over \sqrt{1 +  \cos z } }
  \;    (   1  +  e^{+iz} )
  \;  { C _{2}- iC_{1}   \over   0  }  \;    \; ;
$$
$$
\rho=\pi/2:
\qquad
\Psi_{1} =      \;   \;   C_{1}     -i   e^{-iz} \;
  { C _{2}+ iC_{1} \over   0 }   \; ,
$$
$$
\rho=\pi/2:
\qquad
\Psi_{2} =  e^{i\phi }  e^{-iz}
 [ \; { C _{2}+ iC_{1}  \over   0 }
      -i \;  e^{+iz} \;    C_{1}   \;      ] \; .
\eqno(9.8c)
$$

\noindent
{\em it must be rejected. by continuity reason.
}
There exists another symmetrical variant:
$$
G_{2}(\rho)  = C_{2} = \; const , \qquad    m = -1/2 \; , \;\; k=+1/2 \; ,
\eqno(9.9a)
$$

\noindent then  (7.8)  takes the form
$$
 0  = (\lambda  -1/2 )\; G_{1} (\rho) \; ,
 $$
$$
( \; {d \over d \rho} + 2 \; {\cos 2\rho \over \sin 2\rho } \; )\; G_{1} (\rho) =
i \; (\lambda +3/2  )\; C_{2} \; ,
\eqno(9.9b)
$$

\noindent
Depending on three possibilities
$$
(1) \qquad \lambda \neq + 1/2,-3/2 : \qquad  \qquad   C_{1}(\rho) =0 ,\qquad G_{2}(\rho) =0\; ,
$$
$$
(2) \qquad \lambda = -3/2 : \qquad \qquad  G_{1} (\rho) = 0 , \qquad G_{2} (\rho) =C_{2} \; ,
\eqno(9.9c)
$$
$$
(3) \qquad \lambda = +1/2:  \qquad \qquad
( \; {d \over d \rho} + 2 \; {\cos 2\rho \over \sin 2\rho } \; )\; G_{1}(\rho) =
2i \;  C_{2} \; .
$$

\noindent we can obtain not trivial solutions. When $\lambda = +1/2$
$$
(3) \qquad
G_{1}(\rho)  = { C _{1}- iC_{2} \cos 2\rho  \over  \sin 2\rho } \; .
$$

\noindent
Thus, two solutions are constructed:
$$
(2) \;\;  \underline{\lambda = -3/2 ,} \qquad   m = -1/2 \; , \;\; k=+1/2 \; ,
$$
$$
\qquad  \qquad G_{1} (\rho) = 0 , \qquad G_{2} (\rho) = C_{2}  \; ;
\eqno(9.10a)
$$

$$
(3) \;\;\underline{\lambda = +1/2 \;,}   \qquad m = -1/2 \; , \;\; k=+1/2  \; ,
$$
$$
\lambda = +1/2:  \qquad   G_{2}(\rho)  = C_{2} \; ,
G_{1}(\rho)  = { C _{1}- iC_{2} \cos 2\rho  \over  \sin 2\rho } \; .
\eqno(9.10b)
$$

Only the solution (2) turns to be  finite and continuous in $S_{3}$. Its
cartesian form and behavior in special points is given below:
$$
(2) \qquad   \underline{\lambda = -3/2 \;, \;  m = -1/2 \; , \;\; k=+1/2 \; },
$$
$$
\psi_{1} = { e^{-i\phi }  e^{iz}  \over \sqrt{1 + \cos \rho \cos z } }
  \; ( -i \sin \rho \;  e^{-iz} \;    C_{2}  \; )  \; ,
$$
$$
\psi_{2} = { 1 \over \sqrt{1 + \cos \rho \cos z } }
  \;   (   1  + \cos \rho  \; e^{+iz} )   \;   C_{2}      \; .
\eqno(9.11)
$$
$$
\rho=0: \qquad
\psi_{1} =  0  \; , \qquad
\psi_{2} = { C_{2}   (   1  +  \; e^{+iz} )  \over \sqrt{1 + \cos z } }
  \;    ,
$$
$$
\rho=\pi/2: \qquad
\psi_{1} = -i C_{2}  e^{-i\phi }     \;  , \qquad
\psi_{2} =   \;     C_{2}      \; .
$$

{\em It is a good function  in the sense of continuity. }

\begin{quotation}
The general conclusion may be done:
no solutions with  $\lambda=\pm 1/2$, continuous in  $S_{3}$, exist.
The Dirac  wave functions with   $\lambda = \pm 3/2$ are constructed --
see  (9.7)  and (9.11).
\end{quotation}

\subsection*{10.  Quantization of the Dirac energy levels}

\hspace{5mm}
Let us turn again to the first order equations  (7.3)
$$
({d \over d \rho} \; -\; m\;{\cos \rho  \over \sin \rho} \;-\; k\;
{\sin \rho \over \cos \rho } )\; F_{1} = i\; (\lambda + {1\over 2} - m + k )\; F_{2} \; ,
\eqno(10.1a)
$$

$$
({d \over d \rho} \; + m\;{\cos \rho  \over \sin \rho} \;+\; k\;
{\sin \rho \over \cos \rho } )\; F_{2} = i \; (\lambda + {1\over 2} + m - k )\; F_{1} \; ,
\eqno(10.1b)
$$

\noindent or equivalent to them second order equations (7.4):
$$
[ {d^{2} \over d \rho^{2}} \; - \;
{m(m-1) \over \sin^{2} \rho } -{k(k+1) \over \cos^{2} \rho } \; + \;
(\lambda + {1\over 2} )^{2} ] \; F_{1} (\rho) = 0 \; ,
\eqno(10.2a)
$$
$$
[ {d^{2} \over d \rho^{2}} \; - \;
{m(m+1) \over \sin^{2} \rho } - {k(k-1) \over \cos^{2} \rho } \; + \;
(\lambda + {1\over 2} )^{2} ] \; F_{2} (\rho) = 0 \; .
\eqno(10.2b)
$$

{\em  The values of energy can be  established  if all the function involved are

 polynomials  and the quantum  number $m$ and  $n$ are  taken to be half-integer. }

\noindent
The radial functions $F_{1},\;F_{2}$ are taken in the form (calculation can be done
for one function only)
$$
F_{1} = \sin^{A_{1}}\rho \; \cos^{B_{1}}\rho \; R_{1}(\rho) \; ,\qquad
F_{2} = \sin^{A_{2}} \rho \; \cos^{B_{2}} \rho \; R_{2}(\rho) \; .
\eqno(10.3)
$$

\noindent Substituting (10.3)  into (10.2а), one gets
$$
{d^{2} \over d \rho^{2}} \; R_{1} + \;
( 2A_{1} {\cos \rho \over   \sin \rho } - 2B_{2} {\sin \rho \over \cos \rho } )
{d \over d \rho} \; R_{1} \; + \;
\left [  A_{1}(A_{1}-1) {\cos^{2}\rho  \over \sin^{2}  \rho} -
{m(m-1) \over \sin^{2}\rho }  \;+\;
\right.
\eqno(10.4)
$$
$$
\left.
+ \;  B_{1}(B_{1}-1) {\sin^{2}\rho  \over \cos^{2}  \rho} -
{k(k+1) \over \cos^{2}\rho }   \; - \;
A_{1}(B_{1}+1) - B_{1} (A_{1} + 1)  +  (\lambda + {1\over 2} )^{2} \right ]  \;
 R_{1}(\rho) = 0\; .
$$

\noindent Let $A_{1},B_{1}$ obey the relations:
$$
A_{1}(A_{1} -1) = m(m-1) \qquad \Longrightarrow
\qquad A_{1} = m , - m + 1  =
 \pm (-m+ 1/ 2) +1/ 2 \; ,
$$
$$
B_{1}(B_{1}-1) = k(k+1) \qquad  \Longrightarrow
\qquad
    B_{1} = -k ,  k+1  =
\pm (k+ 1/2) + 1/ 2 \; .
\eqno(10.5)
$$

\noindent Then eq.  (10.4) will be much simpler
$$
{d^{2} \over d \rho^{2}} \; R_{1} \; + \;
2\; ( A_{1}\; {\cos \rho \over   \sin \rho } \; - \; B_{1}\;
 {\sin \rho \over \cos \rho } ) \; {d \over d \rho} \; R_{1} \; + \;
[\; (\lambda + {1\over 2} ) ^{2} \; - \; (A_{1}+B_{1})^{2}\; ]\;R_{1}   = 0 \; .
\eqno(10.6)
$$

\noindent  In the same way, for  $R_{2}$ we have
$$
A_{2}(A_{2} -1) = m(m+1) \qquad \Longrightarrow \qquad
 A_{2} = -m ,  m + 1  =\pm (m+ 1/ 2) + 1/ 2 \; ,
$$
$$
B_{2}(B_{2}-1) = k(k-1) \qquad \Longrightarrow \qquad
 B_{2} =  k ,  -k + 1  = \pm (-k + 1/ 2   ) +1 /2 \; .
\eqno(10.7)
$$

\noindent Introducing a new variable  $\cos^{2}\rho = y$, eq.
(10.6) is changed to the form
$$
y(1-y){d^{2} \over dy^{2}} \;R_{1}\;  + \; [\;B_{1}  + {1\over 2} -
(A_{1} + B_{1} + 1 ) y\;] \;
{d \over dy} \; R_{1} \; + \;
$$
$$
+ {1\over 4}\;
[\; (\lambda + 1 / 2 ) ^{2} \; - \; (A_{1}+B_{1})^{2}\; ]\;R_{1} = 0 \; ,
$$

\noindent that can be compared with the hypergeometric case
$$
R_{1} = C_{1} \;  F(\alpha_{1},\; \beta_{1},\; \gamma_{1}; y) \; , \qquad
\gamma_{1} = B_{1} + 1/2  \;
$$
$$
A_{1} +B_{1} +1 = \alpha_{1} + \beta_{1} + 1 , \qquad
{(\lambda + 1 / 2 ) ^{2}  -  (A_{1}+B_{1})^{2} \over 4}  = - \alpha_{1} \beta_{1} \; ;
\eqno(10.8b)
$$

\noindent with  $\alpha_{1},\beta_{1}$  taken as
$$
\alpha_{1} = {A_{1} + B_{1} + (\lambda+1/2) \over 2} \; , \qquad
\beta_{1}  = {A_{1} + B_{1} - (\lambda+1/2) \over 2} \; .
\eqno(10.8b)
$$

\noindent
Analogously, in the case of  $R_{2}$:
$$
R_{2} = C_{2} \;  F(\alpha_{2},\; \beta_{2},\; \gamma_{2}; y) \; , \;\;
\gamma_{2} = B_{2} + 1/2  \; ,
$$
$$
\alpha_{2} = {A_{2} + B_{2} + (\lambda+1/2) \over 2} \; , \;\;
\beta_{2}  = {A_{2} + B_{2} - (\lambda+1/2) \over 2} \; .
\eqno(10.9)
$$

\vspace{5mm}
Taking in mind (10.5) and  (10.7), instead of
$A_{1},B_{1},A_{2},B_{2}$ one can introduce
$$
A_{1} = a_{1} + {1\over 2}\; , \qquad B_{1} = b_{1} + {1\over 2}\; , \qquad
A_{2} = a_{2} + {1\over 2}\; , \qquad  B_{2} = b_{2} + {1\over 2}\; ,
\eqno(10.10)
$$

\noindent
then
$$
G_{1}(\rho)  = C_{1} \; \sin^{a_{1}} \rho \; \cos^{b_{1}} \rho \;
F(\alpha_{1},\; \beta_{1},\; \gamma_{1}; \cos^{2}\rho )  \; ,
$$
$$
G_{2}(\rho)  = C_{2} \; \sin^{a_{2}} \rho \; \cos^{b_{2}} \rho \;
F(\alpha_{2},\; \beta_{2},\; \gamma_{2}; \cos^{2}\rho)
\; ,
\eqno(10.11a)
$$
$$
a_{1} = \mid m - 1 / 2 \mid  \; , \qquad b_{1}  = \mid k +1/ 2 \mid \; ,
$$
$$
a_{2}  = \mid m + 1/ 2 \mid  \; , \qquad b_{2}  = \mid k -1/ 2 \mid  \; ,
$$
$$
a_{i},b_{i} \in \{\;  0,+1,+2,+3,... \; \}\; .
$$

\noindent  Parameters in hypergeometric functions  (10.11a) are defined by
$$
\qquad \lambda = \pm \; \sqrt{\epsilon ^{2}  - M^{2}}\; ,
\eqno(10.11b)
$$
$$
\alpha_{1} = {a_{1} + b_{1} +1  + (\lambda+1/2) \over 2} \; , \qquad
\beta_{1}  = {a_{1} + b_{1} +1  - (\lambda+1/2) \over 2} \; , \qquad
\gamma_{1} = b_{1} + 1  \; ,
$$
$$
\alpha_{2} = {a_{2} + b_{2} +1  + (\lambda+1/2) \over 2} \; , \qquad
\beta_{2}  = {a_{2} + b_{2}  +1 - (\lambda+1/2) \over 2} \; , \qquad
\gamma_{2} = b_{2} + 1  \;  \; .
$$

\vspace{5mm}
\noindent
Depending on the sign of $\lambda$
the  hypergeometric series are the  polynomials according to:
$$
\lambda  > 0  \;,\qquad
\beta_{1}  = {a_{1} + b_{1} +1  - (\lambda+1/2) \over 2} = -n_{1}\; ,\;\;
\beta_{2}  = {a_{2} + b_{2} +1  - (\lambda+1/2) \over 2} = -n_{2} \; ,
$$
$$
\lambda  < 0 \;  ,\qquad
\alpha _{1}  = {a_{1} + b_{1} +1  + (\lambda+1/2) \over 2} = -n_{1}\; ,\;\;
\alpha_{2}  = {a_{2} + b_{2} +1  + (\lambda+1/2) \over 2} = -n_{2} \; .
$$
$$
\eqno(10.12)
$$

\begin{quotation}

Evidently, in view of existence of first order differential relation between $F_{1}$ and
$F_{2}$ one can examine only one  second order equation, wether  for $F_{1}$ or for
 $F_{2}$.

\end{quotation}

From the second order equation for  $G_{1}(\rho)$ it follows
the quantization rule:

\vspace{5mm}
$
\underline{\lambda  =+ 3/2, +5/2, +7/2, ...  \;} ,
$
$$
\beta_{1}  = {a_{1} + b_{1} +1  - (\lambda+1/2) \over 2} = -n_{1}\; ,
\eqno(10.13a)
$$
$$
\lambda = +( a_{1} + b_{1} +1/2 + 2n_{1} )\; , \qquad  \epsilon^{2} = M^{2} + \lambda ^{2} \; ,
$$
$$
G_{1}(\rho) =
C_{1} \; \sin^{a_{1}} \rho \; \cos^{b_{1}} \rho \;
F(a_{1}+b_{1}+1 +n_{1}, -n_{1}, \; b_{1}+1 ; \cos^{2}\rho )
$$

$
\underline{\lambda  =- 3/2, -5/2, -7/2, ...  \;} ,
$
$$
\alpha _{1}  = {a_{1} + b_{1} +1  + (\lambda+1/2) \over 2} = -n_{1}\; ,
\eqno(10.13b)
$$
$$
\lambda =- ( a_{1} + b_{1} +3/2 + 2n_{1})\;  , \qquad  \epsilon^{2} = M^{2} + \lambda ^{2} \; ,
$$
$$
G_{1}(\rho) =
C_{1} \; \sin^{a_{1}} \rho \; \cos^{b_{1}} \rho \;
F(-n_{1}, a_{1}+b_{1}+1 +n_{1},  \; b_{1}+1 ; \cos^{2}\rho )
$$

All the energy levels are degenerated  in $\lambda,m,k$:
$$
\Psi_{\epsilon}  =  \Psi_{\epsilon, \lambda, m, k  } \;  ,
\qquad
\epsilon = \sqrt{M^{2} + \lambda^{2}}  \; ,
$$
$$
\lambda = \pm 3/2, \pm 5/2, \pm 7/2, ... \; .
\eqno(10.14)
$$

\noindent Quantization of $\lambda$ is given by
$$
\lambda = a_{1} + b_{1} +1/2 + 2n_{1}  \qquad \mbox{or} \qquad
-\lambda = a_{1} + b_{1} +3/2 + 2n_{1} \; .
\eqno(10.15)
$$

\noindent
Some first energy levels  may be readily detailed:
$$
 \underline{\lambda =+  3/2}:  \qquad
1 = a_{1}+ b_{1} + 2n_{1}  \qquad \Longrightarrow
$$
$$
\qquad \qquad \qquad \qquad n_{1} = 0 \; ,  \qquad ( a_{1} , b_{1})=(1,0), (0,1) \; .
 $$
$$
\underline{ \lambda =-  3/2}: \qquad  0 = a_{1} + b_{1} + 2n_{1}  \qquad \Longrightarrow
$$
$$
\qquad \qquad \qquad n_{1}=0 \; ,\qquad ( a_{1} , b_{1})=(0,0)
\; .
$$
$$
 \underline{\lambda =+  5/2}:  \qquad
2 = a_{1}+ b_{1} + 2n_{1}  \qquad \Longrightarrow
$$
$$
\qquad \qquad \qquad \qquad \qquad  n_{1}=0\; , \qquad  (a_{1},b_{1})= (0,2),(1,1),(2,0) \; ,
$$
$$
\qquad \qquad \;\;\;  n_{1}= 1 \; , \qquad(a_{1},b_{1})= (0,0).
$$
$$
\underline{ \lambda =-  5/2}:  \qquad  1 = a_{1} + b_{1} + 2n_{1}  \qquad \Longrightarrow
$$
$$
\qquad \qquad \qquad \qquad  n_{1} = 0 \; ,  \qquad  ( a_{1} , b_{1})=(1,0), (0,1) \; .
 $$
$$
 \underline{\lambda =+  7/2}: \qquad
3 = a_{1}+ b_{1} + 2n_{1}  \qquad \Longrightarrow
$$
$$
\qquad \qquad \qquad \;\;\;  \qquad \qquad \qquad  n_{1}=0\;,
\qquad  (a_{1},b_{1})= (0,3),(1,2),(2,1),(3,0) \; ,
$$
$$
\qquad  \qquad \qquad \qquad  n_{1}=1 \; , \qquad (a_{1},b_{1})= (0,1),(1,0) \; .
$$
$$
\underline{ \lambda =-  7/2}: \qquad  2 = a_{1} + b_{1} + 2n_{1}  \qquad \Longrightarrow
$$
$$
\qquad \qquad \qquad \qquad \qquad  n_{1}=0\; , \qquad (a_{1},b_{1})= (0,2),(1,1),(2,0) \; ,
$$
$$
\qquad \qquad \;\;\;  n_{1}= 1 \; , \qquad  (a_{1},b_{1})= (0,0) .
$$
$$
 \underline{\lambda =+9/2}:  \qquad
4 = a_{1}+ b_{1} + 2n_{1}  \qquad \Longrightarrow
$$
$$
\qquad \qquad \qquad  \qquad \qquad \qquad  n_{1}=0\; ,\qquad (a_{1},b_{1})= (0,4),(1,3),
(3,1),(4,0) \; ,
$$
$$
\qquad  \qquad \qquad \;\;\; \; \qquad  n_{1}=1\; , \qquad (a_{1},b_{1})= (0,2),(1,1),(2,0) \; ,
$$
$$
\qquad  \qquad  \qquad  n_{1}=2 \; ,\qquad  (a_{1},b_{1})= (0,0) \; .
$$
$$
\underline{ \lambda =-  9/2}:  \qquad  3 = a_{1} + b_{1} + 2n_{1}  \qquad \Longrightarrow
$$
$$
\qquad \qquad \qquad \qquad \qquad  n_{1}=0\;, \qquad (a_{1},b_{1})= (0,3),(1,2)(2,1),(3,0) \; ,
$$
$$
\qquad \qquad \;\;\;  n_{2 }= 1 , (a_{1},b_{1})= (0,1),(1,0) \; .
$$

 \noindent and so on.

\subsection*{11.  The wave functions constructed and continuity testing  }

\hspace{5mm}
All physical Dirac wave solutions  must be single-valued and continuous in $S_{3}$.
Solutions with minimal values of energy
$$
\epsilon =\sqrt{M^{2} + (3/2)^{2}} \; , \qquad  \lambda = \pm 3/2
\eqno(11.1a)
$$

\noindent are just those (see \S  9). It remains to consider
solutions with
$$
\epsilon =\sqrt{M^{2} + \lambda ^{2}} \; ,  \qquad  \mid \lambda \mid = 5/2, 7/2, 9/2, ...
\eqno(11.1b)
$$

Half-integer values for $m,k$ will guarantee continuity of the solutions when
 $ 0 < \rho < \pi/2$. Special cases are two closed curves:
 $$
\rho =0 : \qquad n_{3} = \sin z , \qquad n_{4} = \cos z \; ,
\eqno(11.2a)
$$
$$
\rho =\pi /2  : \qquad n_{1} = \cos \phi  , \qquad n_{2} = \sin \phi \; .
\eqno(11.2b)
$$

In conformally flat tetrad the wave functions must
behave themselves in accordance with the rules
$$
\rho =0 : \qquad \Psi_{cart} = \{ \psi (z) \;\; ,or\;\;  \mbox{const}\} \; ,
\eqno(11.2a)
$$
$$
\rho =\pi/2  : \qquad \Psi_{cart} = \{ \psi (\phi) \;\;, or\;\;   \mbox{const} \} \; .
\eqno(11.2b)
$$

\noindent
Two first components of the wave functions (see (8.5c))
$$
\psi_{1} = { e^{i(m-1/2)\phi }  e^{i(k+1/2)z}  \over \sqrt{1 + \cos \rho \cos z } }
  \; \left [   \; (   1  + \cos \rho \;  e^{-iz} )
  \;   G_{1} (\rho)    -i \sin \rho \;  e^{-iz} \;    G_{2} (\rho)  \; \right ] \; ,
$$
$$
\psi_{2} = { e^{i(m+1/2)\phi }  e^{i(k-1/2)z}  \over \sqrt{1 + \cos \rho \cos z } }
  \; \left [ \;   (   1  + \cos \rho  \; e^{+iz} )
  \;   G_{2} (\rho)    -i \sin \rho \;  e^{+iz} \;    G_{1} (\rho)  \;     \right ]
$$

\noindent  will take the form
$$
\rho=0: \qquad
\psi_{1} = { e^{i(m-1/2)\phi }  e^{i(k+1/2)z}  \over \sqrt{1 + \cos z } }
  \;  (   1  +   e^{-iz} )
  \;   G_{1} (0)    \; ,
$$
$$
\rho=0: \qquad
\psi_{2} = { e^{i(m+1/2)\phi }  e^{i(k-1/2)z}  \over \sqrt{1 +  \cos z } }
  \;   (   1  +  e^{+iz} )
  \;   G_{2} (0)     \; ,
\eqno(11.3)
$$
$$
\rho = \pi/2: \qquad
\psi_{1} =  e^{i(m-1/2)\phi }  e^{i(k+1/2)z}
  \; \left [   \;
  \;   G_{1} (\pi/2)    -i \;  e^{-iz} \;    G_{2} (\pi/2)  \; \right ] \; ,
$$
$$
\rho = \pi/2: \qquad
\psi_{2} =  e^{i(m+1/2)\phi }  e^{i(k-1/2)z}
  \; \left [ \;
  \;   G_{2} (\pi/2)     -i \;  e^{+iz} \;    G_{1} (\pi/2)  \;     \right ]
$$

\noindent
where
$$
G_{1}(\rho)  = C_{1} \; \sin^{a_{1}} \rho \; \cos^{b_{1}} \rho \;
F(\alpha_{1},\; \beta_{1},\; \gamma_{1}; \cos^{2}\rho )\; ,
$$
$$
G_{2}(\rho)  = C_{2} \; \sin^{a_{2}} \rho \; \cos^{b_{2}} \rho \;
F(\alpha_{2},\; \beta_{2},\; \gamma_{2}; \cos^{2}\rho )\; ,
$$
$$
a_{1} = \mid m-1/2 \mid \; , \qquad  b_{1} = \mid k+1/2 \mid \; ,\;
a_{2} = \mid m+1/2 \mid \;  , \qquad b_{2} = \mid  k -1/2 \mid  \; .
\eqno(11.4)
$$

\noindent
While  all four parameters  $a_{1},b_{2},b_{1}, b_{2} \neq 0$,
 because of  the presence of the terms $\sin^{a_{1}} \rho  \cos^{b_{1}}  $ and
$\sin^{a_{2}} \rho  \cos^{b_{2}} \rho
$
the functions $G_{1}(\rho),G_{2}(\rho) $ will be equal to zero when $\rho =0, \pi/2$,
Correspon\-dingly, eqs.  (11.3)   read
$$
\rho=0: \qquad \psi_{1}= 0 \; , \qquad \psi_{2}= 0 \; ,
$$
$$
\rho=\pi/2 : \qquad \psi_{1} = 0 \; , \qquad \psi_{2}= 0 \; .
$$

\noindent
All the remaining cases
$$
m=+1/2 \; , \; k=+1/2 \;  ,  \qquad  m=-1/2 \; , \; k=-1/2 \;  ,
$$
$$
m=+1/2 \; , \; k=-1/2 \;  ,  \qquad m=-1/2 \; , \; k=+1/2 \;  ,
\eqno(11.5a)
$$
$$
m=+1/2 \; , \; \mid k \mid > +1/2  ,  \qquad  m=-1/2 \; , \; \mid  k \mid > 1/2  \; ,
$$
$$
\mid m \mid > 1/2 \; , \; k=-1/2 \;  ,  \qquad \mid m \mid > 1/2 \; , \; k=+1/2  \; ,
\eqno(11.5b)
$$

\noindent must be considered separately.
First, let us turn to the four cases in ( 11.5a):

\vspace{5mm}
 $\underline{m=+1/2, k=+1/2}$:
$$
a_{1} = 0\;, \;  b_{1}= 1 \; , \;  a_{2}=1 \; ,  \; b_{2}=0  \; ,
$$
$$
G_{1}(0) = const \; , \qquad  G_{2}(0) = 0 ,
$$
$$
\qquad G_{1}(\pi/2)= 0 \; , \qquad  G_{2}(\pi/2) = \;  const
$$
$$
\rho=0: \qquad
\psi_{1} = {   e^{iz} (   1  +   e^{-iz} ) \over \sqrt{1 + \cos z } }
    \;   const    = \psi(z)  \; ,
$$
$$
\rho=0: \qquad \qquad
\psi_{2} = { e^{i\phi } (   1  +  e^{+iz} )  \over \sqrt{1 +  \cos z } }
    \;   0 = 0     \; .
\eqno(11.6)
$$
$$
\rho = \pi/2: \qquad
\psi_{1} =    e^{iz}
  \; ( \;   0    -i \;  e^{-iz} \;   const  \;)  =-i \; const \; ,
$$
$$
\rho = \pi/2: \qquad
\psi_{2} =  e^{i\phi }
  \; (  \;  const      -i \;  e^{+iz} \;   0  \; )  =const \;  e^{i\phi} \; .
$$

$\underline{m=-1/2, k=-1/2}$:
$$
a_{1} = 1\;, \;  b_{1}= 0 \; , \;  a_{2}=0 \; , \;  b_{2}=1\;  ,
$$
$$
G_{1}(0) = 0 \; , \qquad  G_{2}(0) = const  ,
$$
$$
\qquad G_{1}(\pi/2)= const \; , \qquad  G_{2}(\pi/2) = 0
$$
$$
\rho=0: \qquad
\psi_{1} = { e^{-i\phi }    \over \sqrt{1 + \cos z } }
  \;  (   1  +   e^{-iz} )
  \;   0 =0     \; ,
$$
$$
\rho=0: \qquad
\psi_{2} = {   e^{-iz} (   1  +  e^{+iz} )  \over \sqrt{1 +  \cos z } }
  \;   const      \; .
\eqno(11.7)
$$
$$
\rho = \pi/2: \qquad
\psi_{1} =  e^{-i\phi }    \; (   \;
  \;   const     -i \;  e^{-iz} \;   0  \; )  = const\; e^{-i\phi }\; ,
$$
$$
\rho = \pi/2: \qquad
\psi_{2} =   e^{-iz}
  \; ( \;
  \;  0     -i \;  e^{+iz} \;   const  \;    )  =-i\; const \; .
$$

$\underline{m=+1/2, k=-1/2}$:
$$
a_{1} = 0 \; , \; b_{1}= 0 \; , \; a_{2}=1 \; , \; b_{2}=1  \; ,
$$
$$
G_{1}(0) = const  \; , \qquad  G_{2}(0) = 0\;   ,
$$
$$
\qquad G_{1}(\pi/2)= const \; , \qquad  G_{2}(\pi/2) = 0
$$
$$
\rho=0: \qquad
\psi_{1} = { (   1  +   e^{-iz} ) \over \sqrt{1 + \cos z } }
  \;   const    \; ,
$$
$$
\rho=0: \qquad
\psi_{2} = { e^{i\phi }  e^{-iz} (   1  +  e^{+iz} )  \over \sqrt{1 +  \cos z } }
  \;   0 = 0      \; .
\eqno(11.8)
$$
$$
\rho = \pi/2: \qquad
\psi_{1} =
  \; (   \;
  \;   const    -i \;  e^{-iz} \;   0  \; ) = const  \; ,
$$
$$
\rho = \pi/2: \qquad
\psi_{2} =  e^{i\phi }  e^{-iz}
   (   0   -i   e^{+iz} \;    const     )  =-i \; e^{i\phi } \;  .
$$

$\underline{m=-1/2, k=+1/2}$:
$$
a_{1} = 1\;, \;  b_{1}= 1 \; , \;  a_{2}=0 \; , \;  b_{2}=0\;  ,
$$
$$
G_{1}(0) = 0 \; , \qquad  G_{2}(0) = const  ,
$$
$$
\qquad G_{1}(\pi/2)=  0  \; , \qquad  G_{2}(\pi/2) = const \; ,
$$
$$
\rho=0: \qquad
\psi_{1} = { e^{-i\phi }  e^{iz}  (   1  +   e^{-iz} ) \over \sqrt{1 + \cos z } }
  \;  0 = 0     \; ,
$$
$$
\rho=0: \qquad
\psi_{2} = { (   1  +  e^{+iz} )  \over \sqrt{1 +  \cos z } }
  \;  const      \; .
\eqno(11.9)
$$
$$
\rho = \pi/2: \qquad
\psi_{1} =  e^{-i\phi }  e^{iz}
   (  0   -i  e^{-iz}   const  \; ) = -i \; const e^{-i\phi } \; ,
$$
$$
\rho = \pi/2: \qquad
\psi_{2} =      const      -i   e^{+iz} 0  = const \; .
$$

Now let us consider the cases in (11.5b).

\vspace{5mm}
\underline{$m=+1/2 \; , \; \mid k \mid > +1/2$}:
$$
a_{1} =0 \;, \;  b_{1} \neq 0\; , \; a_{2} =1\; , \; b_{2}\neq 0 \; ,
$$
$$
G_{1}(0) = const \; , \; G_{2}(0) = 0 \; ,
$$
$$
G_{1}(\pi/2) = 0 \; , \; G_{2}(\pi/2) = 0 \; ,
$$
$$
\rho=0: \qquad
\psi_{1} = { e^{i(k+1/2)z}  \over \sqrt{1 + \cos z } }
  \;  (   1  +   e^{-iz} )   \;   const  =\psi (z) \; ,
$$
$$
\rho=0: \qquad
\psi_{2} = { e^{i\phi }  e^{i(k-1/2)z}  \over \sqrt{1 +  \cos z } }
  \;   (   1  +  e^{+iz} )
  \;   0    \; ,
$$
$$
\rho = \pi/2: \qquad
\psi_{1} =    e^{i(k+1/2)z}
  \; \left [   \;
  \;   0    -i \;  e^{-iz} \;   0 \; \right ]  =0 \; ,
$$
$$
\rho = \pi/2: \qquad
\psi_{2} =  e^{i\phi }  e^{i(k-1/2)z}
  \; \left [ \;
  \;  0     -i \;  e^{+iz} \;   0 \;     \right ] =0 \; .
\eqno(11.10)
$$

\underline{$m=-1/2 \; , \; \mid k \mid > +1/2$}:
$$
a_{1} =1 \;, \;  b_{1} \neq 0\; , \; a_{2} =0\; , \; b_{2}\neq 0 \; ,
$$
$$
G_{1}(0) = 0 \; , \; G_{2}(0) = const  \; ,
$$
$$
G_{1}(\pi/2) = 0 \; , \; G_{2}(\pi/2) = 0 \; ,
$$
$$
\rho=0: \qquad
\psi_{1} = { e^{-i \phi}e^{i(k+1/2)z}  \over \sqrt{1 + \cos z } }
  \;  (   1  +   e^{-iz} )   \;  0  = 0 \; ,
$$
$$
\rho=0: \qquad
\psi_{2} = {  e^{i(k-1/2)z}  \over \sqrt{1 +  \cos z } }
  \;   (   1  +  e^{+iz} )
  \;   const = \psi_{2}(z)    \; ,
$$
$$
\rho = \pi/2: \qquad
\psi_{1} =    e^{i(k+1/2)z}
  \; \left [   \;
  \;   0    -i \;  e^{-iz} \;   0 \; \right ] =0  \; ,
$$
$$
\rho = \pi/2: \qquad
\psi_{2} =  e^{i\phi }  e^{i(k-1/2)z}
  \; \left [ \;
 0     -i \;  e^{+iz} \;   0 \;     \right ] =0 \; .
\eqno(11.11)
$$

\underline{$\mid m \mid > 1/2 \; , \;  k = +1/2$}:
$$
a_{1} \neq 0 \;, \;  b_{1} = 1 \; , \; a_{2} \neq 0 \; , \; b_{2} = 0 \; ,
$$
$$
G_{1}(0) = 0 \; , \; G_{2}(0) = 0 \; ,
$$
$$
G_{1}(\pi/2) = 0 \; , \; G_{2}(\pi/2) = const  \; ,
$$
$$
\rho=0: \qquad
\psi_{1} = { e^{i(m-1/2)\phi} e^{iz}  \over \sqrt{1 + \cos z } }
  \;  (   1  +   e^{-iz} )   \;  0  = 0 \; ,
$$
$$
\rho=0: \qquad
\psi_{2} = { e^{i(m+1/2)\phi}   \over \sqrt{1 +  \cos z } }
  \;   (   1  +  e^{+iz} )
  \;  0 =0     \; ,
$$
$$
\rho = \pi/2: \qquad
\psi_{1} =   e^{i(m-1/2)\phi} e^{iz}
  \; \left [   \;
  \;   0    -i \;  e^{-iz} \;   const  \; \right ]  = \psi_{1}(\phi)  \; ,
$$
$$
\rho = \pi/2: \qquad
\psi_{2} = e^{i(m+1/2)\phi}
  \; \left [ \;
  \;  const      -i \;  e^{+iz} \;   0 \;     \right ] =\psi_{2}(\phi)  \; .
\eqno(11.12)
$$

\underline{{$\mid m \mid > 1/2 \; , \;  k = -1/2$}:}
$$
a_{1} \neq 0 \;, \;  b_{1} = 0 \; , \; a_{2} \neq 0 \; , \; b_{2} = 1 \; ,
$$
$$
G_{1}(0) = 0 \; , \; G_{2}(0) = 0 \; ,
$$
$$
G_{1}(\pi/2) = const  \; , \; G_{2}(\pi/2) = 0  \; ,
$$
$$
\rho=0: \qquad
\psi_{1} = { e^{i(m-1/2)\phi} \over \sqrt{1 + \cos z } }
  \;  (   1  +   e^{-iz} )   \;  0  = 0 \; ,
$$
$$
\rho=0: \qquad
\psi_{2} = { e^{i(m+1/2)\phi}  e^{-iz}  \over \sqrt{1 +  \cos z } }
  \;   (   1  +  e^{+iz} )
  \;  0 =0     \; ,
$$
$$
\rho = \pi/2: \qquad
\psi_{1} =   e^{i(m-1/2)\phi}  \; \left [   \;
  \;   const     -i \;  e^{-iz} \;   0  \; \right ]  = \psi_{1}(\phi)  \; ,
$$
$$
\rho = \pi/2: \qquad
\psi_{2} = e^{i(m+1/2)\phi} e^{-iz}
  \; \left [ \;
  \;  0      -i \;  e^{+iz} \;   const  \;     \right ] =\psi_{2}(\phi)  \; .
\eqno(11.13)
$$

\begin{quotation}

{\em

Thus, the general conclusion can be done:
all the constructed  Dirac wave  functions in cylindrical coordinates
$
\Psi_{\epsilon,\lambda,m,k}(t,\rho, \phi,z)
$
provide us with single-valued and continuous functions on the sphere
$S_{3}$. For proving these properties an explicit form of the  wave  function in conformally
flat tetrad basis has been used.
}

\end{quotation}


\begin{center}
{\bf PART III. THE DIRAC EQUATION IN THE ELLIPTICAL SPACE $\tilde{S}_{3}$}
\end{center}

\subsection*{12. Cylindrical coordinates and tetrad in
elliptical space space,\\ spinor gauge transformations}

\hspace{5mm}
Now  we are to examine the  properties
of a fermion particle in elliptical space model
$\tilde{S}_{3}$.   As a first step, some analogue of the cartesian tetrad basis
in  $\tilde{S}_{3}$ should be defined and  the explicit form of Dirac solutions in that basis
must be calculated. After that one can investigate such wave solution from the  point
of vied of their  continuity in the elliptical manifold $\tilde{S}_{3}$.

Like in \S 3 we will employ realizations of spherical and elliptical models
as group manifolds $SU(2)$ and  $SO(3.R)$ respectively:
$$
B = n_{0} - i\; n_{j}\sigma_{j}\;, \qquad  n_{0}^{2} + n_{1}^{2} + n_{2}^{2} +  n_{3}^{2}  =
 +1 \; ,
 \eqno(12.1)
$$
$$
O(\vec{c}) =
 I + 2 {\vec{c}^{\;\times}  + (\vec{c}^{\;\times})^{2} \over 1 +
 \vec{c}^{\;2}}   \;,
 c^{i} = {+n_{i} \over + n_{4}} =  {-n_{i} \over - n_{4}}\; .
  \eqno(12.2)
$$

\noindent
To pair of  vectors
$
\vec{c}^{\;\pm} = +\pm \infty \; \vec{c}_{0} \; , \; \vec{c}^{\;2}_{0} = 1
$
corresponds  one the same  matrix, point in $\tilde{S}_{3}$:
$
O(\vec{c}^{\;\pm }) =  I + 2 (\vec{c}^{\;\times}_{0})^{2} \; .
$
From formal point of view, elliptical model can be obtained by
identification of opposite point on 4-sphere.
However, such a simple view  may be insufficient in applications.
Below, a more detailed treatment of this relating procedure will be  done.
In the following, the coordinates  $(c^{1}, c^{2},  c^{3})$ will be
considered as Cartesian ones in the model  $\tilde{S}_{3}$.
Let us start with the known formulas for the spherical model:
$$
x^{1} = {n_{1} \over 1 + n_{0}}\; , \;\;
x^{2} = {n_{2} \over 1 + n_{0}}\; , \;\;
x^{3} = {n_{3} \over 1 + n_{0}}\; ,
$$
$$
n_{i} = {2 x^{i} \over  1 + x^{2} } \; , \;\;
n_{0} = {1 - x^{2} \over  1 + x^{2} } \; ,
 $$
$$
dS^{2} =dt^{2} -dl^{2}(x), \qquad
dl^{2}(x) = {1 \over f^{2}} \; [\; (dx^{1})^{2} +
(dx^{2})^{2} +    (dx^{3})^{2} \; ] \; ,
$$
$$
 f = { 1 + x^{2}\over 2} \;, \qquad
 e_{(a)\alpha}(x) = \left | \begin{array}{cccc}
1 &  0    &     0  & 0 \\
0 &  f &  0  &  0   \\
0 &  0 &  f  &  0   \\
0 &  0 &  0   & f   \\
\end{array}  \right | \; .
\eqno(12.3a)
$$

\noindent
 Let
$n_{0}= (-1 + \delta)$, where  $\;\delta$ stands for a infinitely small quantity , then
$$
n_{a} = (-1 + \delta,\; 0,\; 0,\; 0 ) \; \Longrightarrow \;
\lim_{\delta \rightarrow 0} x^{i} = {1 \over \delta} \;n_{i}  = \infty \; n_{i} \; .
\eqno(12.3b)
$$

\noindent This means that all the set of finite coordinates  $\{ \; x^{i} = \infty\; n_{i} \; ; \;
\vec{n}^{2} = 1 \; \}$  represents  one the same point
$n_{a} = (-1;0,0,0)$ on the sphere.
Cartesian  $c^{i}$-coordinate in $\tilde{S}_{3}$ can be  connected with  $x^{i}$-coordinates
in the $S_{3}$ as follows (it is a $2 \Longrightarrow 1$  mapping):
$$
 {2 x^{i} \over 1 - x^{2}}  = c^{i}   \; ;
\eqno(12.4a)
$$

\noindent to the sets $x^{i}$ and  $x^{'i} = -x^{i} / x^{ 2}$ corresponds one the same
vector  $c^{i} = c'^{i}$. The inverse map is 2-valued:
$$
\delta= \pm 1, \qquad x^{\;\;i}_{(\delta)}
 = {c^{i} \over (1 + \delta \sqrt{1 + c^{2}}) } \; ;
\eqno(12.4b)
$$

\noindent one should note  the identities
$$
{1 - x^{2}_{(\delta) }  \over 2 } =  {1 \over
1  + \delta \sqrt{1 + c^{2}} } \; , \qquad
{1 + x^{2}_{(\delta)}  \over 2 } =  { \delta \sqrt{1+ c^{2} } \over
1  + \delta \sqrt{1 +  c^{2}} } \; ,
$$
$$
x^{2}_{(+)} < 1 \; , \qquad x^{2}_{(-)} > 1 \; , \qquad
(c^{2} \rightarrow \infty , x^{2} _{(\pm)} \rightarrow 1 \mp 0 )\; ,
$$
$$
 {1 \over \sqrt{1 + c^{\;2}} }   =  \delta \; { 1- x^{\;2}_{(\delta)} \over 1 + x^{\;2}_{(\delta)}} \; .
\eqno(12.4c)
$$

\noindent
Thus, the whole sphere, composed of two parts,
 can be covered by two maps:
$$
n_{0(\delta)} = { \delta  \over \sqrt{1 + c^{\;2}} } :
\qquad \Longrightarrow \qquad
\left.  \begin{array}{c}  n_{0(+1)} \in [\; 0,\; +1\;] \; ,\\[3mm]
 n_{0(-1)} \in [\;-1, \; 0\; ] \; ,
\end{array} \right.
$$
$$
\vec{n}_{(\delta)}  = {2 \vec{x}_{(\delta)} \over 1 +x^{2}_{(\delta)}} =
\delta\; { \vec{c} \over \sqrt{1+ c^{2}}}\; ;
\eqno(12.4d)
$$

\noindent it is convenient to have  their  form  in two limits:
$$
\left. \begin{array}{lll}
\vec{c} \rightarrow 0 \; \vec{c}_{0} :\qquad  \qquad &
 n_{0(+1)} \rightarrow +1  \;, \qquad &
\vec{n}_{(+1)}  \rightarrow  0 \; (+1) \;  \vec{c}_{0}  = 0\; ,\\[2mm]
\vec{c} \rightarrow  \infty \; \vec{c}_{0} : \qquad \qquad &
n_{0(+1)} \rightarrow  +1 / \infty  = 0 \; , \qquad &
\vec{n}_{(+1)}  \rightarrow  + \; \vec{c}_{0} \; , \\[4mm]
\vec{c} \rightarrow 0 \; \vec{c}_{0} : \qquad \qquad &
 n_{0(-1)} \rightarrow -1  \;, \qquad &
\vec{n}_{(-1)}  \rightarrow  0 \; (-1) \;  \vec{c}_{0}  = 0\; ,\\[2mm]
\vec{c} \rightarrow  \infty \; \vec{c}_{0} : \qquad  \qquad &
n_{0(-1)} \rightarrow  -1 /  \infty  = 0 \; , \qquad &
\vec{n}_{(-1)}  \rightarrow  -\; \vec{c}_{0} \; .
\end{array} \right.
\eqno(12.4e)
$$

\begin{quotation}

So, the  mapping with  $\delta=+1$ to the zero vector  $\vec{c}=(0,0,0)$
refers the point $n_{a}= (+1,0,0,0)$; whereas the mapping with   $\delta=-1$ to the zero
vector  corresponds the point $n_{a}=(-1,0,0,0)$.

\end{quotation}

Now  with the use of tensor transformation law
let us calculate the  metric tensor $c_{i}$-coordinates  of elliptical space
Here we face some peculiarities steaming from $2 \Longrightarrow 1$ character of the
mapping from $S_{3}$ to $\tilde{S}_{3}$. Therefore, we must handle every half-space separately:
$$
dc^{i} = {\partial c^{i} \over \partial x^{j(\delta)}} \; dx^{j(\delta)} =
$$
$$
=
[\; {2 \delta_{ij} \over 1 - x^{2}_{(\delta)} }  + c^{i}\;c^{j} \; ] \;dx^{j(\delta)} \; =
[\; \delta_{ij} \;(1+ \delta \sqrt{1+c^{2}}) + c^{i}\;c^{j} \; ]\;dx^{j(\delta)}
 \; ,
\eqno(12.5a)
$$
$$
dc^{i} dc^{i} = [\; {2 \delta_{ij} \over 1 - x^{2}_{(\delta)} }  + c^{i}\;c^{j} \; ]
\;dx^{j(\delta)}
[\; {2 \delta_{ik} \over 1 - x^{2}_{(\delta)} }  + c^{i}\;c^{k} \; ] \;dx^{k(\delta)}=
$$
$$
= \left [\; {4 \delta_{jk} \over (1 - x^{2}_{(\delta)})^{2} }   + {4 c^{j}c^{k}
\over  1 - x^{2}_{(\delta)} }   + c^{2} c^{j}c^{k}   \; \right ]
 \;dx^{j(\delta)} dx^{k(\delta)} \; .
$$
$$
c^{2} = {4x^{2} _{(\delta)} \over (1 - x^{2}_{(\delta)})^{2} }\; ,
\qquad
dc^{i} dc^{i} =  {4 \over (1 - x^{2}_{(\delta)})^{2} } \;  [
\delta_{jk} +c^{j}c^{k}  \; ] \;dx^{j(\delta)} dx^{k(\delta)} \; .
$$

\noindent So one gets
$$
{dc^{i} \; dc^{i} \over 1 + c^{2} } = {4 \over (1 + x^{2}_{(\delta)})^{2} } \;  [
\delta_{jk} +c^{j}c^{k}  \; ] \;dx^{j(\delta)} dx^{k(\delta)} \; .
\eqno(12.5b)
$$

\noindent In the same way, one obtains
$$
{ c^{j}\;c^{k}  dc^{j} dc^{k} \over
(1 +  c^{2})^{2}}
 = {4 \over ( 1 + x^{2}_{(\delta)})^{2}} \;
\; c^{j}c^{k}\; dx^{j(\delta)}\; dx^{k (\delta)} \; .
\eqno(12.5c)
$$

\noindent
Combining two last  relations,
we get to
$$
 dl^{2}(x_{(\delta)}) = {4\;dx^{j(\delta)}\; dx^{k(\delta)} \over
  [\; 1 + x^{2}_{(\delta)}\; ]^{2} } =
\left [ \;  {\delta _{jk} \over 1 + c^{2}} - {c^{j} c^{k} \over
(1 + c^{2})^{2}}  \; \right ] \; dc^{j} \; dc^{k} =
dl^{2}(c) \; .
\eqno(12.5d)
$$

\noindent  In the following we will need the contra-variant metrical tensor:
$$
g^{kj}(c) = (1+c^{2}) \; (\delta^{kj} +  c^{k} c^{j}) \; .
\eqno(12.5e)
$$

 When the metric  $g_{ik}(c)$  is employed one need to fix a certain  tetrad -- by definition
 it obeys
$$
e_{(i)j}(c) \; e_{(i)k} (c) = g_{jk}(c) \; ,  \qquad g_{jk}(c)  =
{\delta _{jk} \over 1 + c^{2}} - {c^{j} c^{k} \over (1 + c^{2})^{2}}
\eqno(12.6a)
$$

\noindent to be  satisfied. There exists one method to fix a tetrad,  preserving
some consistency with  the situation in spherical model. It consists in coordinate
transforming the conformally flat tetrad  $e_{(i)k}(x_{(\delta)})$
 (the domains  $x^{i}_{(+1)}$  and   $x^{i}_{(-1)}$ must be treated separately)
 to $c^{i}$-coordinates of the  elliptical model:
$$
e_{(i)k}(x_{(\delta)}) = {1\over f(x_{(\delta)})} \;  \delta_{ik} \; ,
\qquad f = { 1 +  x^{2}_{(\delta)} \over 2} \; ,
$$

$$
e^{(\delta)}_{(i)j}(c) = {\partial  x^{k} _{(\delta)} \over  \partial c^{j} }
 \; e_{(i)k}(x) = {1 \over f(x_{(\delta)}) } \;
 {\partial  x^{i}_{(\delta)}  \over  \partial c^{j} }   =
 {  1  + \delta \sqrt{1 +  c^{2}} \over  \delta \sqrt{1+ c^{2} }
 } \;\; {\partial  x^{i}_{(\delta)}  \over  \partial c^{j} }  \; .
\eqno(12.6b)
$$

\noindent In this way we are able to construct two tetrads
satisfying eq.   (12.6). The problem is reduced to rather simple calculating:
$$
{\partial  x^{i}_{(\delta)} \over  \partial c^{j} } = {\partial   \over  \partial c^{j} }\;
{c^{i} \over 1 + \delta \sqrt{1+c^{2}}} =
{ \delta _{ij} \over 1 + \delta \sqrt{1+c^{2}}}-
{ \delta \; c^{i} c^{j} \over  \sqrt{1+c^{2}}\; (1 + \delta \sqrt{1+c^{2} } )^{2} } \; ,
\eqno(12.6c)
$$

\noindent and further
$$
e^{(\delta)}_{(i)j}(c)   ={ \delta_{ij} \over  \delta \sqrt{1+ c^{2} }} -
{  c^{i} c^{j} \over (1+c^{2}) \;  (1 + \delta \sqrt{1+c^{2} } ) } \;.
\eqno(12.6d)
 $$

\noindent
These two tetrads can be written down as follows
$$
e^{(\delta)}_{(i)j}(c) =
\left |  \begin{array}{lll}
 a +  b \; c^{1}c^{1}    &   b \; c^{1}\; c^{2} &   b \; c^{1} \;  c^{3}   \\
 b \; c^{1}\;  c^{2}  & a +  b \; c^{2}c^{2}     &    b \; c^{2}\;  c^{3}  \\
  b \; c^{1}\;  c^{3}  &   b \; c^{2}\;  c^{3}   &  a +  b \; c^{3}c^{3}
\end{array}  \right | \; ,
$$
$$
a = { \delta \over \sqrt{ 1 +  c^{2}} } \; ,\;\;
b =  -\; { 1 \over 1 +  c^{2}}\;{1 \over 1 + \delta \sqrt{ 1 + c^{2}} } \; .
\eqno(12.7a)
$$

One may  perform independent calculation which will lead us to the same result.
Indeed, let us  look for a solution to tetrad equation in the form
$$
e_{i(j)} (c) \; e_{(j)k } (c)  = g_{ik } (c) \; , \qquad
e_{i(j)} (c) = A \;\delta _{ij} + B \;c^{i} c^{j} \; ,
$$

\noindent  from where it follow two equations
$$
A^{2} = {1 \over 1+ c^{2}} \; , \qquad  2AB + B^{2}c^{2} = - {1 \over (1+c^{2})^{2} } \; .
$$

\noindent  that is
$$
A = -{ \mu  \over \sqrt{1+c^{2}}  } \; , \qquad \mu = \mp 1 \; , \qquad
B = {1 \over c^{2} (1+c^{2} )} \; ( \mu \;  \sqrt{1+c^{2}}   +   \nu \; ) \; .
$$

\noindent
With the use of identity
$$
{(\mu \sqrt{1+c^{2}} +\nu) \over c^{2}  } = { 1 \over ( \mu \sqrt{1+c^{2}} -\nu ) } \; ,
$$

\noindent  $B$  can be taken to the form
$$
B =  {1 \over 1+c^{2}} \; { 1 \over ( \mu \sqrt{1+c^{2}} -\nu ) } = -\nu \;
{1 \over 1+c^{2}} \; { 1 \over ( 1 - \nu \mu \sqrt{1+c^{2}}  ) } \; .
$$

\noindent Thus, four different variants are possible:
$$
\left. \begin{array}{lll}
(1) \qquad  & (\mu,\nu)=(+1,+1) : \qquad A = -{ 1  \over \sqrt{1+c^{2}}  } \; ,\;&
B = - \; {1 \over 1+c^{2}} \; { 1 \over ( 1 -  \sqrt{1+c^{2}}  ) } \; ,\\[2mm]
(2) \qquad  & (\mu,\nu)=(-1,+1) : \qquad A =  + { 1  \over \sqrt{1+c^{2}}  } \;, \; &
B = - \; {1 \over 1+c^{2}} \; { 1 \over ( 1 + \sqrt{1+c^{2}}  ) } \; ,\\[2mm]
(3)\qquad & (\mu,\nu)=(-1,-1) : \qquad A = + { 1  \over \sqrt{1+c^{2}}  }\;, \;&
B = {1 \over 1+c^{2}} \; { 1 \over ( 1 -  \sqrt{1+c^{2}}  ) } \; ,\\[2mm]
(4)\qquad & (\mu,\nu)=(+1,-1) : \qquad A = -{ 1 \over \sqrt{1+c^{2}}  } \;,\;&
B = {1 \over 1+c^{2}} \; { 1 \over ( 1 +  \sqrt{1+c^{2}}  ) } \; .
\end{array} \right.
\eqno(12.7b)
$$

\begin{quotation}
The variants (1) and  (2) coincide with listed above (12.7a). variants
 (3) and (4) correspond to tetrads, $P$-reflected to (1) and (2).
 These additional possibilities can be omitted.

\end{quotation}

 These tetrads are substantially different which become evident on looking in
 their form in limiting points:
 $$
\vec{c} \;\; \rightarrow \;\; \epsilon ( \vec{c}^{\;0})  \; , \qquad
\vec{c}  \;\; \rightarrow \; \; \pm \;  \epsilon^{-1}  (\vec{c}^{\;0}) \; ,
\qquad \epsilon \rightarrow 0\; .
 $$

\noindent  Indeed, the  metric tensor being
$
g_{ik} (0) = \delta_{ik} \;  , \;
g_{ik} (\infty) =  \epsilon^{2}  \;
 (\delta_{ij} -c^{0}_{i} c_{j}^{0})  \rightarrow 0
$
whereas tetrads behave in accordance with
$$
e^{(+1)}_{(i)j} (0) =
 \delta_{ij}  \; , \qquad
  e^{(+1)}_{(i)j} (\infty)  =
\left | \begin{array}{cc}
1 & 0 \\
0 & \pm \epsilon (\delta_{ij} -c_{i}^{0}c_{j}^{0})
\end{array} \right |
 \; ,
$$
$$
   e^{(-1)}_{(i)j} (0) =  \left | \begin{array}{cc}
1 & 0 \\
0 &  -\delta_{ij} +2 c_{i}^{0}c_{j}^{0}
\end{array} \right | \;  ,   \qquad
  e^{(-1)(a)}_{\alpha} (\infty)  =
\left | \begin{array}{cc}
1 & 0 \\
0 & \mp \epsilon (\delta_{ij} -c_{i}^{0}c_{j}^{0})
\end{array} \right | \; .
$$

On general considerations, two tetrads  (12.7a) must be connected by a local 3-rotation
$$
e^{(-1)}_{(i)k}(c)  = O_{ij} \; e^{(+1)}_{(j)k} (c)  \;,
\eqno(12.7)
$$

\noindent which can be calculated explicitly (the notation
$\sqrt{1+c^{2}} = s$ is used):

$$
O_{k}^{\;\;l}(c)  =  O_{k}^{\;\;l}(c)  =
e^{(-1)}_{(k)i}(c) \; g^{ij}(c) \;  e^{(+1)}_{j(l)}(c) =
$$
$$
={1\over s^{2} } \; (-s  \;\delta_{ki} -  {c^{k}c^{i} \over 1-s })\;
 (\delta_{ij} +c^{i}c^{j}) \; ( s \;\delta _{jl} - {c^{j}c^{l} \over 1+s})=
$$
$$
= -\delta_{kl} + 2 {c^{k}c^{l}\over c^{2}} =  -\delta_{kl} + 2 c^{k}_{0}c^{l}_{0} \; ,
 \qquad \vec{c}_{0}^{\;2} = 1 \; .
\eqno(12.8a)
$$

\noindent
It is a local gauge rotation of the above  special type (12.2b):
$$
O(\pm \infty \;\vec{c}_{0}) = I + 2 (\vec{c}_{0}^{\times})^{2} =
\eqno(12.8b)
$$
$$
= \left | \begin{array}{ccc}
1 -2(c^{2}_{0} c^{2}_{0} + c^{3}_{0}c_{0}  ^{3}) & 2c^{1}_{0}c^{2}_{0} & 2c^{1}_{0}c^{3}_{0} \\
2 c^{1}_{0} c^{2}_{0} & 1 -2(c^{1} _{0}  c^{1}_{0} + c^{3}_{0}c^{3}_{0})& 2c^{2}_{0}c^{3}_{0} \\
2c^{3}_{0}c^{1}_{0} & 2c^{3}_{0}c^{2}_{0} & 1 -2(c^{1}_{0} c^{1}_{0} + c^{2}_{0} c^{2}_{0})
\end{array} \right | = -\delta_{kl} + 2 c^{k}_{0} c^{l} _{0} \; .
$$

\vspace{5mm}

Now in the elliptical space
let us introduce cylindrical coordinates and tetrad. Two half-space of spherical
space are given by
$$
n_{0} = \cos  \rho  \cos  z \; , \;\;  n_{3} = \cos  \rho  \sin  z
\; , \;\; n_{1} = \sin  \rho  \cos  \phi  \; , \;\; n_{2} = \sin  \rho
\sin  \phi \; ,
$$
$$
\underline{\delta = +1 \;}
 , \qquad  \qquad  n_{0} = \cos  \rho  \cos  z  \in [\; 0, +1 \; ]\; , \qquad
z \in [-{\pi \over 2}, \; {\pi \over 2}\; ] \; ;
\eqno(12.9a)
$$
$$
\underline{\delta = -1 \;} ,
\qquad n_{0} = \cos  \rho  \cos  z  \in [\; -1,\; \; 0 \; ]\; , \qquad
z \in [-\pi, \; -{\pi \over 2} \; ] \oplus [\; {\pi \over 2}, \pi \; \; ] \; .
$$

\noindent So one can use any of these two domains to cover the elliptical space model:
$$
\left. \begin{array}{ll}
\tilde{G }^{(+1)}(\phi,z)  \; : \qquad &   \phi  \in  [-
\pi  ,\;  + \pi  ]\;  ,  \; z \in  [-\pi /2, +\pi /2 ] \;  ,\\[3mm]
\tilde{G }^{(-1)}(\phi,z)  : \qquad    &   \phi  \in  [-
\pi  ,\;  + \pi  ]\;  ,  \; z \in  [-\pi , -\pi/2 ] \oplus [+\pi /2, +\pi  ] \; ;
\end{array} \right .
\eqno(12.9b)
$$

\noindent
The latter can be illustrated be the Fig 24:
$$
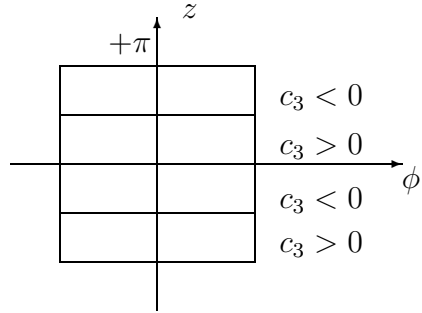
\mbox{Fig} 24   \qquad \mbox{Domains}\qquad
\tilde{G}^{(+1)}(\phi,z) - \tilde{G}^{(-1)}(\phi,z)
$$

\unitlength=0.65 mm
\begin{picture}(160,60)(-95,-25)
\special{em:linewidth 0.4pt} \linethickness{0.4pt}

\put(-30,0){\vector(+1,0){80}}    \put(+50,-5){$\phi $}
\put(0,-30){\vector(0,+1){60}}    \put(+5,+30){$z$}
\put(-20,+10){\line(+1,0){40}}

\put(-20,+20){\line(+1,0){40}}    \put(-20,-20){\line(0,+1){40}}
\put(-20,-10){\line(+1,0){40}}
\put(+20,-20){\line(0,+1){40}}
\put(-20,-20){\line(+1,0){40}}

\put(-10,+23){$+\pi $}

\put(+25,+2){$c_{3} >0$}  \put(+25,-18){$c_{3} >0$}
\put(+25,-9){$c_{3} <0$} \put(+25,+12){$c_{3} <0$}

\end{picture}
\vspace{5mm}

It should be noted that the domain  $\tilde{G}^{(-1)}$ consisting of two parts
can be transformed to the domain  $\tilde{G}^{(+1)}$ through the special
change of variables:
$$
z \in [{\pi\over 2}, \pi] \; , \qquad  z = \pi - Z \; ,  \qquad
 Z \in [ 0, +{\pi \over 2} ] \; ,
$$
$$
z \in [ -\pi, -{\pi \over 2} ] \;, \qquad  z = -\pi -Z \;, \qquad
 Z \in [ -{\pi\over 2} , 0 ]  \; ,
\eqno(12.10a)
$$

\noindent Here the half-space $n_{0}^{(-1)} \in [ \;-1,\;  0 \; ]$ is covered by
$$
n_{0} =  -
\cos  \rho  \cos Z   \; , \qquad
n_{3} = \cos  \rho  \sin  Z \; ,
$$
$$
n_{1} = \sin  \rho  \cos  \phi  \; , \qquad n_{2} = \sin  \rho \sin  \phi \; ,
\eqno(12.10b)
$$

\noindent in $c^{i}$-coordinates this looks as
$$
c_{1} =  -{\tan  \rho \over cos Z}\;  \cos  \phi  \; , \qquad
c_{2} = - {\tan  \rho  \over cos Z}\; \sin  \phi \; ,
c_{3} = - \tan  Z \; ;
\eqno(12.10c)
$$

This referring of two domains
$$
\tilde{G}^{(\delta=-1)} (\rho, \phi,z) \qquad
\stackrel{z \rightarrow Z}{\Longrightarrow}   \qquad \tilde{G}^{(\delta=+1)} (\rho,\phi,Z)
$$

\noindent
permits to establish identification rules
for $\tilde{G}^{(\delta=-1)} (\rho, \phi,z)$.  Indeed, from the scheme and relations
\vspace{3mm}
$$
 \mbox{Fig } \; 25\;\;\;\; \tilde{G}^{(\delta=+1)}_{1}(\phi,Z) : \;  \rho \neq 0, \pi/2
$$


\unitlength=0.5 mm
\begin{picture}(160,60)(-140,-30)
\special{em:linewidth 0.4pt} \linethickness{0.4pt}

\put(-60,0){\vector(+1,0){120}}
\put(+60,-5){$\phi $} \put(0,-30){\vector(0,+1){60}}
\put(+5,+30){$Z$}

\put(-40,-20){\line(+1,0){80}}
\put(-40,-20){\line(0,+1){40}} \put(+40,+20){\line(-1,0){80}}
\put(+40,+20){\line(0,-1){40}} \put(+40,+20){\line(-1,-1){40}}
\put(-40,-20){\line(+1,+1){40}} \put(-20,-20){\line(+1,+1){40}}
\put(-40,+20){\line(+1,-1){40}}
\put(-20,+20){\line(+1,-1){40}} \put(0,+20){\line(+1,-1){40}}
\put(-40,-10){\line(+1,0){80}}
\put(-40,+10){\line(+1,0){80}}


\put(+30,0){\circle{2}}   \put(+10,0){\circle{2}}
\put(-30,0){\circle{2}}   \put(-10,0){\circle{2}}

\end{picture}
\vspace{-5mm}

$$
\left. \begin{array}{llll}
\tilde{G}_{2}(0,\phi,Z): \qquad  & \rho =0 \; , \qquad  & \vec{c} = -(0,0, \tan Z ),
\qquad  & \\[2mm]
\tilde{G}_{3}(0,\phi,Z): \qquad  & \rho =\pi/2\;  , \qquad  &
\vec{c} = -\infty \; (\cos \phi, \sin \phi, 0 ),
\qquad &
  .
 \end{array} \right.
\eqno(12.11a)
$$

it follows the corresponding scheme and relations  for
$\tilde{G}^{(\delta=-1)}_{1}(\phi,z) $:
\vspace{3mm}
$$
\mbox{Fig}  \; 26\;\;\;\;
\tilde{G}^{(\delta=-1)}_{1}(\phi,z) : \;  \rho \neq 0, \pi/2
$$

\vspace{10mm}

\unitlength=0.5 mm
\begin{picture}(160,60)(-140,-30)
\special{em:linewidth 0.4pt} \linethickness{0.4pt}

\put(-80,0){\vector(+1,0){160}}
\put(+60,-5){$\phi $} \put(0,-50){\vector(0,+1){100}}
\put(+5,+50){$Z$}

\put(-40,-20){\line(+1,0){80}}

\put(-40,-20){\line(0,-1){20}}    \put(-40,+20){\line(0,+1){20}}     \put(+40,+20){\line(-1,0){80}}
\put(+40,+20){\line(0,+1){20}}    \put(+40,-20){\line(0,-1){20}}
\put(+40,+19){\line(-1,-1){38}}

\put(-39,-19){\line(+1,+1){38}} \put(-20,-19){\line(+1,+1){38}}
\put(-40,+19){\line(+1,-1){38}}
\put(-20,+19){\line(+1,-1){38}} \put(0,+19){\line(+1,-1){38}}
\put(-40,-40){\line(+1,0){80}}
\put(-40,+40){\line(+1,0){80}}


\put(+30,+40){\circle{2}} \put(+10,+40){\circle{2}} \put(-30,40){\circle{2}} \put(-10,40){\circle{2}}
\put(+30,-40){\circle{2}} \put(+10,-40){\circle{2}} \put(-30,-40){\circle{2}} \put(-10,-40){\circle{2}}

\put(-40,30){\line(+1,0){80}} \put(-40,-30){\line(+1,0){80}}

\end{picture}

\vspace{10mm}

$$
\left. \begin{array}{llll}
\tilde{G}_{2}(0,\phi,z): \qquad  & \rho =0 \; , \qquad  & \vec{c} = (0,0, \tan z ),
\qquad  &  ;\\[2mm]
\tilde{G}_{3}(0,\phi,z): \qquad  & \rho =\pi/2\;  , \qquad  &\vec{c} =
\infty \; (\cos \phi, \sin \phi, 0 ),
\qquad &
  .
 \end{array} \right.
\eqno(12.11b)
$$

\begin{quotation}

{\em
We will see below that two other variants combining previous ones  $\delta=+1,-1$
may be of interest. In the Fig 24 they  correspond to the domains
 $(A),\; z \in [0, + \pi ]$  and
 $(B),\; z \in [-\pi,  0]$  }.

\end{quotation}

For definiteness, let us follow the variant $\delta = \pm 1$ and
correspondingly two tetrads:
$$
(\phi,z) \in \tilde{G}^{(+1)}(\phi,z)\; , \qquad
 (\phi,z) \in  \tilde{G}^{(-1)}(\phi,z)
$$
$$
dl^{2} (y^{(\delta)}) =  d\rho^{2} + \sin^{2} \rho  \; d\phi^{2} +
\cos^{2} \; dz^{2}  \; ,
$$
$$
e^{(\delta)}_{(i)j}(y) = \left | \begin{array}{lll}
  1  &  0  &  0  \\
  0  & \sin  \rho  &   0  \\
  0  &  0  & \cos \rho
\end{array}  \right |  \; .
\eqno(12.12)
$$

One should find an explicit form of the gauge matrix relating
cartesian  $e^{(\delta)(b)}_{\beta}(c)$ and cylindrical   $e_{(a)}^{(\delta)\alpha}(y)$
tetrad bases in elliptical space.
$$
L_{a}^{(\delta) \;b}(y) = e_{(a)}^{\alpha}(y) \;
{\partial c^{\beta} \over \partial y^{\alpha}} \; e^{(\delta)(b)}_{\beta}(c)\; ;
\eqno(12.13a)
$$

\noindent
With the use of (3.2), for
 $(\partial c^{\beta} /  \partial y^{\alpha})$ we have
$$
{\partial c^{1} \over \partial \rho } = {\cos \phi  \over \cos^{2}\rho \cos z} \; , \qquad
{\partial c^{2} \over \partial \rho } = {\sin \phi  \over \cos^{2}\rho \cos z} \; , \qquad
{\partial c^{3} \over \partial \rho } = 0 \; ,
$$
$$
{\partial c^{1} \over \partial \phi } = - {  \tan \rho \over \cos z} \sin \phi \; , \qquad
{\partial c^{2} \over \partial \phi } = + {  \tan \rho \over \cos z} \cos \phi \; , \qquad
{\partial c^{3} \over \partial \phi } = 0 \; ,
$$
$$
{\partial c^{1} \over \partial z } =
{\tan \rho \sin z  \over \cos^{2} z} \cos \phi \; , \qquad
{\partial c^{2} \over \partial z } =
{\tan \rho \sin z  \over \cos^{2} z} \sin \phi \; , \qquad
{\partial c^{3} \over \partial z } =
{1  \over \cos^{2} z}  \; ,
$$

\noindent and thee matrix  $L_{a}^{(\delta) \;b}(y) $
reads s  (the notation is used: $(1 + \delta \cos \rho \cos z)^{-1}= \varphi $)
$$
L_{a}^{(\delta) \;b}(\rho,\phi,z) = \left | \begin{array}{ll}
1 & \qquad 0 \\
0 & \qquad\delta\;\varphi \; (\cos \rho + \delta \cos z)\cos \phi \\
0 &\qquad-\delta\; \sin \phi                                 \\
0 & \qquad\varphi \;\sin \rho \;\sin z \; \cos  \phi
\end{array} \right.
$$
$$
\left. \begin{array}{ll}
0 & 0 \\
\delta\;\varphi\; (\cos \rho + \delta \cos z)\sin \phi &
-\delta\;\varphi \;\sin \rho \;\sin z   \\
\delta\; \cos \phi  & 0  \\
\varphi \;  \sin \rho \;\sin z \;\sin  \phi  &
\delta\;\varphi \; (\cos z + \delta \cos \rho )
\end{array} \right | \; .
\eqno(12.13b)
$$

\begin{quotation}
By definition, the matrix $L_{a}^{(\delta= +1) \;b}(\rho, \phi,z)$
must be considered in the domain  $\tilde{G}^{(+1)}(\rho,\phi,z)$,
whereas the matrix  $L_{a}^{(\delta = -1) \;b}(\rho,\phi,z)$  should be taken
in $\tilde{G}^{(-1)}(\rho,\phi,z)$.
\end{quotation}

In isotropic form (for more detail see in [23])  the matrix  $L^{(\delta)}$ is given by
$$
U_{(\delta)} (\rho,\phi,z) = {1\over 2}
\left | \begin{array}{ll}
( 1 + \delta \; \varphi \; (\cos z + \delta \cos \rho) ) &
( 1 - \delta \; \varphi \; (\cos z + \delta \cos \rho) ) \\
( 1 - \delta \; \varphi \; (\cos z + \delta \cos \rho))  &
( 1 + \delta \; \varphi \; (\cos z + \delta \cos \rho) ) \\
-\delta \; \varphi \; \sin \rho \; \sin z &
+\delta \; \varphi \; \sin \rho \; \sin z \\
-\delta \; \varphi \; \sin \rho \; \sin z &
+\delta \; \varphi \; \sin \rho \; \sin
\end{array} \right.
$$
$$
\left. \begin{array}{rr}
\varphi \; \sin \rho \; \sin z \; e^{-i\phi}  &
\varphi \; \sin \rho \; \sin z \; e^{+i\phi}  \\
- \varphi \; \sin \rho \; \sin z \; e^{-i\phi}  &
- \varphi \; \sin \rho \; \sin z \; e^{+i\phi}  \\
\delta \; (1 + \delta \; \varphi \; (\cos z + \delta \cos \rho)
\; e^{-i\phi}  &
-\delta \; (1 - \delta \; \varphi \; (\cos z + \delta \cos \rho))
\; e^{+i\phi}  \\
-\delta \; (1 - \delta \; \varphi \; (\cos z + \delta \cos \rho))
\; e^{-i\phi}  &
+\delta \; (1 + \delta \; \varphi \; (\cos z + \delta \cos \rho))
\; e^{+i\phi}
\end{array} \right |
\eqno(12.15a)
$$

\noindent    The isotropic matrix (12.15a) must be  seeing as parameterized
in accordance with
$$
U(\rho ,\; \phi ,\; z) =  \left | \begin{array}{cccc}
a a^{*}   &  c c^{*}  &  a c^{*}  & a^{*} c  \\
c c^{*}   &  a a^{*}  & -a c^{*} & -a^{*} c \\
- a c     &  a c      &   a  a   &  - c  c  \\
- a^{*} c^{*}     &  a^{*} c^{*}    &  -c^{*} c^{*}   & a^{*} a^{*}
\end{array} \right | \; ;
\eqno(12.15b)
$$

\noindent where   $a,b$ stand for the elements of spinor  matrix  $ B_{(\delta)}$:
$$
B_{(\delta)}(\rho,\phi,z) =
\left | \begin{array}{cc}
a  & c \\-c^{*}  & a^{*} \end{array} \right | =
\left | \begin{array}{cc}
A(y) \; e^{i\alpha(y)} & C(y)\;e^{is(y)} \\
- C(y)\;e^{-is(y)} &     A(y) \; e^{-i\alpha(y)}
\end{array} \right |
\eqno(12.15c)
$$

\noindent
As a first step, comparing (12.15a) and (12.15b),   $A$ and  $C $ are established
$$
A^{2} = {1\over 2} ( 1 + \delta \; \varphi \; (\cos z + \delta \cos \rho) ) = {(1+\cos \rho)(1+\delta\cos z)
\over 1+ \delta \cos \rho \cos z} \; ,
$$
$$
C^{2} = {1\over 2} ( 1 - \delta \; \varphi \; (\cos z + \delta \cos \rho) ) =
{(1-\cos \rho)(1 -\delta\cos z) \over 1 + \delta \cos \rho \cos z} \; ,
$$

From the given  $A^{2}$ and  $a^{2}$ one finds  $e^{-i\alpha}$:
$$
  e^{2i\alpha} \; A^{2} = a^{2} =  \delta  e^{-i\phi}
\; [1 + \delta \; \varphi \; (\cos z + \delta \cos \rho)  ] : \qquad  \Longrightarrow
$$
$$
e^{i\alpha} =  \sigma \; \sqrt{\delta} \;  e^{-i\phi/2}   \; , \qquad
e^{-i\alpha} = \sigma \; (\sqrt{\delta} \;) ^{*} \;  e^{+i\phi/2}  \; , \;\; \sigma = \pm 1 \;  .
$$

\noindent Now, from given  $-a^{*}c$, it follows expression for  $e^{is}$:
$$
 - A e^{-i\alpha} C e^{is} =- \varphi \; \sin \rho \; \sin z \; e^{+i\phi}
 \qquad \Longrightarrow
$$
$$
e^{is} = { \varphi \; \sin \rho \sin z \over AC} \; e^{i\phi} \; e^{i\alpha} =
{\sin z \over \sqrt{\sin^{2} z}} \; (\sigma \sqrt{\delta} e^{i\phi/2} )\; .
$$

\noindent
Thus, the spinor gauge  matrix  $B^{(\delta)}(\rho,\phi,z)$ is
$$
A(y) = + {\sqrt{(1+\cos \rho)(1+\delta\cos z)\over
2(1 + \delta \cos \rho \cos z)}} \; ,
$$
$$
C(y) = + {\sqrt{(1-\cos \rho)(1-\delta\cos z)\over
2(1 + \delta \cos \rho \cos z)}} \; , \;
$$
$$
e^{i\alpha(y)} = \sigma \; \sqrt{\delta}\; e^{-i\phi/2} \; , \qquad
e^{-i\alpha(y)} = \sigma \; (\sqrt{\delta}\;)^{*} \; e^{+i\phi/2} \; ,
$$
$$
e^{is(y)} =
 \sigma \sqrt{\delta} e^{i\phi/2} \; {\sin z \over \sqrt{\sin^{2} z}}  ,
\qquad
e^{-is} =
\sigma (\sqrt{\delta}\; )^{*} e^{-i\phi/2} \; {\sin z \over \sqrt{\sin^{2} z}} \;  \; ,
$$

\noindent

From  (12.13a,b),  taking into account identity
$$
 + \sqrt{{1-\cos z \over 2}} \; ( \;{+\sqrt{\sin^{2} z} \over \sin z}\;)
= \sin  {z \over 2} \; ,
$$

\noindent
for  $B^{(+1)}(\rho,\phi,z) $ and $B^{(-1)}(\rho,\phi,z)$  we get
$$
B_{(+1)}(\rho,\phi,z) =
{1 \over \sqrt{1 +  \cos \rho \cos z }   } \; \times
$$
$$
\times
\left | \begin{array}{rr}
\sqrt{1 + \cos \rho} \; \cos{z\over 2}\; e^{+i\phi/2} &
\sqrt{1 - \cos \rho} \; \sin{z\over 2}\; e^{-i\phi/2} \\
-\sqrt{1 - \cos \rho} \; \sin{z\over 2}\; e^{+i\phi/2} &
\sqrt{1 + \cos \rho} \; \cos{z\over 2}\; e^{-i\phi/2}
\end{array} \right |     \; ,
\eqno(12.16a)
$$
$$
B_{(-1)}(\rho,\phi,z) =
{1 \over \sqrt{1 -  \cos \rho \cos z }   } \;
( {+\sqrt{\sin^{2} z} \over \sin z}) \times
$$
$$
\times
\left | \begin{array}{rr}
\sqrt{1 + \cos \rho} \; \sin{z\over 2}\;i\; e^{+i\phi/2} &
\sqrt{1 - \cos \rho} \; \cos {z\over 2}\;i\; e^{-i\phi/2} \\
\sqrt{1 - \cos \rho} \; \cos{z\over 2}\;i \; e^{+i\phi/2} &
- \sqrt{1 + \cos \rho} \; \sin {z\over 2}\; i\;e^{-i\phi/2}
\end{array} \right | \; ;
\eqno(12.16b)
$$

\noindent
Below it will be used notation
$ {+\sqrt{\sin^{2} z} /  \sin z}  = \mbox{Sgn}\; z  \in \{ +1,-1 \} \; .
$

\begin{quotation}
By procedure used, the matrix  $B_{(\delta= +1)}(\rho, \phi,z)$
should be considered in the domain  $\tilde{G}^{(+1)}(\rho,\phi,z)$,
and the matrix  $B_{(\delta = -1)}(\rho,\phi,z)$
should be taken in the  domain  $\tilde{G}^{(-1)}(\rho,\phi,z)$.
In the same time, one might employ them both in one the same domain, in such a context
these two matrices would be connected by a local spinor gauge  transformation  associated with
vector matrix (12.8).
\end{quotation}

In the vicinity of the point  \underline{$(n_{0}=+1,0,0,0)$} (the limit of flat space)
 the matrix $B_{(+1)}$ will look
$$
\rho  \;\; \rightarrow \;\; 0, \; z  \;\; \rightarrow \;\; 0: \qquad
B_{(+1)}(\rho,\phi,z) \rightarrow
\left | \begin{array}{rr}
 e^{+i\phi/2} & 0  \\
0 &  e^{-i\phi/2}
\end{array} \right |     \; .
\eqno(12.17a)
$$

\noindent
As for the matrix $B_{(-1)}$, the vicinity of opposite point
 \underline{$(n_{0} = -1,0,0,0)$} should be taken:
$$
\rho \rightarrow 0, \qquad z = \pi - Z , \qquad  Z \rightarrow +0 \; ,\qquad
 z = -\pi - Z , \qquad  Z \rightarrow -0
$$

\noindent and
$$
B_{(-1)}(\rho,\phi,z)  \;\; \rightarrow \;\;
\left | \begin{array}{rr}
\;i\; e^{+i\phi/2} & 0\\
0 & -  i\;e^{-i\phi/2}
\end{array} \right | \; ;
\eqno(12.17b)
$$

\subsection*{13. On continuity condition for a fermion in elliptical space }

\hspace{5mm}
Now we are ready to solve the main problem -- analyzing the
continuity condition for a fermion in elliptical space.
There are two different bases playing the role of Cartesian in elliptical space:
$$
\psi_{cart} = B ^{-1} (\rho, \phi, z) \; \psi_{cyl} \; ;
\eqno(13.1)
$$

\noindent each variant with $\delta =+1$ and  $\delta=-1$  presupposes
its own domain  $\tilde{G}^{(\delta)}(\rho,\phi,z)$ with respective identification
 rules on the boundary:

 \vspace{5mm}
\underline{$(\delta=+1)$}
$$
\psi^{(\delta=+1)}_{1} = { e^{i(m-1/2)\phi }  e^{i(k+1/2)z}  \over \sqrt{1 + \cos \rho \cos z } }
  \; \left [   \; (   1  + \cos \rho \;  e^{-iz} )
  \;   G_{1} (\rho)    -i \sin \rho \;  e^{-iz} \;    G_{2} (\rho)  \; \right ] \; ,
$$
$$
\psi^{(\delta=+1)}_{2} = { e^{i(m+1/2)\phi }  e^{i(k-1/2)z}  \over \sqrt{1 + \cos \rho \cos z } }
  \; \left [ \;      -i \sin \rho \;  e^{+iz} \;    G_{1} (\rho)  +(   1  + \cos \rho  \; e^{+iz} )
  \;   G_{2} (\rho)  \;     \right ] \; .
$$
$$
\eqno(13.2)
$$

\underline{$(\delta=-1)$}
$$
\psi^{(\delta=-1)}_{1} = ( \mbox{Sgn} \; z)\; { e^{i(m-1/2)\phi }  e^{i(k+1/2)z}  \over \sqrt{1 -\cos \rho \cos z } }
  \; \left [   \; (  - 1  + \cos \rho \;  e^{-iz} )
  \;   G_{1} (\rho)    -i \sin \rho \;  e^{-iz} \;    G_{2} (\rho)  \; \right ] \; ,
$$
$$
\psi^{(\delta=-1)}_{2} = ( \mbox{Sgn}\; z) \; { e^{i(m+1/2)\phi }  e^{i(k-1/2)z}  \over \sqrt{1 - \cos \rho \cos z } }
  \; \left [ \;     -i \sin \rho \;  e^{+iz} \;    G_{1} (\rho)
  + (  - 1  + \cos \rho  \; e^{+iz} )
  \;   G_{2} (\rho)  \;     \right ] \; .
$$
$$
\eqno(13.3)
$$

\begin{quotation}

{\em
Because the requirement of continuity includes
identity
$$
\phi  \Longrightarrow \phi \pm \pi \; ,
\qquad z \Longrightarrow  z \pm \pi \; .
$$

\noindent
it is easily seen that both variants,  (13.2) and  (13.3),
lead us to the same result: no fermion solutions continuous in elliptical space exist.
}
\end{quotation}

\begin{quotation}

{\em
One could try to construct continuous in $\tilde{S}_{3}$  solutions
through the  combination  of variants  (13.2) and  (13.3).
Let  us examine this possibility in detail.
}

\end{quotation}

\underline{Variant (A):}

\vspace{3mm}
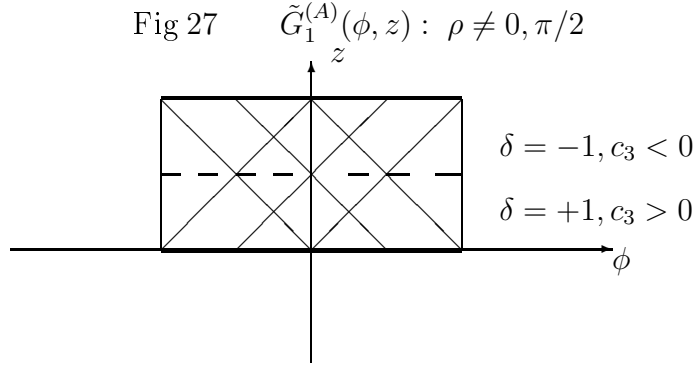
$$
 \mbox{Fig}  \; 27\qquad \tilde{G}^{(A)}_{1}(\phi,z) : \;  \rho \neq 0, \pi/2
$$


\unitlength=0.5 mm
\begin{picture}(160,60)(-140,-30)
\special{em:linewidth 0.4pt} \linethickness{0.4pt}

\put(-80,-20){\vector(+1,0){160}}
\put(+80,-25){$\phi $} \put(0,-50){\vector(0,+1){80}}
\put(+5,+30){$z$}

\put(+50,-12){$\delta = +1, c_{3} >0 $}  \put(+50,+5){$\delta = -1, c_{3} <0$}

\put(-40,-20){\line(+1,0){80}} \put(-40,-20.4){\line(+1,0){80}} \put(-40,-20.6){\line(+1,0){80}}
\put(-40,+20){\line(+1,0){80}} \put(-40,+20.4){\line(+1,0){80}} \put(-40,+20.6){\line(+1,0){80}}

\put(-40,-20){\line(0,+1){40}} \put(+40,+20){\line(-1,0){80}}
\put(+40,+20){\line(0,-1){40}} \put(+40,+20){\line(-1,-1){40}}
\put(-40,-20){\line(+1,+1){40}} \put(-20,-20){\line(+1,+1){40}}
\put(-40,+20){\line(+1,-1){40}}
\put(-20,+20){\line(+1,-1){40}} \put(0,+20){\line(+1,-1){40}}

\put(-40,0){\line(+1,0){5}}
\put(-30,0){\line(+1,0){5}}
\put(-20,0){\line(+1,0){5}}
\put(-10,0){\line(+1,0){5}}
\put(+10,0){\line(+1,0){5}}
\put(+20,0){\line(+1,0){6}}
\put(+33,0){\line(+1,0){7}}

\end{picture}

\vspace{15mm}

Identification in horizontal  boundaries is made in accordance with
$$
(\phi, \qquad z= +0)\; : \;\;\;\qquad c_{1} = + \tan \rho \; \cos \phi\; , \;  c_{2} = + \tan \rho \; \sin \phi \;, \;
c_{3} = 0 \; , \qquad  \;\;
$$
$$
(\phi+\pi, z= \pi - 0)\; : \qquad
c_{1} = + \tan \rho \; \cos \phi \; ,\;  c_{2} = +\tan \rho \; \sin \phi \; ,\;  c_{3} = 0 \; ;
\eqno(13.4a)
$$

\noindent  the points of the internal dashed line  in Fig 27 are associated with the
infinite length vectors
$$
(\phi, z = \pi/2 -0)\;: \qquad c_{1} = + \infty\; \tan \rho \; \cos \phi \; , \;
c_{2} = + \infty \tan \rho \; \sin \phi \; , \; c_{3} = + \infty \; ,
$$
$$
(\phi, z = \pi/2 +0)\;: \qquad c_{1} = - \infty\; \tan \rho \; \cos \phi \; , \;
c_{2} = - \infty \tan \rho \; \sin \phi \; , \; c_{3} = - \infty \; .
$$
$$
\eqno(13.4b)
$$

\underline{Variant  (B):}

\vspace{3mm}
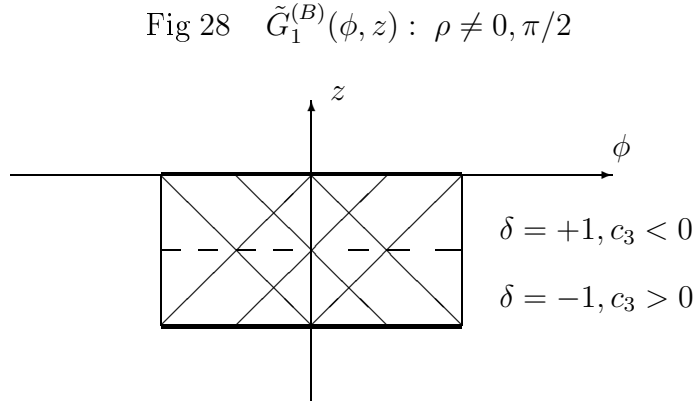
$$
\mbox{Fig}  \; 28\;\;\;\;
\tilde{G}^{(B)}_{1}(\phi,z) : \;  \rho \neq 0, \pi/2
$$

\vspace{10mm}

\unitlength=0.5 mm
\begin{picture}(160,60)(-140,-30)
\special{em:linewidth 0.4pt} \linethickness{0.4pt}

\put(-80,+20){\vector(+1,0){160}}
\put(+80,+25){$\phi $} \put(0,-40){\vector(0,+1){80}}
\put(+5,+40){$z$}

\put(-40,-20){\line(+1,0){80}} \put(-40,-20.4){\line(+1,0){80}} \put(-40,-20.6){\line(+1,0){80}}
\put(-40,+20){\line(+1,0){80}} \put(-40,+20.4){\line(+1,0){80}} \put(-40,+20.6){\line(+1,0){80}}

\put(+50,+3){$\delta = +1, c_{3} <0 $}  \put(+50,-15){$\delta = -1, c_{3} >0$}

\put(-40,-20){\line(0,+1){40}} \put(+40,+20){\line(-1,0){80}}
\put(+40,+20){\line(0,-1){40}} \put(+40,+20){\line(-1,-1){40}}
\put(-40,-20){\line(+1,+1){40}} \put(-20,-20){\line(+1,+1){40}}
\put(-40,+20){\line(+1,-1){40}}
\put(-20,+20){\line(+1,-1){40}} \put(0,+20){\line(+1,-1){40}}

\put(-40,0){\line(+1,0){5}}
\put(-30,0){\line(+1,0){5}}
\put(-20,0){\line(+1,0){5}}
\put(-10,0){\line(+1,0){5}}
\put(+10,0){\line(+1,0){5}}
\put(+20,0){\line(+1,0){6}}
\put(+33,0){\line(+1,0){7}}

\end{picture}

\vspace{10mm}
Identification in horizontal  boundaries is made in accordance with
$$
(\phi, \qquad z= -0)\; : \;\;\;\qquad c_{1} = + \tan \rho \; \cos \phi\; , \;  c_{2} = + \tan \rho \; \sin \phi \;, \;
c_{3} = 0 \; ,
$$
$$
(\phi+\pi, z= -\pi + 0)\; : \qquad
c_{1} = + \tan \rho \; \cos \phi \; ,\;  c_{2} = +\tan \rho \; \sin \phi \; ,\;  c_{3} = 0 \; ;
$$
$$
\eqno(13.5a)
$$

\noindent  the points of the internal dashed line  in Fig 27 are associated with the
infinite length vectors
$$
(\phi, z = -\pi/2 +0)\;: \qquad c_{1} = + \infty\; \tan \rho \; \cos \phi \; , \;
c_{2} = + \infty \tan \rho \; \sin \phi \; , \; c_{3} = - \infty \; ,
$$
$$
(\phi, z = -\pi/2 - 0)\;: \qquad c_{1} = - \infty\; \tan \rho \; \cos \phi \; , \;
c_{2} = - \infty \tan \rho \; \sin \phi \; , \; c_{3} = + \infty \; ,
$$
$$
\eqno(13.5b)
$$

The Variants  (A) and (B)   lead to the fermion functions of the form:
\vspace{5mm}
\underline{ Variant  (A)}
$$
z\in [0, +{\pi \over 2} ]: \qquad
\psi^{(\delta=+1)}_{1} = { e^{i(m-1/2)\phi }  e^{i(k+1/2)z}  \over \sqrt{1 + \cos \rho \cos z } }
  \left [   (   1  + \cos \rho   e^{-iz} )
    G_{1}     -i \sin \rho  e^{-iz}     G_{2}   \right ] \; ,
$$
$$
z \in [{\pi \over 2}  , +\pi]: \qquad
\psi^{(\delta=-1)}_{1} =  { e^{i(m-1/2)\phi }  e^{i(k+1/2)z}  \over \sqrt{1 - \cos \rho \cos z } }
\left [    (   1  - \cos \rho   e^{-iz} )
G_{1}   +i \sin \rho   e^{-iz}     G_{2}    \right ] \; ;
$$
$$
z\in [0, +{\pi\over 2} ]: \qquad
\psi^{(\delta=+1)}_{2} = { e^{i(m+1/2)\phi }  e^{i(k-1/2)z}  \over \sqrt{1 + \cos \rho \cos z } }
  \left [ -i \sin \rho   e^{+iz}    G_{1}   +(   1  + \cos \rho   e^{+iz} )
 G_{2}       \right ] \; ,
$$
$$
 z\in [{\pi \over 2}  , +\pi]:  \qquad
\psi^{(\delta=-1)}_{2} = { e^{i(m+1/2)\phi }  e^{i(k-1/2)z}  \over \sqrt{1 - \cos \rho \cos z } }
\left [   +i \sin \rho  e^{+iz}     G_{1}
  + (   1  - \cos \rho  e^{+iz} )
   G_{2}      \right ] \; .
$$
$$
\eqno(13.6)
$$

\underline{ Variant  (B)}
$$
z\in [0, -{\pi \over 2} ]: \qquad
\psi^{(\delta=+1)}_{1} = { e^{i(m-1/2)\phi }  e^{i(k+1/2)z}  \over \sqrt{1 + \cos \rho \cos z } }
  \left [    (   1  + \cos \rho   e^{-iz} )
     G_{1}     -i \sin \rho   e^{-iz}     G_{2}    \right ] \; ,
$$
$$
z\in [-\pi, -{\pi \over 2} ]: \qquad
\psi^{(\delta=-1)}_{1} =  { e^{i(m-1/2)\phi }  e^{i(k+1/2)z}  \over \sqrt{1 - \cos \rho \cos z } }
   \left [    (  - 1  + \cos \rho   e^{-iz} )
  G_{1}     -i \sin \rho   e^{-iz}     G_{2}  \right ] \; ,
$$
$$
z\in [0, - {\pi \over 2} ]: \qquad
\psi^{(\delta=+1)}_{2} = { e^{i(m+1/2)\phi }  e^{i(k-1/2)z}  \over \sqrt{1 + \cos \rho \cos z } }
 \left [       -i \sin \rho   e^{+iz} \;    G_{1}   +(   1  + \cos \rho   e^{+iz} )
 G_{2}       \right ]  .
$$
$$
z\in [-\pi, -{\pi \over 2} ]: \qquad
\psi^{(\delta=-1)}_{2} =  { e^{i(m+1/2)\phi }  e^{i(k-1/2)z}  \over \sqrt{1 - \cos \rho \cos z } }
  \left [      -i \sin \rho   e^{+iz}     G_{1}
  + (  - 1  + \cos \rho   e^{+iz} )
   G_{2}        \right ] \; ,
$$
$$
\eqno(13.7)
$$

\begin{quotation}

{\em
It is easily seen that both in  (A)  and  (B),  one can satisfy
$\psi (M) = \psi (M')$, where  $M$ and $M'$ belong to vertical and horizontal boundaries,
however such solutions would take  different values in the identified
point $\pm \infty \vec{c}_{0}$  of the internal dashed lines.
Therefore, the variants (A) and (B)  give the same result: any Dirac solutions continuous in
 elliptical space model do not exist.
 }

\end{quotation}

\subsection*{Supplement A.  Cartesian and cylindrical tetrads in
$S_{3}$.}

\hspace{5mm}
In $S_{3}$
$$
 n_{0}^{2} + n_{1}^{2} + n_{2}^{2} +  n_{3}^{2}  = \rho ^{2} \; ,
\eqno(A.1)
$$

\noindent one  may  introduce (quasi) cartesian coordinates and tetrad
$$
x^{1} = {n_{1} \over 1 + n_{4}}\; , \;\;
x^{2} = {n_{2} \over 1 + n_{4}}\; , \;\;
x^{3} = {n_{3} \over 1 + n_{4}}\; ;
\eqno(A.2a)
$$
$$
n_{i} = {2 x^{i} \over  1 + x^{2} } \; , \;\;
n_{4} = {1 - x^{2} \over  1 + x^{2} } \; , \;\;
 \; , \;\;x^{2} =  x^{1} x^{1} + x^{2} x^{2}  + x^{3} x^{3}  \; ;
\eqno(A.2b)
$$
$$
dl^{2}(x) = {1 \over f^{2}} \; [\; (dx^{1})^{2} +
(dx^{2})^{2} +    (dx^{3})^{2} \; ] \; , \;\; f = { 1 + x^{2}\over 2} \; ;
\eqno(A.2c)
$$
$$
e_{(a)}^{\;\;\alpha}(x) = \left | \begin{array}{cccc}
1 &  0    &     0  & 0 \\
0 &  1/f &  0  &  0   \\
0 &  0 &  1/f  &  0   \\
0 &  0 &  0   & 1/f   \\
\end{array}  \right | \; ;
\eqno(A.2d)
$$

\noindent here  $a$ and  $\alpha$ take values  $0,1,2,3$; $(a)$ stands for a tetrad index;
 $\alpha$ stands for a  generally covariant index.
Also, in the spherical space one cam introduce cylindrical coordinates and tetrad:
$$
n_{1} = \sin \rho \; \cos \phi \; , \;\;
n_{2} = \sin \rho \; \sin \phi \; , \;\;
n_{3} = \cos \rho \; \sin z  \;   , \;\;
n_{4} = \cos \rho \; \cos z  \; ,
$$
$$
dl^{2} (y) = ( d\rho^{2} + \sin^{2} \rho  \; d\phi^{2} +
\cos^{2} \; dz^{2} ) \; ,
\;
e_{(a)}^{\alpha}(y) = \left | \begin{array}{llll}
1  &  0  &  0  &  0  \\
0  &  1  &  0  &  0  \\
0  &  0  & \sin^{-1} \rho  &   0  \\
0  &  0  &  0  & \cos^{-1}\rho
\end{array}  \right |  \; ;
\eqno(A.3)
$$
$$
G(\rho,\; \phi,\; z ) =  \{
\rho \in [ 0, \; +\pi /2 ] \; , \;\;
\phi \in [-\pi,\; +\pi ] \; , \;\;
z    \in [-\pi,\;+\pi ] \; \} \; .
$$

It is the matter of simple calculation to find a vector and spinor gauge matrices
relating Cartesian and cylindrical bases in space $S_{3}$. The vector matrix obeys the equation
$$
e_{(a)}^{\alpha}(y) = {\partial y^{\alpha} \over \partial x^{\beta}}\;
L_{a}^{\;\;b}(y) \;e_{(b)}^{\beta}(x)   \; ;
\eqno(A.4a)
$$

\noindent  from this it follows $$
L_{a}^{\;\;b}(y) = e_{(a)}^{\alpha}(y) \;
{\partial x^{\beta} \over \partial y^{\alpha}}\;  e_{\beta}^{(b)}(x) \; .
\eqno(A.4b)
$$

\noindent
With the use of the formulas
$$
x^{1} = {\sin \rho \; \cos \phi \over 1 + \cos \rho \cos z }\; , \;\;
x^{2} = {\sin \rho \; \sin \phi \over 1 + \cos \rho \cos z }\; , \;\;
x^{3} = {\cos \rho \; \sin z \over 1 + \cos \rho \cos z }\; ;
\eqno(A.5)
$$
$$
{ \partial x^{1} \over \partial \rho } =
{(\cos \rho + \cos z ) \cos \phi \over (1 + \cos \rho \cos z )^{2}}\; , \;\;
{ \partial x^{2} \over \partial \rho } =
{(\cos \rho + \cos z ) \sin \phi \over (1 + \cos \rho \cos z )^{2}}\; , \;\;
{ \partial x^{3} \over \partial \rho } = -\;
{\sin  \rho \;  \sin z  \over (1 + \cos \rho \cos z )^{2}}\; ,
$$
$$
{ \partial x^{1} \over \partial \phi } = - \;
{\sin \rho \; \sin \phi  \over (1 + \cos \rho \cos z )}\; , \qquad
{ \partial x^{2} \over \partial \phi  } =
{\sin \rho \;  \cos \phi \over (1 + \cos \rho \cos z )}\; , \qquad
{ \partial x^{3} \over \partial \phi  } = 0 \; ,
$$
$$
{ \partial x^{1} \over \partial z } =
{\sin \rho \; \cos \rho \; \sin z  \over (1 + \cos \rho \cos z )^{2}}\;
\cos \phi \; , \;
{ \partial x^{2} \over \partial z } =
{\sin \rho \; \cos \rho \; \sin z  \over (1 + \cos \rho \cos z )^{2}}\;
\sin \phi \; , \;
{ \partial x^{3} \over \partial z } =
{(\cos \rho + \cos z ) \cos \rho  \over (1 + \cos \rho \cos z )^{2}}\; .
$$

\noindent
one can  determine  the explicit form of $ L_{a}^{\;\;b}(y)$
(below  $ \varphi = (1 + \cos \rho \cos z )^{-1}$)
$$
L_{a}^{\;\;b}(y)   =
\left | \begin{array}{cccc}
1  &  0  &  0  &  0  \\
0  &  \varphi (\cos \rho + \cos z ) \cos \phi  &
\varphi (\cos \rho + \cos z ) \sin \phi        &
- \varphi \sin  \rho \;  \sin z                \\
0  &  -\sin \phi  & \cos \phi  & 0  \\
0  & \varphi \sin \rho \; \sin z \; \cos \phi  &
\varphi \sin \rho \; \sin z \; \sin \phi       &
\varphi ( \cos \rho +  \cos z )
\end{array} \right |\; .
\eqno(A.6)
$$

\noindent By  the given vector matrix  $L_{a}^{\;\;b}(\rho, \phi, z)$ one should
determine  a corresponding spinor matrix
$$
B (\rho, \phi, z) = \left | \begin{array}{rr}
a  &  c  \\ - c^{*}  &  a^{*}  \end{array} \right |\; , \;\;
B \in   SU(2)   \; ,\; \; a \; a^{*} + c \; c^{*} = + 1 \; .
$$

\noindent It is convenient to employ isotropic representation
for the above  vector  matrix $L(y) \; \Longrightarrow \; U(y) = S \; L(y) S^{-1}$:
$$
S = {1\over \sqrt{2}} \left | \begin{array}{rrrr}
1   &   0  &  0   &  1  \\
1   &   0  &  0   & -1  \\
0   &   1  & -i   &  0  \\
0   &   1  & +i   &  0      \end{array} \right | \; , \;
S^{-1} = {1\over \sqrt{2}} \left | \begin{array}{rrrr}
1   &   1  &  0   &  0  \\
0   &   0  &  1   &  1  \\
0   &   0  &  i   & -i  \\
1   &  -1  &  0   & 0       \end{array} \right |\; .
\eqno(A.7)
$$

\noindent In isotropic basis the Lorentz matrix must be parameterized in accordance  with
\footnote{For more detail see  [23].}
 $$
U(\rho ,\; \phi ,\; z) =  \left | \begin{array}{cccc}
a a^{*}   &  c c^{*}  &  a c^{*}  & a^{*} c  \\
c c^{*}   &  a a^{*}  & -a c^{*} & -a^{*} c \\
- a c     &  a c      &   a  a   &  - c  c  \\
- a^{*} c^{*}     &  a^{*} c^{*}    &  -c^{*} c^{*}   & a^{*} a^{*}
\end{array} \right | \;  .
\eqno(A.8)
$$

In this way for the spinor matrix one find
the expression (it suffices to follow the 2-component spinor)
$$
\xi_{cyl} = B (\rho,\phi,z) \xi_{cart} \; : \;\;
B (\rho,\phi,z) =  \sigma \; \left | \begin{array}{rr}
A \; e^{+i\phi/2}   & C\;e^{-i\phi/2}     \\
- C\;e^{+i\phi/2}   &  A \; e^{-i\phi/2}  \end{array} \right | \; ,
\eqno(A.9)
$$
$$
A =  \sqrt{{ 1 + \cos \rho  \over
1 + \cos \rho \cos z  }} \; \cos {z\over 2} \;, \;\;\;
C =  \sqrt{{ 1 - \cos \rho  \over
1 + \cos \rho \cos z  }} \; \sin  {z\over 2} \; .
$$

More details on the properties of that spinor transformation see in [23].

\newpage

REFERENCES

\noindent
1. Peter F., Weyl H.  Die Vollst\"{a}ndigkeit der  primitiven Darstellungen einer
geschlossenen  kontinuierlichen Gruppe. // Math. Ann. 1927. Bd. 97. S. 737-755.
Russian translation in: On completeness of  primitive representations of a compact continuous
group. Yspexi matem. nauk. 1936. Tome  2. 144-160; Weyl H. Harmonics on  homogeneous   manifolds. Math. Ann.
1934.  Vol. 35. P. 486-499.

\noindent
2. Barut A., Ronchka R. Theory of the representations on the groups ant its
appllications. Vol. 1.  1980 (in Russian).

\noindent
3. Tretyakova N.N. Harmonic analysis in homogeneous spaces. Minsk. 1996 (in Russian).

\noindent
3.  Wigner E Е.  Group theory and its applications to the quantum theory of atomic
spectra.  New York. 1959; Moscow. 1961 (in Russian)

\noindent
4. Fano U., Racah G. Irreducible tensorial sets. New York. 1959.

\noindent
5. Edmonds A. Angular momentums in quantum mechanics. Prinston. 1957;  See in:
Deformation of atomic nucleuses. Moscow. 1958. Pages  305-351 (in Russian);

\noindent
6. Brink D.M., Satchler G.R. Angular momentum. Oxford. 1962.

\noindent
12. Bopp F., Haag.  Z.  Naturf. 1950. Bd. 5a.  S. 644.

\noindent
11. Gelfand I.M.,  Shapiro E.Ya. Represenrtations of the rotation group of the 3-dimensional space and its
application. Uspexi matem. nauk. 1952. Tome 7. Pages 2-... (in Russian).

\noindent
11.  Godement R.  A theory of spherical functions. Trans. Amer. math. Soc. 1952.
 Vol. 73. pages 496-556.

\noindent
 14. Kahan T.  Theorie des groupes en physique classique  et quantique. Tome 1. Paris, 1960.

\noindent
 16. Lyubarski G.Ya. The application of the group theory in physics. Oxford. 1960.

\noindent
 15.
Yutsis A.P., Levinson I.B., Vanagas V.V.  The theory of  angular momentum.
Vilnus. 1960 (in Russian).

\noindent
13. Schwinger J.  Quantum theory of angular momentum. ed. by
 L.C. Biedenharn and  H. van Dam. new York. 1965.

\noindent
16. Vilenkin N.Ya. Special function and the the theory  of representations of the groups.
Moscow.  1965(in Russian).

\noindent
17. Varshalovich D.A., Moskalev A.N., Xersonskiy V.K. Quantum theory of angular
momentum.  Leningrad.  1975 (in Russian).

\noindent
20. Biedenharn L.C., Lauck J.D.  Angular momentum in quantum  physics.
Volumes 1-2. Moscow.  1964 (in Russian).

\noindent
21. Junker J. E., De Vries E. A note on orthogonality and completness
of the rotation matrices.  Nucl. Phys. 1967. Vol A105. P. 621-626.

\noindent
21. Taylor M.E. Fourier series on compact Lie groups. Proc. Amer. Math. Soc.
1968. Vol. 19. P. 110301105.

\noindent
7.  Davydov A.S. Quantum mechanics. Moscow. 1973 (in Russian).

\noindent
8. Landau L.D., Lifshits  E.M. Quantum mechanics, non-relativistic theory. Moxcow. 1974 (in Rissian).

\noindent
23. Lee Beers, Richard S. Millman.  Analytic vector harmonic expansions on $SU(2)$
 and $S^{2}$. J. math. Phys. 1975. Vol. 16. P. 11-19.

\noindent
23. Vernon D. Sandberg, Tensor spherical  harmonics on $S^{2}$ and $S^{3}$
as eigenvalue problem. J. Math. Phys. 1978. Vol 19. P. 2441-2446.

\noindent
9. Gibson W., Pollard B. Symmetry principles in
particle physics. Moscow.  1979 )in Russian).

\noindent
22. Madelung E. Mathematical apparatus in physics. Moscow.  1968 (in Russian).

\noindent
10. Fedorov F.I. The Lorentz group. Moscow.  1979 (in Russian).

\noindent
23.

\noindent
23. Olevskiy M.N.  Three-orthogonal systems in spaces of constant curvature in which equation
$\Delta_{2}U + \lambda U)=0$ permits the full separation of variables.
Matem. Sbornik. 1950 P. 379-426.

\noindent
23. Red'kov V.M. Gauge transformations for a fermion function in spherical space. Riemanian
manyfold of constant positive curvature with spinor structure. Preprint   669. Institute
of Physics, Academy of Sciences of Belarus. Minsk. 1993. 35 pages.

\noindent
24.
    Red'kov V.M. On spinor structure of the pseudo Riemanian space-time and property of global
    continuity for fermionic wave functions. //Vesti AN BSSR. ser. fiz.-mat. nauk.
    1994.  3.  Pages 49-55.

\newpage

\end{document}